\documentclass[aps,prl,amsmath,amssymb,twocolumn, showpacs, superscriptaddress,10pt]{revtex4-1}

\usepackage[utf8]{inputenc}

\usepackage{amsmath}
\usepackage{hyperref}
\usepackage{graphicx}
\usepackage{amsfonts}
\usepackage{amsthm}
\usepackage{cases}
\usepackage{bm}
\usepackage{comment}

\usepackage{color}
\definecolor{Blue}{rgb}{0.00, 0.00, 1.00}
\definecolor{Red}{rgb}{1.00, 0.00, 0.00}

\newcommand{\blue}{\color{Blue}}

\hypersetup{
    colorlinks=true,       
    linkcolor=red,          
    citecolor=blue,        
    filecolor=magenta,      
    urlcolor=cyan           
}

\newcommand{\nn}{\nonumber}
\newcommand{\be}{\begin{equation}}
\newcommand{\ee}{\end{equation}}
\newcommand{\bea}{\begin{eqnarray}}
\newcommand{\eea}{\end{eqnarray}}
\def\rf#1{(\ref{#1})}

\newcommand{\grad}{{\bm{\nabla}}}

\newcommand{\beq}{\begin{equation}}
\newcommand{\eeq}{\end{equation}}
\newcommand{\beqn}{\begin{eqnarray}}
\newcommand{\eeqn}{\end{eqnarray}}

\newcommand{\fpi}{{(4\pi)^{D/2}}}

\newcommand{\g}{{\Gamma}}

\newcommand{\intqq}{{\int_{\Lambda_\ell e^{-d\ell}}^{\Lambda_\ell}{d^Dq\over(2\pi)^D}}}

\newcommand{\xb}{{\bf x}}

\newcommand{\qb}{{\bf q}}
\newcommand{\kb}{{\bf k}}
\newcommand{\pb}{{\bf p}}
\newcommand{\q}{{\bf q}}
\newcommand{\ks}{{\bf k}}

\newcommand{\rv}{{\vec r}}

\newcommand{\eps}{\epsilon}

\newcommand{\Gc}{{\cal G}}

\newcommand{\p}{{\bf p}}

\newcommand{\cH}{{\cal H}}

\def\q{\hbox{\bf q}}
\def\k{\hbox{\bf k}}

\begin{document}

\title{``Tattered'' membrane}
\author{Pierre Le Doussal}
\affiliation{Laboratoire de Physique de l'Ecole Normale Sup\'erieure, ENS, Universit\'e PSL, CNRS, Sorbonne Universit\'e, Universit\'e de Paris, 75005 Paris, France}
\email{ledou@lpt.ens.fr}
\author{Leo Radzihovsky} 
\affiliation{Department of Physics,
  University of Colorado, Boulder, CO 80309}
\email{radzihov@colorado.edu}

\date{\today}

\begin{abstract}
   Ideal crystalline membranes, realized by graphene and other atomic
  monolayers, exhibit rich physics -- a universal anomalous elasticity
  of the critical ``flat'' phase characterized by a negative Poisson
  ratio, universally singular elastic moduli, order-from-disorder and
  a crumpling transition.  We formulate a generalized $D$-dimensional
  field theory, parameterized by an $O(d)\times O(D)$ tensor field
  with an {\it energetic} longitudinal constraint.  For a soft constraint
  the resulting field theory describes a new class of a fluctuating
  ``tattered'' membranes, exhibiting a nonzero density of topological
  connectivity defects - slits, cracks and faults at an effective
  medium level. For hard, infinite-coupling constraint, the model
  reproduces the conventional crystalline membrane and its
  crumpling transition,  
  and thereby demonstrates the essence of the
  difference between an elastic membrane and conventional field
  theories. Two additional fixed points emerge within the
  critical manifold, (i) globally attractive, "isotropic"
  $O(d)\times O(D)$, and (ii) "transverse", which in $D=2$ is the exact "dual" of the
  elastic membrane. Their properties are obtained in general $D,d$
  from the renormalization group and the self-consistent screening analyses. 
\end{abstract}



\maketitle

{\bf Introduction, background and motivation}.
Fluctuating elastic membranes, first explored in the context of soft
and biological matter more than three decades ago
\cite{NP,AL,CrumplingBucklingGuitter,GDLP,LRprl,LRrapid,GuitterMC,Jerusalem,Bensimon,
  NRlett, RNpra, ML, RTtubule,LRReview},
have seen a renaissance following experimental realization of freely
suspended graphene \cite{Geim2004, GeimMacDonald,reviewRMPGraphene,
  suspendGrapheneNature2007} and other exfoliated two-dimensional
atomic monolayers.

Spectacularly, tensionless {\em crystalline} membranes exhibit a novel
``flat'' phase\cite{NP}, that spontaneously breaks the rotational symmetry
of the embedding space, in stark contrast to canonical two-dimensional
field theories for which the Hohenberg-Mermin-Wagner
theorems\cite{Hohenberg,MerminWagner,Coleman} preclude spontaneous
breaking of a continuous symmetry in two dimensions.  Such {\em
  critical} ordered state\cite{criticalMatterLR} is made possible
through a striking phenomenon of order-from-disorder, where
destabilizing thermal height fluctuations with roughness
$h_{rms}\sim L^{\zeta}$ stiffen the long-wavelength ($k^{-1}$) bending
rigidity $\kappa_0\rightarrow \kappa(k)\sim k^{-\eta}$ and soften 
{\blue the}
Young modulus, $K_0\rightarrow K(k)\sim k^{\eta_u}$.  
For a membrane of internal dimension $D$, embedded in $d$ dimensions, the universal
exponents: $\eta$, $\eta_u=4-D-2 \eta$, related exactly by the
underlying rotational invariance, the roughness exponent
$\zeta = (4-D-\eta)/2$ and the universal Poisson ratio 
$\sigma \approx -1/3$ \cite{LRprl,LRReview}  are controled by an infrared attractive fixed
point, nontrivial for $D<4$. 
This flat phase 
has been studied by a variety of
complementary methods\cite{NP,AL,GDLP,LRprl,LRReview,Gazit,MouhannaTwoLoopFlat,Burmistrovlarged,MouhannaThreeLoopFlat}, verified in
numerical simulations \cite{simulationsGraphene} and continues to be
explored experimentally \cite{experimentElasticModuli}. For a physical
membrane, $D=2, d=3$, early estimates give $\eta\approx 0.82$
and $\zeta \approx 0.59$ \cite{LRprl,LRReview}, supported by higher order
calculations with small corrections
\cite{MouhannaTwoLoopFlat,Gazit,MouhannaThreeLoopFlat}.

The other, related feature of an elastic membrane is its {\it crumpling
transition} 
from the flat phase  -- at mean-field level, the
transition is akin to a ferromagnet-to-paramagnet transition with
normals playing the role of a spin (magnetization) 
\cite{NelsonKantorCrumplingTransition,KantorNelson} -- that appears to
survive only in a ``phantom''
(non-self-avoiding)\cite{PKN,CrumplingBucklingGuitter,ALGolubovic,PaczuskiCrumplingLarged,
LRprl,LRReview,Mouhanna1,MouhannaCrumpling,OurBuckling} or
extensively perforated\cite{perforatedCrumple} membrane.  The
corresponding Landau theory is formulated in terms of a tangent tensor
field $\partial_\alpha\vec r$, encoding a ``shift'' symmetry in
addition to internal and embedding space rotational
symmetries. 
Crucially, it is the $d$-component embedding field
$\vec r(\xb)$ of the $D$-dimensional atomic reference positions
${\bf x}$, and {\em not} the tangent field itself, that is the
independent degree of freedom of the corresponding statistical
mechanics.  We will demonstrate that it is this feature that is at the
heart of all striking characteristics of the elastic membrane,
distinguishing it from a conventional $\phi^4$ field theory of an
unconstrained tensor field $\vec t_\alpha$.

To elucidate this distinction, provide new efficient calculational approach
to membrane field theory,
 and to introduce a new
fascinating class of
membranes, that we dub ``tattered'', in this Letter we study a
$D$-dimensional field theory of a $O(d)\times O(D)$ tensor order
parameter, $t_{i\alpha}$. To make contact with a membrane, we
supplement the model with an {\it energetically}-imposed constraint, 
controlled by a coupling $g$, that
for the physical 2D case tends to suppress 
transverse fluctuations
$\grad\times {\bf t}_i$, to constraint it to a tangent field, $\partial_\alpha \vec r(\xb)$.

For a non-infinite coupling $g$ of this energetic constraint the resulting
field theory describes a new class of a fluctuating ``tattered''
$D$-dimensional ``membrane'' embedded in $d$-dimensions. At an
effective medium level it encodes a nonzero density of topological
connectivity defects - slits, cracks and faults, that violate the
longitudinal constraint. 

This extension of degrees of freedom is akin
to allowing e.g., vortices in a superfluid and dislocations in a
crystal, that distinguish them from their topologically disordered
non-superfluid state and non-crystalline fluid counterparts,
respectively 
For the hard, infinite-coupling constraint $g=+\infty$ the
topological defects are excluded and the model reduces to a
conventional crystalline membrane with its well-known rich anomalous
properties\cite{PKN,CrumplingBucklingGuitter,LRReview}, reviewed
above.

We also relate our energetically constrained "spin-orbit" coupled
$O(d)\times O(D)$ model to a number of other systems studied in the
literature. These include the $O(n)\times O(m)$ symmetric
generalization of the Heisenberg field theory with no
constraint \cite{VicariNM2001}, of relevance to a large variety of physical
systems, from spinor condensates\cite{spinorBEC,MolecularBECChoi} and unconventional
superconductors\cite{He3book,FFLOlr}, to magnets with non-collinear
spin order\cite{magnets,SpinSmectic}. The Heisenberg model with dipolar interactions
effectively (asymptotically) imposing spin-transversality constraint
$\grad\cdot{\bf S}\rightarrow 0$ and leading to modified
criticality \cite{AharonyFisher,FreyIsotropicDipolar,NovelClarkFerroNematic}
is another system of close relevance.

Here we introduce and study the criticality of the tattered membrane
model using a complementary combination of the renormalization group
(RG) and self-consistent screening approximation (SCSA) 
\cite{Bray,LRprl}. We uncover its several critical points, including
the conventional crumpling transition \cite{PKN,LRprl}, the isotropic
``chiral'' Heisenberg critical point, as well as the "dual"
of the crumpling transition, a multi-component 
generalization of the  "dipolar-Heisenberg" critical point.

%

\medskip

{\bf ``Tattered'' membrane model}.
We begin by introducing a $D$-dimensional field theory of a
$O(d)\times O(D)$ tensor order parameter field, $t^i_{\alpha}({\bf x})$, with
${\bf x} \in \mathbb{R}^D$, 
$\alpha = 1,\ldots,D$ and $i = 1,\ldots,d$.
We take the Landau-Ginzburg-Wilson effective Hamiltonian
$H = \int d^Dx\cH[\vec t_{\alpha}]$, to be a power-law expansion in
the order parameter $\vec t_{\alpha}$, with energy density,
\begin{widetext}
\be
\cH = \frac{\kappa}{2} (\partial_\alpha \vec t_\alpha)^2 + \frac{g}{2}
(\partial_\gamma\vec t_\alpha)\cdot P^T_{\alpha \beta}\cdot
(\partial_\gamma\vec t_\beta)
+ \frac{\tau}{2} (\vec t_\alpha)^2
+ \frac{u}{4d} (\vec t_{\alpha} \cdot \vec t_{\beta})^{2} +
\frac{v}{4d} (\vec t_{\alpha} \cdot \vec t_{\alpha})^2+\ldots \; , 
\label{H}
\ee
\end{widetext}
where $\kappa$ is the bending rigidity and $\tau \sim T - T_c$ is the reduced temperature
that controls the transition (setting $T=1$ everywhere else, i.e. measuring 
couplings in units of temperature). For convenience we have taken our quartic couplings (related to
those of Ref.~\cite{PKN} via $u_P=u/(4 d)=\mu/(4 d)$,
$v_P=v/(4 d)=\lambda/(8 d)$) to be proportional to the Lam\'e elastic
moduli of the flat phase and to be scaled down by $d$, for the large
$d$ and SCSA expansions.\cite{NP,LRReview}. To make contact with a
model of an elastic membrane we supplemented our model with an
energetically-imposed longitudinal constraint term $g$, where
$P^T_{\alpha\beta}$ is a projection operator
$P^T_{\alpha\beta} = \delta_{\alpha\beta} -
\partial_\alpha\partial_\beta/\nabla^2$ transverse to wavevector
$k_\alpha$, penalizing fluctuations of $\vec t_\alpha$ with spatial modulation
transverse to $k_\alpha$, i.e., in 2D ($D=2$) those with a non zero
$\grad\times {\bf t}_i \equiv
\epsilon_{\alpha\beta} k_\alpha\vec t_\beta$. 

The hard, infinite-coupling $g\rightarrow\infty$ constraint,
\be
P^T_{\alpha\beta}\vec t_\beta = 0,
\label{constraint_hard}
\ee
is solved by the $d\times D$ tensor order parameter given by
\be
\vec t_\alpha = \partial_\alpha\vec r\;,
\label{t_ia}
\ee
with $\vec r(\xb) \in \mathbb{R}^d$, and atoms (in the continuum limit) labeled by their
position ${\bf x} \in \mathbb{R}^D$. With this, the
$g\rightarrow\infty$ limit of the model \rf{H} thereby reduces to that of a
conventional crystalline membrane, with its properties summarized in
the Introduction.\cite{NP,AL,PKN,CrumplingBucklingGuitter,LRReview}. We 
note that even outside of elastic membranes the $g\rightarrow\infty$ constraint (as well as its
``dual'' $\kappa\rightarrow\infty$) can naturally arise as a result of
long-range interactions, where such coupling has a singular
dependence in momentum
$g\rightarrow g(k)\sim k^{-\alpha}$, diverging as
$k\rightarrow 0$, as for example for dipolar interactions where
$\kappa(k) \sim k^{- \alpha}$ with 
$\alpha = 2$.

For a finite constraint coupling $g$ the field theory \rf{H} is
qualitatively distinct, allowing additional transverse degrees of
freedom, in 2D,
\bea \epsilon_{\alpha \beta} \partial_{\alpha} \vec
t_\beta = \grad\times\grad\vec r(\xb) &=& \vec b(\xb)\;,
\label{bdefects}
\eea
that corresponds to a non-single-valued embedding $\rv(\xb)$. At an
effective medium level $\vec b(\xb)$ encodes a nonzero density of
topological connectivity defects - slits, cracks and
faults\cite{LRslits}, depending on its orientation, that violate the
longitudinal constraint akin to e.g., vortices in a superfluid and
dislocations in a crystal.\cite{KT,HNmelting,Youngmelting} Thus, model \rf{H}
describes a new class of fluctuating ``tattered'' -- liquid-crystal-like --
$D$-dimensional membranes' embedded in $d$-dimensions.

Next we turn to the RG and SCSA analyses of the tattered membrane model, \rf{H},
focussing on the ordering phase transition, at $\tau_R = 0$ criticality,
controlled by the RG flow of the couplings $u, v$ and of the ratio 

%
\be
s\equiv \kappa/g \;.
\ee

As detailed below, we find that
 (i) The $s\rightarrow 0$
(i.e., $g\rightarrow\infty$) limit corresponds to the crumpled-to-flat
phase transition\cite{PKN} of a conventional elastic membrane
and provides valuable checks for the calculations in this
new formulation. (ii)
The $s\rightarrow + \infty$ limit is the ``dual'' of the crumpling
critical point (in a sense discussed below) which is also universal
and 
describes spin systems
asymptotically constrained, e.g by dipolar interactions.
 (iii) For a
non-zero and non-infinite $s$, we have a more general criticality as a
function of $s$, that to lowest order in $\epsilon = 4-D$ is a fixed
line. (iv) For $s=1$ the criticality is spatially isotropic
and is related to the ``chiral'' critical point of the $O(n)\times O(m)$
model \cite{VicariNM2001}, 
with $n=d$, but with the additional spin-orbital constraint
that $m=D$ the space dimensionality. However,
more generally, as we will detail below, $s(\ell)$ flows as a function of the
RG coarse-graining scale $b=e^\ell$, with its fate controlled by a
critical point $s_*$. We find that $s_*=1$ describes the criticality 
at large scales, unless the bare moduli $\kappa,g$ 
have singular small $k$ behavior, as alluded to in the Introduction.


\medskip

{\bf Renormalization-group analysis of the ``tattered'' 
  membrane criticality}. We now analyze the tattered membrane model \rf{H}, $H = H_0 + H_v$,
focussing on the critical point $\tau_R=0$. To this end, it is
convenient to write the harmonic part
$H_0=\int_\kb\cH_0[\vec t_\alpha(\kb)]$ in $D$-dimensional Fourier space
$\kb$, and the quartic energy 
$H_v = \int d^D x \cH_v[\vec t_\alpha(\xb)]$ in real space, with
\bea
&& \cH_0
=\frac{1}{2}G^{-1}_{\alpha\beta}(\kb) \vec t_\alpha(\kb)\cdot\vec
t_\beta(-\kb)\;, \\
&& \label{HR} 
\cH_v = \frac{1}{4 d} R_{\alpha\beta\gamma\delta}~
{\vec t}_\alpha\cdot{\vec t}_\beta~
{\vec t}_\gamma\cdot{\vec t}_\delta
\eea
where $\int_\kb$ denotes $\int d^Dk/(2 \pi)^D$, and we
introduced the bare inverse propagator and four-point 
tensorial vertex,
\bea 
 G^{-1}_{\alpha\beta}(\kb)
&=&\left( \kappa P^L_{\alpha \beta}(\kb) 
  + g P^T_{\alpha \beta}(\kb) \right) k^2\;,
\label{invG}\\
R_{\alpha \beta,\gamma \delta}
&=&\frac{u}{2}(\delta_{\alpha\gamma}\delta_{\beta\delta}
+\delta_{\alpha\delta}\delta_{\beta\gamma}) + v
\delta_{\alpha\beta}\delta_{\gamma\delta}\; .
\label{vertexR}
\eea
The bare correlator of the vector
field $t^i_\alpha$ is given by
\be
\langle t^i_\alpha(\kb)  t^j_\beta(\kb) \rangle_0 = \delta_{ij}
G_{\alpha \beta}(\kb) (2 \pi)^d \delta^d(\kb+\kb')\;,
\label{G}
\ee 
with the bare propagator
\be
G_{\alpha \beta}(\kb) = \frac{P_{\alpha \beta}^L(\kb)}{\kappa k^2}
+ \frac{P_{\alpha \beta}^T(\kb)}{g k^2}\;.
\ee
In the limit $g\rightarrow +\infty$ the transverse component is
suppressed, $\vec t_\alpha = \partial_\alpha\vec r $ solves the
constraint, and one recovers the standard defect-free elastic membrane
model\cite{PKN,LRReview}, as discussed in the Introduction.

Standard momentum shell RG analysis (which involves integrating out
short-scale fields in a shell of momenta, $\Lambda_\ell e^{-d\ell} < k < \Lambda_\ell$
where $\Lambda_\ell=\Lambda e^{-\ell}$ is the running UV cutoff) 
gives a one-loop vertex correction at zero vertex momentum,
\begin{widetext}
\bea 
\delta R_{\alpha\beta\gamma\delta}
&=&-\frac{1}{d}\left[d R_{\alpha\beta\sigma\kappa}R_{\sigma'\kappa'\gamma\delta}
+ 4 R_{\alpha\beta\sigma\kappa}R_{\sigma'\gamma,\kappa'\delta}
+ 2 R_{\alpha\sigma\kappa\gamma}R_{\beta\sigma'\kappa'\delta}
+ 2 R_{\alpha\sigma\kappa\gamma}R_{\beta\kappa'\sigma'\delta}\right]
\int_\kb G_{\sigma\sigma'}(\kb)G_{\kappa\kappa'}(-\kb).
\label{deltaV}
\eea

As a check, the above form agrees with a standard $4(N+8)$ symmetry factor
in the $O(N=d)$ model, corresponding to a scalar vertex and an
isotropic propagator.  Another check is that for $u=0$ and $g =\kappa$
the theory exhibits $O(N=d D)$ symmetry and thus \rf{deltaV} reduces
to the $d D + 8$ symmetry factor.

\end{widetext}
  

As in a conventional $O(N)$ model, because the vertex is independent
of momentum, a singular momentum-dependent correction to the inverse
propagator only appears at two-loop order, i.e.,
$\eta_\kappa = \eta_g = 0$ to one-loop order that we focus on here.
The SCSA of the next section will give nonzero anomalous exponents
since it is computed for an arbitrary $D$.

At a vanishing vertex momentum, the loop integral is rotationally
invariant and can be easily computed over the spherical momentum
shell. Performing the angular averages
$\langle\hat k_\alpha \hat k_\beta\ldots\rangle$ appearing in \rf{deltaV}
and the index contractions, we obtain the corrections
to the couplings, $\delta u$, $\delta v$. This allows one to obtain the RG flows 
on the critical, $\tau_R=0$, $u-v$ surface. As usual for the quartic model, 
the upper critical dimension
is $D=4$, and thus we expand in $\epsilon=4-D$. Defining the 
dimensionless couplings
$\hat u = C_D \Lambda_\ell^{-\epsilon} \, u/\kappa^2$,
$\hat v = C_D \Lambda_\ell^{-\epsilon}  \, v/\kappa^2$ 
with $C_4=1/8 \pi^2$ we 
obtain the RG flows:
\bea \label{rg01} 
d\hat u/d\ell &=& \epsilon \hat u - a_1 \hat u^2 - a_2 \hat v^2 - a_3 \hat u \hat v\;,\\
d \hat v/d\ell &=& \epsilon \hat v - b_1 \hat u^2 - b_2 \hat v^2 - b_3
\hat u \hat v\;, \label{rg02}  \\
d s/d\ell &=& [\eta_\kappa(\hat u,\hat v,s) - \eta_g(\hat u,\hat
v,s)]s\;, \label{rgeta}  \eea
The coefficients
$a_1, a_2, a_3, b_1, b_2, b_3$ are functions of $s$ and $d$ given in 
\cite{SM}. Here $\eta_{\kappa,g}(\hat u,\hat v)$ are the coarse-graining
corrections to $\kappa,g$, that become anomalous exponents at the
fixed point $\hat u_*, \hat v_*$, $s_*$.  As alluded to above, to
one-loop order $\eta_{\kappa,g}(\hat u_*,\hat v_*,s) =0$ vanish at
$s> 0$ and thus $s$ does not flow, giving a fixed line of critical
points labelled by $s$.

%

These RG equations pass a few simple checks. For $s=1$, i.e. $\kappa= g$,
$a_2$ vanishes so that $\hat u=0$ is preserved, which 
corresponds to an $O(d D)$ symmetric model, with $\hat v^*=d/(8+4 d)$ 
the Wilson-Fisher critical point. For $d=1$ the $u$ and $v$ terms in 
\eqref{vertexR} cannot be distinguished, and \eqref{rg01},\eqref{rg02}
reduce to a flow fo the combined
coupling 
\be 
d(\hat u+\hat v)/d\ell = \epsilon (\hat u+\hat v) 
-\frac{1}{2} (s (17 s+2)+5) (\hat u+\hat v)^2
\ee
which always admits an attractive fixed point 
$\hat u+\hat v = \frac{2 \epsilon}{s (17 s+2)+5}$
In the transverse only limit $\kappa \to +\infty$, i.e. $s \to +\infty$
one can check \cite{SM} that it recovers the RG equation
of the Heisenberg model with isotropic dipolar interactions 
\cite{FreyIsotropicDipolar}, which corresponds to a $q$-dependent bare $\kappa(q) = g + c/q^2$.
The non-analytic structure of $\kappa(q)$ forbids graphical corrections to the 
amplitude $c$, explaining the connection to the $\kappa = +\infty$ limit of our RG equations.
Our result for $d=1$ are however valid for a general $s$.

For general $d$, insight is gained by first considering the large $d$ limit of these equations,
which read
\bea  \label{RGlargeds0} 
 d\hat u/d\ell  &=& \epsilon \hat  u -\frac{7 s^2+4 s+1}{12}  \hat u^2 \\
  d\hat v/d\ell  &=& \epsilon \hat  v -\frac{(s-1)^2 \hat u^2}{24}
    - (3
   s^2+1) \hat v^2 
-\frac{3 s^2 +1}{2}  \hat u \hat v \nn
   \eea 
For any given $s$ there are four fixed points (FP). The (fully unstable) Gaussian FP, 
$P_1: \hat u=\hat v=0$, and
\bea  \label{FPs0} 
&& P_2: ~ \hat u=0 \quad , \quad \hat v= \frac{\epsilon}{1 + 3 s^2}  \\
&& P_3 : ~ \hat u= \frac{12 \epsilon}{s (7 s+4)+1}    
\quad , \quad \hat v=\frac{-2 (s-1)^2}{\left(3
   s^2+1\right) \left(7 s^2+4
   s+1\right)}
 \nn  \\
&& P_4: ~  \hat u= \frac{12 \epsilon}{s (7 s+4)+1}  \quad , \quad \hat v= \frac{-3 \epsilon}{s (7 s+4)+1} \nn
\eea 
where only $P_3$ is fully stable with eigenvalues $(-\epsilon,-\epsilon)$,
while $P_2$ and $P_4$ have eigenvalues $(\pm \epsilon, \mp \epsilon)$
respectively. A study of the finite $d$ RG equations 
\eqref{rg01},\eqref{rg02} shows \cite{SM}
that this fixed point structure remains qualitatively
the same for $d$ large enough. 
The FP $P_3$ moves but remains stable for $d>d_c(s)$
when it annihilates with the finite $d$ extension of $P_4$,
so that there is no stable FP for $d_c(s) > d > d_c^-(s)$
(these special embedding dimensions $d_c(s)$ and $d_c^-(s)$
are computed in \cite{SM}, see Fig. \ref{dcsFig} there. Since for $d=1$ there is
always a stable FP, 
discussed above, it implies $d_c^-(s)>1$,
and in fact we find $d_c^-(s) \gtrsim 1$.

Let us now study in more details the three cases $s=0,1,+\infty$.
In some cases they connect to previously studied models.


{\it Elastic membrane}. For $s=0$ ($g\rightarrow\infty$), a single-value surface constraint on
the tangent vector $\vec t_\alpha$, requires the defect density $\vec b$
in \rf{bdefects} to vanish, and the flow equations \eqref{rg01},\eqref{rg02} 
reduce to that for the
crumpling transition of a polymerized defect-free membrane:
\bea
d\hat u/d\ell &=& \epsilon\hat u 
- \frac{d+21}{12 d} \hat u^2 
- \frac{1}{3 d} \hat v^2
- \frac{5}{3 d}\hat u \hat v,\\
d\hat v/d\ell &=& \epsilon \hat v
- \frac{d+15}{24 d}\hat u^2 
- \frac{6d+7}{6 d}\hat v^2 
- \frac{3d+17}{6 d}\hat u \hat v\;,\;\;\;
\eea
These equations are equivalent to those obtained
by Paczuski, Kardar and Nelson\cite{PKN} 
(within our definitions of $u,v$ given above in \eqref{HR},
when reexpressed in terms of their $4(\hat v-\hat u/D), 4 \hat u$ couplings).
In the $d=+\infty$ limit the fixed points are those given in \eqref{FPs0}
setting $s=0$. These FP's are the counterparts, within the
critical manifold, of those obtained in the $D=4-\epsilon$ RG 
for the flat ordered phase \cite{AL}. 
The only fully stable FP $P_3: \hat u^*= 12 \epsilon,\hat v^* = -2\epsilon$
describes the crumpling transition of crystalline membranes. 
At the FP $P_2: \hat u^* = 0, \hat v^* = \epsilon$ 
(unstable to $P_3$) 
the quartic interaction in \eqref{H} 
has the enhanced $O(dD)$ Heisenberg symmetry,
but with an anisotropic propagator corresponding to $g=+\infty$.
Hence it can be termed "dual" of the "dipolar $O(dD)$ Heisenberg". 
It also represents the crumpling transition
of a "fixed connectivity fluid" with zero 
renormalized shear modulus $\mu=0$,
a situation which occurs in nematic elastomers
\cite{LRelastomer,LubenskyElastomer}. 
Finally, $P_4: \hat u^*= 12 \epsilon,\hat v^* = -3\epsilon$
(unstable to $P_3$)
is also at the boundary of physical stability,
which requires a positive bulk modulus $2 \mu + D \lambda \geq 0$, i.e.
$\hat u + 4 \hat v \geq 0$.
At finite $d$ the structure of these FP's remains similar
as long as $d> d_c=d_c(0)= 218.206$ (see 
\cite{SM} for a precise calculation of $d_c$).
For $d<d_c$ the RG flow has a runaway, often
interpreted as a first-order transition \cite{PKN}.
We also find $d^-_c(0)= 1.12657$ below which
a critical point reemerges, consistent
with our analysis for $d=1$.

{\it Isotropic "tattered membrane"}
For $s = 1$ ($g = \kappa$) the propagator is isotropic
and our RG equations become
\bea \label{RGs1} 
&& \partial_\ell \hat u = \epsilon \hat u - \frac{\hat  u}{d} \left(  (d+8) \hat  u+12
   \hat  v \right) \, , \\
&&  \partial_\ell \hat v = \epsilon \hat v - \frac{1}{d} 
(\hat  u+2 \hat  v) (2 (d+2)
   \hat  v+3 \hat  u) \, .
\eea 
For $d \to +\infty$ the stable fixed point $P_3$ has $\hat u=\epsilon$ and
$\hat v=0$. It exists for $d \geq d_c(1)= 22+12 \sqrt{3} = 46.1284$,
see \cite{SM} (and for $d<d_c^-(1)=22-12 \sqrt{3} = 1.21539$). 
For $s=1$ our tattered membrane model corresponds to a
$O(n) \times O(m)$ magnet, with $n=d$ and  where the number of components $m$ 
is tied to the space dimension $D$. The $O(n) \times O(m)$ magnet
has been much studied, see e.g. 
\cite{VicariNM2001} for a three-loop order study,
and the isotropic fixed point $P_3$ that we obtain here for $s=1$ 
corresponds to the "chiral" fixed point in \cite{VicariNM2001}
(while $P_2$ corresponds to the 
Heisenberg $O(m n)$ fixed point, and $P_4$ to
the "anti-chiral" FP). One can check that our RG equations \eqref{RGs1},
fixed points and results for $d_c(1)$, $d^-_c(1)$, are consistent with those obtained in
\cite{VicariNM2001} setting $m=4$ there. That work can further
be used to predict, to next order in $\epsilon$, 
$\eta = \frac{5}{4 d} \epsilon^2 + O(1/d^2)$,
which we will also obtain via SCSA below.

{\it Transverse, "dual'' of  elastic membrane}. For $s \to +\infty$, i.e.
for $\kappa \to +\infty$, the $\vec t_\alpha$ vector field is
purely transverse (its divergence is constrained to vanish, $\partial_\alpha \vec{t}_\alpha=0$). 
To study this limit the natural dimensionless couplings are 
$\bar u = C_4 \Lambda_\ell^{-\epsilon} \, u/g^2 = s^2 \hat u$, and
similarly $\bar v  = s^2 \hat v$, and Eqs.
 \eqref{rg01},\eqref{rg02} then lead to the flow
\bea
\partial_\ell \bar u &=& \epsilon \bar u
-\frac{(7 d+69) \bar u^2}{12 d}-\frac{23 \bar u \bar v}{3 d}-\frac{\bar v^2}{3 d}
\\
\partial_\ell \bar v & = &  \epsilon \bar v  
-\frac{(d+51) \bar u^2}{24 d}-\frac{(9 d+47) \bar u \bar v}{6
   d}- \left(\frac{31}{6 d}+3\right) \bar v^2 \nn
\eea 
For $d=+\infty$ the stable fixed point 
$P_3:  \bar u= \frac{12 \epsilon  }{7 } , \bar v= - \frac{2}{21} \epsilon$
describes the "dual" of the crumpling transition of membranes. 
The FP $P_2: \bar u=0, \bar v= \frac{\epsilon}{3}$  
(unstable to $P_3$) has again the enhanced $O(dD)$ Heisenberg symmetry,
but with an anisotropic propagator corresponding to $\kappa=+\infty$.
For $d>1$, it can be seen as a multi-component extension of the 
dipolar Heisenberg model of \cite{FreyIsotropicDipolar} discussed above for $d=1$. 
Finally, $P_4:  \bar u= \frac{12 \epsilon  }{7 }    \quad , \quad \bar v= -\frac{3 \epsilon}{7}$
(instable to $P_3$) 
is dual to the $P_4$ FP for membranes.
For arbitrary $d$, we find that the stable FP $P_3$ 
exists only for $d> d_c(+\infty) = 61.31$
or $d< d_c^-(+\infty)=1.042$.


Beyond one-loop, below $D=4$, we expect $s(\ell)$ to flow under the RG
and thus the criticality is controlled by $s_*$.  On symmetry grounds
it is natural to expect that for non-zero and non-infinite $s$, it flows to 
the isotropic fixed point value $s_*=1$. To derive this flow 
within a $1/d$ expansion and for any $D<4$, we next analyze the same tattered membrane
$O(d)\times O(D)$ transition, using the self-consistent screening
approximation (SCSA) applied to the
Hamiltonian in \rf{H}. 

{\bf SCSA analysis of the ``tattered'' membrane: crumpling transition}.
The SCSA 
is a self-consistent
resummation of the expansion in large embedding space dimension $d$,
and is exact for any $D$ to order $O(1/d)$ \cite{Bray,LRprl}. It aims 
to determine the exact field correlator
${\cal G}_{\alpha \beta}(\kb)$ 
\be 
\langle t^i_\alpha(\kb)  t^j_\beta(\kb) \rangle = \delta_{ij}
{\cal G}_{\alpha \beta}(\kb) (2 \pi)^d \delta^d(\kb+\kb')\;,
\label{Gexact}
\ee 
a counterpart of the bare correlator ${G}_{\alpha \beta}(\kb)$
 defined in \eqref{G}. Focusing on the
critical point $\tau_R=0$, it gives two coupled closed integral 
equations for the field correlator ${\cal G}_{\alpha \beta}(\kb)$
(and its associated self-energy ${\cal G}^{-1}_{\alpha \beta}(\kb) - G^{-1}_{\alpha \beta}(\kb)$)
%
and for the dressed (i.e. renormalized) quartic vertex $\tilde R_{\alpha \beta, \gamma \delta}(\q)$,
which acquires momentum dependence. These coupled SCSA integral equations are given by
\bea
&& {\cal G}^{-1}_{\alpha \beta}(\kb) - G^{-1}_{\alpha \beta}(\kb) = {2 \over d} \int_q 
\tilde{R}_{\alpha \beta, \gamma \delta}(\q) 
{\cal G}_{\beta \delta}(\kb - \q)\;, \label{sigma}\\
&& \tilde{R}_{\alpha \beta, \gamma \delta}(\q)  R_{\alpha \beta, \gamma \delta}
- R_{\alpha \beta, \alpha' \beta'}  \Pi_{\alpha' \beta',\gamma',\delta'}(\q) 
\tilde{R}_{\gamma',\delta', \gamma \delta}(\q) ,  \nn
\eea
where $R_{\alpha \beta, \gamma \delta}$ is the bare vertex defined
in \eqref{vertexR}.
The tensor $\Pi_{\alpha \beta, \gamma \delta}(\q)$ is the "vacuum polarization 
bubble" which encodes the screening of the interactions
by the fluctuations
\be \Pi_{\alpha \beta,
  \gamma \delta}(\q) = {\rm sym}_{\alpha \beta} {\rm sym}_{\delta
  \gamma} \int_p \Gc_{\alpha \gamma}(\pb) \Gc_{\beta \delta}(\q-\pb) \, .
  \label{VP} 
\ee
Since we tune to the critical point $\tau_R=0$,
the integral in the r.h.s. of the first equation in \eqref{sigma} 
is to be understood with its value at $\k=0$ subtracted (for the
same reason, the tadpole term does not appear in \eqref{sigma}).
Although that self-consistent 
closure is not fully controlled, it gives accurate results 
for membranes \cite{LRReview} as it includes the physics of screening.

We now look for a solution which describes the
critical manifold, i.e., for a propagator which behaves as a power law at small momentum $k \to 0$
\be
{\cal G}_{\alpha \beta}(\kb) \simeq \frac{P_{\alpha \beta}^L(\kb)}{Z_\kappa k^{2-\eta}} 
+ \frac{P_{\alpha \beta}^T(\kb)}{Z_g k^{2-\eta}} \label{formGtext} 
\ee 
and with two a priori distinct amplitudes $Z_\kappa$ and $Z_g$ for each component.
While each is non-universal their ratio may be, and in this part we
again denote $s= \frac{Z_\kappa}{Z_g}$, a dressed (or renormalized) 
version of $\kappa/g$. Inserting the form \eqref{formGtext} 
into \eqref{VP} and performing the integrals leads to
a divergent form $\Pi(\qb) \propto q^{-(4-D-2\eta)}$,
which upon inversion of the second SCSA equation
in \eqref{sigma} yields the screened
interaction $\tilde R(\qb) \propto q^{4-D-2\eta}$.
Both tensors acquire a complicated momentum 
structure, with non trivial amplitudes, worked out in \cite{SM},
and here we will only display the result. 
Inserting the obtained $\tilde R(\qb)$, together
with \eqref{formGtext} into the r.h.s of the first SCSA equation
in \eqref{sigma}, and performing all
integrals yields a self energy $\propto k^{2-\eta}$.
Equating the longitudinal and the transverse components
on both sides of \eqref{sigma},
respectively yields two self-consistency conditions 
\bea \label{scF0} 
 \frac{d}{2} = F(\eta, D, s)  \quad , \quad  \frac{d}{2} = F_1(\eta, D, s) \, .
\eea 
The functions $F$ and $F_1$
are given explicitly in \cite{SM}. The system \eqref{scF0} determines the 
the possible fixed point values of the pair $(\eta,s)$ for any given $D<4$.
In $D=2$ it obeys a remarkable duality relation $F_1(\eta,D=2,s) = F(\eta,D=2,1/s)$,
which originates from the exact duality in momentum space, $\q \leftrightarrow \q^T
\equiv \hat {\bf z} \times {\bf q}$,
leading to $P^T(\q) \leftrightarrow P^L(\q)$ (since in $D=2$ one has 
$P^T_{\alpha,\beta}(\q)=\q^T_\alpha \q^T_\beta$). Using this duality in $D=2$
we predict and find \cite{SM} that the only solution of \eqref{scF0} at finite $s>0$ is at
$s^*=1$. In fact the latter result extends to any $D<4$ \cite{SM}, hence we find from SCSA that the
only fixed points are at $s^*=0,1,+\infty$. Furthermore, from the SCSA we 
derive the RG flow, exact to $O(1/d)$, for the ratio $s$
\be 
\partial_\ell s =  \frac{2}{d} \left[ f(D,s) - f_1(D,s) \right] \, s  \, , \label{flows0} 
\ee 
where $f_1(D,s)  = \lim_{\eta \to 0}  \eta F_1(\eta,D,s)$, 
$f(D,s)  = \lim_{\eta \to 0}  \eta F(\eta,D,s)$. Analysis of this equation
confirms that $s^*=1$ is the only stable FP and $s^*=0$, $=+\infty$ 
are unstable to $s^*=1$ under a small perturbation of $s$. Let us 
now give further results about these FP. 

{\it Crumpling transition}. For $s=0$ only
the first equation in \eqref{scF0} holds (see \cite{SM}),
i.e. $\frac{d}{2} = F(\eta, D, 0)$, which 
recovers the SCSA self-consistency condition
for the crumpling transition \cite{LRprl}.
The predictions were much detailed in \cite{LRReview},
e.g. one finds $\eta=\eta_{cr}(D,d)$ with
$\eta_{cr}(D=3,d=2) =0.5352$. From
\eqref{flows0} we find within SCSA that the FP $s^*=0$ 
is unstable, i.e. at small $s$, $\partial_\ell s = \omega_0 s$ 
where $\omega_0=\eta_{cr}$. 

{\it Isotropic critical point, "tattered"}. For $s=1$ one finds
that $F_1(\eta,D,s=1) = F(\eta,D,s=1)$, hence the two
equations in \eqref{scF0} reduce to a single one for $\eta$,
displayed and analyzed in \cite{SM}
(e.g. the SCSA gives 
$\eta_{D=3,d=3} \approx  0.136$,
$\eta_{D=3,d=4} \approx  0.11$).
Expanding, one finds
$\eta \simeq \frac{2}{d} C(D) + O(d^{-2})$,
where $C(D)$ vanishes for $D=2,4$, 
with $C(3)=\frac{8}{3 \pi^2}$ and $C(4-\epsilon) \simeq
    \frac{5}{8} \epsilon^2$. 
    Again, this agrees with the results of \cite{VicariNM2001}
    for the "chiral" critical fixed point of the $O(n) \times O(m)$ model, 
    identifying $n=d$ and $m=D$.
    The vanishing of $\eta$ in $D=2$ to this order is consistent
with a lower critical dimension $D_{lc}=2$ for the tattered membrane 
critical point. However, the stability of the ordered phase
remains an open problem.
Finally, we find that the isotropic FP at $s^*=1$ is stable,
i.e. $s-1 \sim e^{\omega_1 \ell}\sim k^{-\omega_1}$ 
with $\omega_{1,D=3,d} \simeq - 44/(15 d \pi^2)$
and $\omega_{1,4-\epsilon,d} \simeq - 7 \epsilon^2/(9 d)$
($\omega_1=0$ for $D=2$), hence $s(\ell)$ flows to $1$ at large scale.

{\it Transverse critical fixed point}.
The fixed point at $s^*=+\infty$ corresponds
to $\kappa \to +\infty$ and $\vec t_\alpha({\bf k})$ 
transverse to $k_\alpha$. In $D=2$, the
aforementioned exact duality in momentum space
maps this case to a purely longitudinal
$\vec t_\alpha({\bf k})$, i.e. onto a membrane and
the crumpling transition, with $\eta_{tr}(2,d)=\eta_{cr}(2,d)$. This mapping extends
to the low-temperature phase, and thus predicts that this "transverse" model
exhibits an ordered phase in $D=2$, dual of the flat phase 
of crystalline membranes (with identical exponents). 
The SCSA self-consistency condition reduces 
to the second equation in \eqref{scF0},
i.e. $\frac{d}{2} = F_1(\eta, D, +\infty)$.
For general $D$ there is no duality mapping to crumpling,
and the SCSA predicts an $\eta_{\rm tr}$ exponent generally smaller than $\eta_{cr}$,
except very near $D=4$,
e.g. $\eta_{tr,D=3,d=3} =0.192$, $\eta_{tr,D=3,d=4} =0.163$
while 
$\eta_{cr,D=3,d=3} =0.236$, $\eta_{cr,D=3,d=4} =0.209$.
At large $d$ we obtain 
$\eta \simeq \frac{2}{d} C_{\rm tr}(D)$, where $C_{\rm tr}(2)=C_{\rm cr}(2)=1$ as for crumpling,
while $C_{\rm tr}(3)=176/(35 \pi^2) \approx 0.51$, significantly smaller than
$C_{cr}(3)=48/(5 \pi^2) \approx 0.97$ for the crumpling transition.
By contrast, $\eta_{tr,D=4-\epsilon,d} \simeq  \frac{16 \epsilon^2}{9 d}
> \eta_{cr,D=4-\epsilon,d} \simeq \frac{25}{3 d} \epsilon^3$.
The SCSA predicts that the lower critical dimension 
(solution of $2 - \eta(D_{lc},d)=D_{lc}$)
is the root $D=D_{lc}$ of
\be 
d = \frac{D \left(D^4-D^3-7 D^2+5
   D+8\right)}{2 (D-2) (D-1)
   \left(D^2-D-1\right)} \, , 
\ee 
leading to $D_{lc}(d) \simeq  2 -\frac{2}{d} +O(d^{-2})$ (exact), and e.g.
$D_{lc}(3) = 1.677 ~,~
D_{lc}(2) = 1.672 ~,~
D_{lc}(1) = 1.667$
while it predicts $D_{lc}(d)= 2 d/(1+d)$ for crumpling
\cite{LRReview}.
Finally, this transverse FP is unstable at
finite $s$, i.e. for $s \gg 1$, 
$\partial_\ell (s^{-1}) = - \omega_{\infty} s^{-1}$
with $\omega_{\infty,D=2,d}=\omega_{0,D=2,d}$
and for general $D$, $\omega_{\infty,D,d}
= - \frac{2}{d} C_{\rm tr}(D) + O(d^{-2})$. 

{\bf Summary and conclusion.}
We introduced a generalized $O(d)\times O(D)$ field-theoretic description of the 
statistical mechanics of fluctuating manifolds --"tattered"
membranes, that allow for a finite density of unbound connectivity
defects, such as slits, cracks and faults,
in an elastic sheet. The model is formulated in
terms of a tensor  field $\vec t_\alpha$ controlled by an energetic constraint,
which, for a hard constraint recover known results for the crumpling
transition of an elastic membrane. 

For a soft constraint, i.e. $g=O(1)$,
we have shown that 
the criticality of the tattered membrane ultimately behaves at large scale
as a critical $O(d) \times O(D)$ magnet. This suggests 
$D_{lc}=2$ as the lower-critical dimension, implying that 
the unbound connectivity defects destroy the crumpling
transition for physical 2D membranes at large scale,
while there is a non trivial critical point for $2<D<4$. 
In addition, our study has unveiled an interesting transverse "dual" 
analog of elastic membranes,
which can be thought as a multi-component generalization
of the dipolar Heisenberg model. We calculated
its non trivial critical exponent for any $D$,
and predicted that it exhibits an ordered phase
that extends down to a lower critical dimension $D_{lc}(d)<2$
that we computed.  Finally, we
have shown that for $g \ll1$ or $\kappa \ll 1$ the defected 
membrane, or its dual, survives until a scale
that we determined. We hope these predictions
stimulate numerical and experimental studies on this
new class of defected
membranes.

{\em Note Added:} 
After this work was completed, a related paper
appeared, arXiv:2307.00600 by L. Delzescaux, C. Duclut,
D. Mouhanna, M. Tissier \cite{Mouhanna}, with
an impressive three-loop calculation. The idea of parameterizing a
membrane with a $O(d)\times O(D)$ tensor field with a longitudinal
constraint (though constraint is applied differently) is common
between the two works. 
On the other hand, our work extends the model to a "soft"
energetic constraint, and thereby introduces a new class
of "tattered" membranes with connectivity defects.

{\it Acknowledgments} We thank D.R. Nelson for
stimulating discussions about tattered membranes.
LR acknowledges support the Simons Investigator
Fellowship through the James Simons Foundation, and thanks \'Ecole
Normale Sup\'erieure for hospitality. 
Both authors thank
KITP for hospitality, supported in part by the National Science
Foundation under Grant No. NSF PHY-1748958 and PHY-2309135.

{}


\newpage

.

\newpage

\begin{widetext}

\begin{large}
\begin{center}

Supplementary Material for {\it Tattered membranes}

\end{center}
\end{large}

\bigskip


We give the principal details of the calculations described in the
main text of the Letter.

\section{I. Renormalization group analysis around $D=4$} 

In this section we give the details of the one-loop RG analysis 
of a $D=4-\epsilon$ dimensional tattered membrane, described by the effective Hamiltonian 
\eqref{H}.

\subsection{Derivation of the RG equations}
We will analyze the model introduced in the text at the critical point $\tau_R=0$ (assuming
that it exists and that one can tune to it). We do not write the bare $\tau$ term since
it does not play a role in the calculation. The effective Hamiltonian reads 
\bea
{H}[\vec t_\alpha]&=&\frac{1}{2} \int_{\kb}
G^{-1}_{\alpha\beta}(\kb) \vec t_\alpha(\kb)\cdot\vec t_\beta(-\kb) 
+\frac{1}{4 d}\int_{\bf x} R_{\alpha\beta\gamma\delta}~
{\vec t}_\alpha\cdot{\vec t}_\beta~
{\vec t}_\gamma\cdot{\vec t}_\delta,\\
&=&\frac{1}{2} \int_{\kb}
G^{-1}_{\alpha\beta}(\kb) \vec t_\alpha(\kb)\cdot\vec t_\beta(-\kb) 
+\frac{1}{4 d}\int_{\bf x} \tilde R^{ijkl}_{\alpha\beta\gamma\delta}~
t^i_\alpha t^j_\beta~
t^k_\gamma t^l_\delta,\\
\eea
where we use $\int_\kb$ to denote $\int d^Dk/(2 \pi)^D$, and the four-point unsymmetrized and symmetrized
tensorial quartic vertex have the form:
\bea 
R_{\alpha \beta,\gamma \delta}
&=&\frac{u}{2}(\delta_{\alpha\gamma}\delta_{\beta\delta}
+\delta_{\alpha\delta}\delta_{\beta\gamma})
+ v \delta_{\alpha\beta}\delta_{\gamma\delta},\\
\hat R^{ijkl}_{\alpha \beta,\gamma \delta}
&=&\frac{1}{3}(R_{\alpha \beta,\gamma \delta}\delta_{ij}\delta_{kl}
+R_{\alpha \gamma,\beta \delta}\delta_{ik}\delta_{jl}
+R_{\alpha \delta,\gamma \beta}\delta_{il}\delta_{kj}).
\label{Vparam}
\eea

The bare tangent-tangent correlator
is given by
\be
\langle t^i_\alpha(\kb)  t^j_\beta(\kb) \rangle_0 = \delta_{ij} G_{\alpha \beta}(\kb) (2 \pi)^d \delta^d(\kb+\kb') \quad , \quad 
\ee 
with the bare propagator
\be
G^{-1}_{\alpha \beta}(\kb) = \left( \kappa P^L_{\alpha \beta}(\kb) + g P^T_{\alpha \beta}(\kb) \right) k^2 \quad , \quad
G_{\alpha \beta}(\kb) = \frac{P_{\alpha \beta}^L(\kb)}{\kappa k^2} + \frac{P_{\alpha \beta}^T(\kb)}{g k^2}.
\ee

Standard momentum shell RG analysis (which involves integrating out
short-scale fields in a shell of momenta, $\Lambda_\ell e^{-d\ell} < k < \Lambda_\ell$
where $\Lambda_\ell=\Lambda e^{-\ell}$ is the running UV cutoff) 
gives a one-loop vertex correction at zero vertex momentum,
illustrated in Fig. \ref{dRgraphFig} (the calculation 
is most efficiently done using the symmetrized vertex $\hat R$
where all the channels are automatically taken into account)

\bea \label{Vcorrection} 
\delta R_{\alpha\beta\gamma\delta}
&=&-\left[d R_{\alpha\beta\sigma\kappa}R_{\sigma'\kappa'\gamma\delta}
+ 4 R_{\alpha\beta\sigma\kappa}R_{\sigma'\gamma,\kappa'\delta}
+ 2 R_{\alpha\sigma\kappa\gamma}R_{\beta\sigma'\kappa'\delta}
+ 2 R_{\alpha\sigma\kappa\gamma}R_{\beta\kappa'\sigma'\delta}\right]
\int_\kb G_{\sigma\sigma'}(\kb)G_{\kappa\kappa'}(-\kb).
\eea
As a check, structurally this agrees with the standard $4(N+8)$ factor
in the $O(N)$ model, with here $N \to d$, corresponding to scalar vertex and isotropic
propagator.  Another check is that for $u=0$ and $g =\kappa$ the
theory exhibits $O(d D)$ symmetry and so should reduce to a $d D + 8$
factor. Alternatively, these factors can be generated by using the symetrized
vertex $\hat R^{ijkl}_{\alpha \beta,\gamma \delta}$, defined above, from a
single diagram (with symmetry factor $(4 \times 3)(4 \times 3)/2$) producing all
the terms above.

 \begin{figure}
    \centering
  \includegraphics[width=0.3\linewidth]{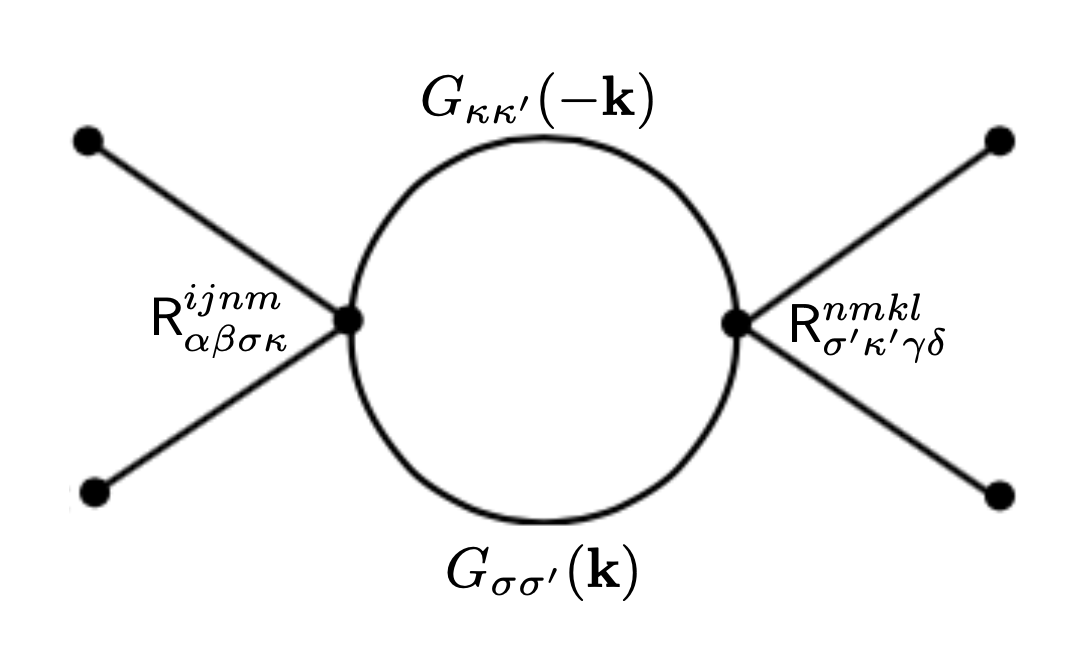}
  \caption{One-loop graphical correction to the symmetrized quartic coupling
    tensor, $\hat R^{ijkl}_{\alpha \beta,\gamma \delta}$.}
    \label{dRgraphFig}
  \end{figure}

As usual, because the vertex is independent of momentum, a singular
momentum-dependent correction to the inverse propagator only appears
at two loops, i.e., the exponent $\eta = 0$ within the one-loop calculation
on which we focus on here. Henceforth the ratio $\kappa/g$ is not
flowing to this order.

At vanishing vertex momentum, the loop integral is rotationally
invariant and thus can be easily computed over the spherical momentum
shell. We first perform angular average using identities for $\langle
\hat k_\alpha \hat k_\beta\rangle = \frac{1}{D}\delta_{\alpha\beta}$
and $\langle \hat k_\alpha \hat k_\beta \hat k_\gamma \hat
k_\delta\rangle =
\frac{1}{D(D+2)}\left(\delta_{\alpha\beta}\delta_{\gamma\delta}+\delta_{\alpha\gamma}\delta_{\beta\delta}+\delta_{\alpha\delta}\delta_{\gamma\beta}\right)$,
which give:

\bea
\langle P^L_{\alpha\beta}P^L_{\gamma\delta}\rangle
&=&\frac{1}{D(D+2)}\left(\delta_{\alpha\beta}\delta_{\gamma\delta}
+\delta_{\alpha\gamma}\delta_{\beta\delta}
+\delta_{\alpha\delta}\delta_{\gamma\beta}\right),\\
\langle P^L_{\alpha\beta}P^T_{\gamma\delta}\rangle
&=&\frac{1}{D}\delta_{\alpha\beta}\delta_{\gamma\delta}
-\frac{1}{D(D+2)}\left(\delta_{\alpha\beta}\delta_{\gamma\delta}
+\delta_{\alpha\gamma}\delta_{\beta\delta}
+\delta_{\alpha\delta}\delta_{\gamma\beta}\right),\\
\langle P^T_{\alpha\beta}P^T_{\gamma\delta}\rangle
&=&\left(1-\frac{2}{D}\right)\delta_{\alpha\beta}\delta_{\gamma\delta}
+\frac{1}{D(D+2)}\left(\delta_{\alpha\beta}\delta_{\gamma\delta}
+\delta_{\alpha\gamma}\delta_{\beta\delta}
+\delta_{\alpha\delta}\delta_{\gamma\beta}\right).
\eea
We calculate the remaining integral using momentum shell $\int \frac{d^D q}{(2 \pi)^D} \frac{1}{k^4} 
\to \intqq \frac{1}{k^4}= C_D \frac{1}{\epsilon} (e^{\eps d \ell}-1) \Lambda_\ell^{-\epsilon}
= C_D \Lambda_\ell^{D-4} d\ell$, where $d\ell$ is the momentum shell width and $C_D =
S_D/(2\pi)^D$, with $S_D = 2\pi^{D/2}/\Gamma(D/2)$ surface area of a
$D$-dimensional ball. Putting the above ingredients together and performing the tensor contractions with
Mathematica, we find that the correction to the vertex \eqref{Vcorrection} 
can be accounted for within the parametrization \eqref{Vparam} by the
following corrections to the couplings $u,v$
\bea
\delta u &=& (A_1 u^2 + A_2 v^2 + A_3 u v) C_D \Lambda_\ell^{D-4} d\ell  ,\\
\delta v &=& ( B_1 u^2 + B_2 v^2 + B_3 u v) C_D \Lambda_\ell^{D-4} d\ell  ,
\eea
where,
 
\bea A_1 &=& -\frac{1}{d D(D+2)}\left[(2d + D^2 + 4D +
  10)\kappa^{-2}
  + 2 ((d+2)D-2)g^{-1}\kappa^{-1} + (d(D^2-2)+D^3+5D^2-6)g^{-2}\right],\\
A_2 &=& -\frac{8}{d D(D+2)} (g^{-1}-\kappa^{-1})^2,\\
A_3 &=& -\frac{4}{d D(D+2)}\left[(D + 6)\kappa^{-2}
  +  4D g^{-1}\kappa^{-1} + (3D^2+D-6)g^{-2}\right],\\
B_1 &=& -\frac{1}{d D(D+2)}\left[(d+2D+7)\kappa^{-2}
  - 2 (d-D+1)g^{-1}\kappa^{-1} + (d+3D^2+2D-5)g^{-2}\right],\\
B_2 &=& -\frac{1}{d D(D+2)}\left[(d D^2+2(d+2)D + 12)\kappa^{-2} +
  8(D+1) g^{-1}\kappa^{-1} + (d D^3+ (d+8)D^2 - 2(d-2)D
  -20)g^{-2}\right],
\;\;\;\;\;\;\;\;\;\\
B_3 &=& -\frac{2}{d D(D+2)}\left[(d(D+2)+D^2+3D+6)\kappa^{-2} - 8
  g^{-1}\kappa^{-1} + (d(D^2+D-2) + D^3 + 2D^2-D+2)g^{-2}\right]. \eea

There are a few simple checks that above result passes. The case of $\kappa
= g$ and $u=0$ corresponds to an $O(d D)$ symmetric model. The 
correction to the $O(d D)$ symmetry breaking quartic
interaction $u$ vanishes with $u$, i.e., $a_2 = 0$ for $\kappa=g$, consistent
with the existence of the $O(d D)$ symmetric Wilson-Fisher critical point. 
The resulting correction $\delta v \propto -(d D + 8) v^2/\kappa^2$ is consistent with the $N+8$
factor of an $O(N)$ model (with here $N=d D$). Another check is for $d=1$, in which case
the $u$ and $v$ terms in \eqref{Vparam} cannot be distinguished from each others, and so they reduce to a single combined
coupling constant $u+v$. Consistent with this, we find $\delta(u+v) 
\propto - (u+v)^2$ (see below for a more detailed discussion of this case).

We now examine the RG flows on the critical, $\tau_R=0$, $u-v$
surface. As usual for the quartic model, the upper critical dimension
is $D=4$, and thus we expand in $\epsilon=4-D$. We define the
dimensionless couplings
\be \label{defuhat} 
\hat u = C_D \Lambda_\ell^{-\epsilon} \, u/\kappa^2 \, \quad , \quad 
\hat v = C_D \Lambda_\ell^{-\epsilon}  \, v/\kappa^2 \,  \quad , \quad C_4=\frac{1}{8 \pi^2}.
\ee 
We also introduce the important parameter
\be 
s = \kappa/g ,
\ee 
whose domain of variation is $s \in [0,+\infty]$,
such as $s=0$ corresponds to the membrane (purely longitudinal model) and to the usual crumpling transition,
$s=1$ is the isotropic case, and $s=+\infty$ is the purely transverse model. 

To derive the flow equations we calculate $\partial_\ell (\hat u,\hat v)$ taking into account 
(i) the rescaling (ii) the above corrections to the interaction vertex. Since there
are no corrections to the self energy to this one-loop order, there are no other contributions.
This leads to
the one-loop RG flow equations
\bea \label{fullRG1} 
\partial_\ell \hat u &=& \epsilon \hat  u + a_1 \hat  u^2 + a_2 \hat  v^2 + a_3 \hat  u \hat  v,\\
\partial_\ell \hat  v &=& \epsilon \hat  v + b_1 \hat  u^2 + b_2 \hat  v^2 + b_3 \hat  u \hat  v, \nn
\eea
with 
\bea \label{fullRG2} 
&& a_1= 
    \frac{-1}{12 d} (   21 + d + (6 + 4 d) s + (69 + 7 d) s^2 ) 
    \quad , \quad a_2 = -\frac{1}{3 d}  (s-1)^2 \quad , \quad
   a_3 = \frac{-1}{3 d} (s (23   s+8)+5) \\
&& b_1 = \frac{-1}{24 d} \left((d+51) s^2+2 (3-d) s+d+15\right)
\quad , \quad 
b_2= \frac{-1}{6 d} \left((18 d+31) s^2+10 s + 6 d+7\right) \nn \\
   &&
   b_3= \frac{-1}{6 d} \left((9 d+47) s^2 -4 s+3 d + 17\right) \nn
\eea 
and we recall that $s=\kappa/g$, to this one-loop order, does not flow.
In the large $d$ limit these equations reduce to
\bea  \label{RGlargeds} 
 \partial_\ell \hat u &=& \epsilon \hat  u -\frac{1}{12} \left(7 s^2+4 s+1\right) \hat u^2 \\
 \partial_\ell \hat  v &=& \epsilon \hat  v -\frac{1}{24}
   (s-1)^2 \hat u^2 -\left(3
   s^2+1\right) \hat v^2 
-\frac{1}{2}  \left(3 s^2 +1\right) \hat u \hat v \nn
   \eea 
   
\subsection{Analysis of the RG equations} 

Let us now analyse the RG equations, starting by examining some particular cases.\\

{\bf Crumpling transition}. The case $s=0$ corresponds to the limit of $g\rightarrow\infty$, i.e. to a single-valued surface
constraint on the tangent vector $\vec t_\alpha$. In that case the flow equations
reduce to that for the crumpling transition of a polymerized membrane:
\bea \label{CrumplingRG} 
\partial_\ell \hat u &=& \epsilon\hat u 
- \frac{d+21}{12 d} \hat u^2 
- \frac{1}{3 d} \hat v^2
- \frac{5}{3 d}\hat u \hat v,\\
\partial_\ell \hat v &=& \epsilon \hat v
- \frac{d+15}{24 d}\hat u^2 
- \frac{6d+7}{6 d}\hat v^2 
- \frac{3d+17}{6 d}\hat u \hat v, \nn
\eea
 Simple algebra shows that these equations are
equivalent to those obtained by Paczuski, Kardar and Nelson in \cite{PKN},
when reexpressed in terms of the
$4(v-u/D), 4u$ couplings. In the large $d$ limit they reduce to
\bea
\partial_\ell \hat u &=& \epsilon\hat u 
- \frac{1}{12} \hat u^2,\\
\partial_\ell \hat v &=& \epsilon \hat v
- \frac{1}{24}\hat u^2 
- \hat v^2 
- \frac{1}{2}\hat u \hat v, \nn
\eea
In that limit these equations have four fixed points (FP):
\bea \label{FP0} 
\hat u &=& 0,\;\;\hat v = 0,\;\;\mbox{Gaussian},\\
\hat u &=& 0,\;\;\hat v = \epsilon ,\;\;\mbox{O($dD$) critical Heisenberg},\\
\hat u &=& 12\epsilon,\;\,
v = -2\epsilon, \;\;\mbox{Crumpling transition}, \\
\hat u &=& 12\epsilon,\;\;\hat v = -3\epsilon, \;\;\mbox{Marginal}.
\eea
The Gaussian FP in unstable for $\epsilon>0$. The Heisenberg FP has one unstable 
direction. The crumpling transition
fixed point is fully attractive. The last fixed point has also one unstable direction
and lies at the boundary of the physical stability domain,
which requires $2 \mu + D \lambda \geq 0$, 
which here means $\hat u + 4 \hat v \geq 0$ (we call it "Marginal"
for that reason).
This fixed point structure has a counterpart in the $D=4-\epsilon$ RG for the flat phase \cite{AL}. 
As noted in \cite{PKN}, for $d$ finite, the equations \eqref{CrumplingRG} have no attractive fixed point for $d<d_c \approx 219$. These equations were recently extended to 2 loops (i.e. order $O(\epsilon^2)$ and it was found that $d_c$ decreases with $\epsilon$ \cite{Mouhanna}.
\\

{\bf General finite $s>0$}.
At large $d$ the equations \eqref{RGlargeds} have the following fixed points,
apart from the Gaussian FP which is fully unstable at $\epsilon>0$

\bea  \label{FPs} 
&& \hat u=0 \quad , \quad \hat v= \frac{\epsilon}{1 + 3 s^2} \\
&& \hat u= \frac{12 \epsilon}{s (7 s+4)+1}    
\quad , \quad \hat v= \frac{\epsilon}{3 s^2+1}-\frac{3 \epsilon}{s (7 s+4)+1}  \\
&& \hat u= \frac{12 \epsilon}{s (7 s+4)+1}  \quad , \quad \hat v= -\frac{3 \epsilon}{s (7 s+4)+1} 
\eea 
The Hessian matrix is triangular. 
The first FP has eigenvalues $(+1,-1)\epsilon $ along $(u,v)$ directions respectively
hence it is unstable along $u$. The second FP has eigenvalues $(-1,-1) \epsilon$
and is the only fully stable one. The third FP has eigenvalues
$(-1,1) \epsilon$ and is at the boundary of the physical domain (it has $\hat u=-3 \hat v$). 
Hence we see that for large $d$ the structure of the FP's does not
qualitatively depend on $s$. 

  \begin{figure}
    \centering
   \includegraphics[width=0.7\linewidth]{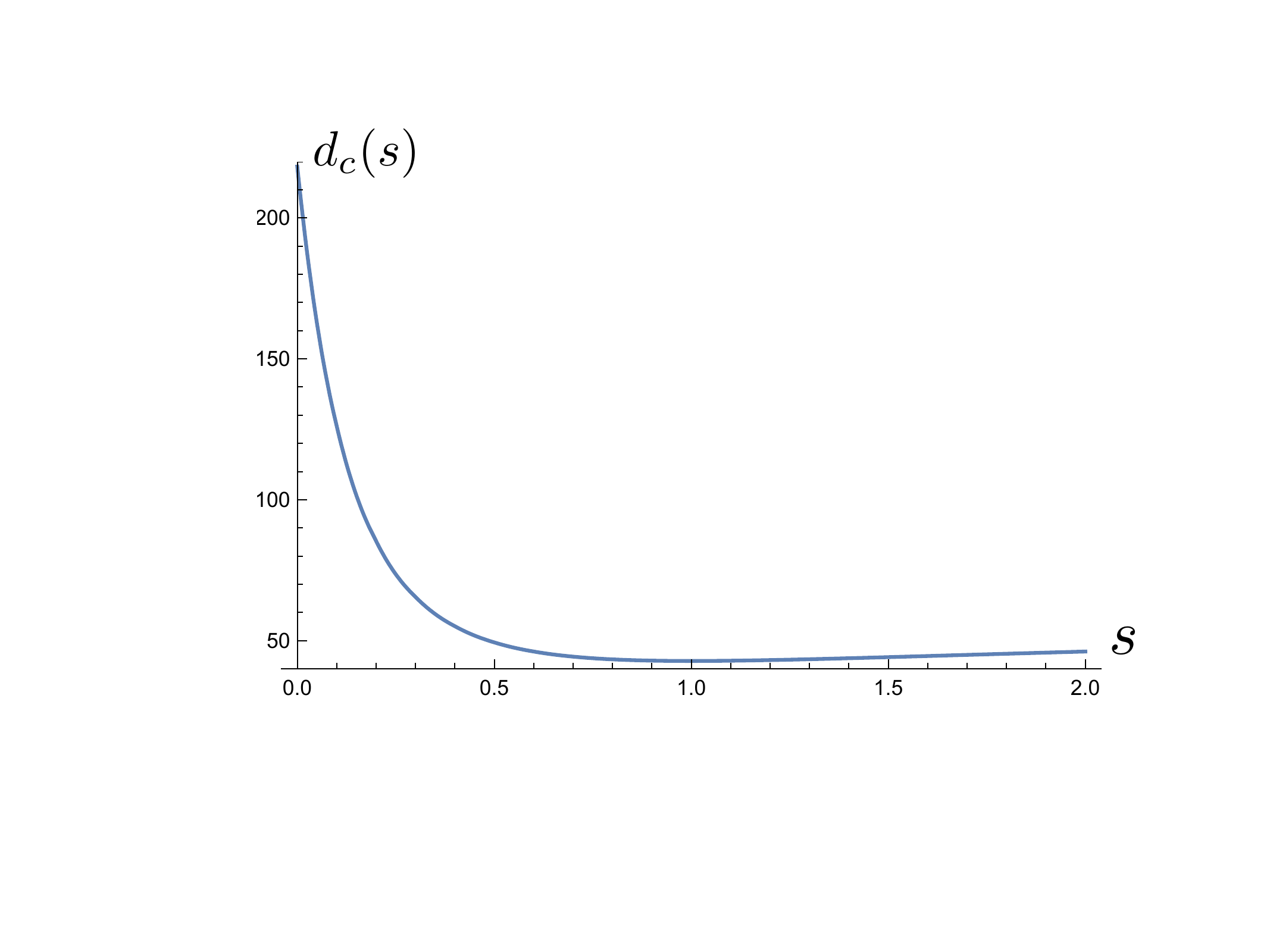}
   \caption{A plot of a critical value of the embedding space dimension
     $d_c(s)$ above which there is a stable critical fixed point 
     to the RG flow to one-loop, as a function of $s=\kappa/g$.
     This fixed point describes a continuous transition to an ordered
     phase. For $s=0$ it describes the crumpling transition.} 
    \label{dcsFig}
\end{figure}

Going back to the full RG equations \eqref{fullRG1},\eqref{fullRG2},
we find that for $d$ finite and large enough, the structure of the
FP's is qualitatively the same as for $d=+\infty$.  We find that for
any fixed $s$ there is an embedding dimension $d_c(s)$ (plotted in
Fig. \ref{dcsFig}) such that for $d > d_c(s)$ there is a fully stable
FP, and for $d<d_c(s)$ there is not. To determine $d_c(s)$ one notes
that at this bifurcation two FP's merge and hence the determinant of
the Hessian of the RG equations vanishes. Imposing this condition
leads to three non-linear equations for $\hat u, \hat v, d$ which are
easy to solve numerically. They yield $d_c(s)$, and also the position
of the two FP $(\hat u(s), \hat v(s))$ when they merge (and move off
to the complex plane).  One finds that $d_c(s)$ decreases fast with
$s$ at small $s$, from its value at $s=0$. One finds for instance that
(see also Fig. \ref{dcsFig}) 
\bea
&& s=0 \quad , \quad  d_c=d_c(0)= 218.206 \quad , \quad (\hat u(s), \hat v(s)) =  (11.1348,-2.35165) \epsilon \\
&& s= 1/4 \quad , \quad d_c(s) = 73.7747 \quad , \quad (\hat u(s), \hat v(s)) =  (4.40623,-0.755289) \epsilon  \\
&& s = 1/2 \quad , \quad d_c(s) = 49.3649
\quad , \quad (\hat u(s), \hat v(s)) = (2.21725,-0.322653) \epsilon \\
&& s = 1 \quad , \quad d_c(s) = 42.7846 \quad , \quad (\hat u(s), \hat
v(s)) = (0.870014,-0.11656) \epsilon
\\
&& s = 2 \quad , \quad d_c(s) = 46.1284
\quad , \quad (\hat u(s), \hat v(s)) = (0.283511,-0.0397547)  \epsilon  \\
&& s = 10 \quad , \quad d_c(s) = 56.7704 \quad , \quad (\hat u(s),
\hat v(s)) = (0.0143203,-0.00222716) \epsilon \eea For $s=1$ one finds
that $d_c(s=1)$ admits an analytic expression \be \label{dc1} d_c(1)=
22+12 \sqrt{3} \quad , \quad (\hat u(1), \hat v(1)) = \left( \frac{7+5
    \sqrt{3}}{18} , \frac{1-3 \sqrt{3}}{36} \right) \epsilon \ee

Interestingly the same condition of vanishing of the Hessian reveals that there is another special dimension $d_c^{-}(s)$ where two fixed points merge, and 
a stable FP again appears for $d<d_c^-(s)$. One finds 
\bea \label{dcm1} 
&& s=0 \quad , \quad  d^-_c(s)= 1.12657 \quad , \quad (\hat u^-(s), \hat v^-(s)) = (1.48178,-1.13827) 
\epsilon \\
&& s=1 \quad , \quad 
d^-_c(s)=  22-12 \sqrt{3} = 1.21539,
~~,~~ (\hat u^-(s), \hat v^-(s)) = 
\left( \frac{7-5 \sqrt{3}}{18} , 
\frac{1+3 \sqrt{3}}{36} \right)  \epsilon
= ( -0.0922363,0.172115) \epsilon \nn 
\eea
This critical dimension is very close to $d=1$ (slightly above), we will
not study it in details here (see specific discussion below for $s=1$ and 
also for $d=1$). 
\\

{\bf Special case $s=+\infty$}. One can study the limit $s=\kappa/g \to +\infty$ 
or equivalently $\kappa \to +\infty$, in which case the $\vec t_\alpha$ vector field is
purely transverse (i.e. the divergence is constrained to vanish, $\partial_\alpha \vec{t}_\alpha=0$). 
To this aim
one must rescale the couplings, and define
\be \label{defubar} 
\bar u = C_4 \Lambda_\ell^{-\epsilon} \, u/g^2 = s^2 \hat u 
\quad , \quad \bar v  = C_4 \Lambda_\ell^{-\epsilon}  \, v/g^2 = s^2 \hat v 
\ee 
One then obtains in the limit $s \to +\infty$,
\bea
\partial_\ell \bar u &=& \epsilon \bar u
-\frac{(7 d+69) \bar u^2}{12 d}-\frac{23 \bar u \bar v}{3 d}-\frac{\bar v^2}{3 d}
\\
\partial_\ell \bar v & = &  \epsilon \bar v  
-\frac{(d+51) \bar u^2}{24 d}-\frac{(9 d+47) \bar u \bar v}{6
   d}- \left(\frac{31}{6 d}+3\right) \bar v^2
\eea 

At large $d$ the above fixed points become
\bea 
&& \bar u=0 \quad , \quad \bar v= \frac{\epsilon}{3} \\
&& \bar u= \frac{12 \epsilon  }{7 }    
\quad , \quad \bar v= - \frac{2}{21} \epsilon  \\
&& \bar u= \frac{12 \epsilon  }{7 }    \quad , \quad \bar v= -\frac{3 \epsilon}{7} 
\eea 
We know from the discussion above, valid for any $s$, that only the second one is
fully stable. For arbitrary $d$, a stable FP exists only for $d> d_c(s=+\infty)$ with
\bea 
&& s= +\infty \quad , \quad d_c(s) = 61.3112
\quad , \quad (\hat u(s), \hat v(s)) = (1.52165,-0.244306) \epsilon 
\eea 
and the second special
dimension is $d_c^-(+\infty)=1.04247$, with $(\hat u^-(+\infty), \hat v^-(+\infty)) =(-0.411147,0.503785) \epsilon$. 
\\

{\bf The special case $d=1$}. For $d=1$, there is no difference 
in the action of the initial model between the terms $u,v$.
Instead, one obtains an equation for the sum $u+v$. Simply adding
the two RG equations in \eqref{fullRG1}
and setting $d=1$ one obtains
\be 
\partial_\ell (\hat u+\hat v) = \epsilon (\hat u+\hat v) 
-\frac{1}{2} (s (17 s+2)+5) (\hat u+\hat v)^2
\ee
and then there is always an attractive fixed point at
\be 
\hat u+\hat v = \frac{2 \epsilon}{s (17 s+2)+5} 
\ee 

In the transverse only limit $\kappa \to +\infty$, i.e. $s \to +\infty$ in terms of
the couplings $\bar u, \bar v$ defined above in \eqref{defubar} 
we obtain
\be 
\partial_\ell (\bar u+\bar v) = \epsilon (\bar u+\bar v) 
-\frac{17}{2} (\bar u+\bar v)^2
\ee 
One can check that 
we obtain the same RG equation
as in \cite{FreyIsotropicDipolar}  
(see (4.15) there taking $\ell \to +\infty$, with the identification that $u(\ell)$ there
equals $6 (\bar u+\bar v)$ here, as obtained comparing 
their action (2.1) with ours - which leads to $u_0 \equiv 6(u+v)$).
In that work they study a Heisenberg model with isotropic dipolar interactions,
which corresponds to a $q$-dependent bare $\kappa(q) = g + c/q^2$.
Its non analytic structure forbids graphical corrections to the 
amplitude $c$. Here we find that it coincides (beyond some crossover scale) 
with the limit $\kappa = +\infty$ of our RG equations, which is reasonable
to expect.
\\

{\bf The special case $s=1$}. In that case the propagator is isotropic $\kappa=g$.
Our RG equations then become
\bea 
&& \partial_\ell \hat u = \epsilon \hat u - \frac{\hat  u}{d} \left(  (d+8) \hat  u+12
   \hat  v \right) \\
&&  \partial_\ell \hat v = \epsilon \hat v - \frac{1}{d} 
(\hat  u+2 \hat  v) (2 (d+2)
   \hat  v+3 \hat  u)
\eea 
For $d \to +\infty$ the stable fixed point is $\hat u=\epsilon$ and
$\hat v=0$. 

One can check that these equations are equivalent to those obtained in
\cite{VicariNM2001} (denoting $u'$ and $v'$ their couplings,
one has $v'=6\hat u/d$ and $u'=6(\hat u+\hat v)/d$). In that paper the standard
$O(n) \times O(m)$ model (with $\kappa=g$) 
is studied to three loops (order $O(\epsilon^3)$).
To make contact with our model for $s=1$ we 
set $m=D=4$ in their equations. Our model for $s=1$
thus corresponds to 
$O(n=d) \times O(m=D)$, but where the
number of components $m$ is tied to the space dimension $D$. 
The isotropic fixed point that we obtain here for $s=1$ corresponds to what they
call {\it the chiral fixed point}. They have a similar structure of FP's as we discussed
above. They also have the Heisenberg $O(m n)$ fixed point, 
together with an "anti-chiral FP" which can be compared to our
FP that we called marginal. They also note that there
are two special values of $n=n^\pm(m)$ such that
their chiral FP exists and is stable only for $n>n_+$ and $n<n_-$. 
This maps onto the special dimensions $d_c(1)$, $d_c^-(1)$ calculated
above. To one-loop order, their result reads
\be 
n^\pm(m) = 5 m +2  \pm 2 \sqrt{6 (m-1) (m+2)} 
\ee 
and setting $n=d$ and $m=D=4$ we recover our results above
\eqref{dc1} and \eqref{dcm1} for $s=1$. They also compute
the correction to $n^\pm(m)$ to the next order in $\epsilon$. Within
their model this correction reads
\be 
 \epsilon \left(  - 5 m -2 \mp \frac{1}{2 \sqrt{6 (m-1) (m+2)} } ( 25 m^2 + 22 m - 32) \right) 
 = -(22+\frac{38}{\sqrt{3}}) \epsilon \approx -43.9393 \epsilon
\ee
where we evaluated only the $+$ branch and set $m=4$, so
that $n^+(4)= 42.7846
-43.9393 \epsilon$. We will refrain from using that result for our model,
since in our model $D$ and $m$ are constrained to be equal. Still
it is quite analogous to the result obtained to two loops for the crumpling
transition in \cite{Mouhanna}. Finally note that
there is an additional critical value of $n$ identified in
\cite{VicariNM2001} such that for 
$n<n_H(m)= \frac{1}{m}(4 - 2 \epsilon)$ the Heisenberg FP becomes the stable one.
Finally, the authors of \cite{VicariNM2001} also compute the $\eta$ exponent. 
From their result we can infer, setting $m=4$
\be 
\eta = \frac{4 d +2}{72} (u')^2 + 3 (d-1) v' ( \frac{v'}{48} - \frac{u'}{36}) + h.o.t
\ee 
To leading order at large $d$ the stable fixed point is at $u'=v'=6 \epsilon/d$ 
and one finds 
\be \label{etaVicari} 
\eta = \frac{5}{4 d} \epsilon^2 + O(1/d^2) 
\ee 
which we will compare with our results below.
Finally, Ref. \cite{VicariNM2001} also computes the exponent $\nu$, which we give here for completeness. 
In our notations
\be 
\frac{1}{\nu} = 2 - \frac{4 d + 2}{6} u' + \frac{d-1}{2} v' + h.o.t = 2 - \frac{d+5}{d} \epsilon + O(\epsilon^2)
\ee

\section{II. SCSA analysis and large $d$ expansion}

In this section we give some of the details of the SCSA calculation, as well as of
the large-$d$ expansion. Some more technical details will be given in
following sections. The method was pioneered for the
$O(N)$ model by Bray in \cite{Bray}. For membranes it was
introduced in \cite{LRprl} to describe the flat phase and 
the crumpling transition. We will use similar methods and notations
as in Refs. \cite{LRReview} and \cite{OurBuckling} to which
we refer for more details on the method (see in particular
the Appendices and Supp. Mat there). 

\subsection{Starting model: bare propagator and bare vertex}

We will consider again the model at the critical point $\tau_R=0$.
Since in the SCSA the (dressed) quartic vertex acquires a momentum dependence,
we extend the notations and write the bare effective Hamiltonian as
\be
H[\vec{r}]= \frac{1}{2} \int_{\kb} \left( \kappa P^L_{\alpha \beta}(\kb) + g P^T_{\alpha \beta}(\kb) \right) k^2 t_\alpha(\kb) t_\beta(-\kb) +
{1 \over {4 d} } \int_{\kb_1,\kb_2,\kb_3} R_{\alpha \beta, \gamma \delta}(\q)~
{\vec t}_\alpha(\kb_1)\cdot{\vec t}_\beta(\kb_2 )~
{\vec t}_\gamma(\kb_3)\cdot{\vec t}_\delta(\kb_4 )
\ee
with $\q=\ks_1+\ks_2$ and $\ks_1+\ks_2+\ks_3+\ks_4={\bf 0}$, and we
use $\int_{\kb}$ to denote $\int d^Dk/(2 \pi)^D$. As before the bare tangent correlator reads
\bea
&& \langle t^i_\alpha(\kb)  t^j_\beta(\kb) \rangle_0 = \delta_{ij} G_{\alpha \beta}(\kb) (2 \pi)^d \delta^d(\kb+\kb') \\
&& 
G^{-1}_{\alpha \beta}(\kb) = \left( \kappa P^L_{\alpha \beta}(\kb) + g P^T_{\alpha \beta}(\kb) \right) k^2 \quad , \quad
G_{\alpha \beta}(\kb) = \frac{P_{\alpha \beta}^L(\kb)}{\kappa k^2} + \frac{P_{\alpha \beta}^T(\kb)}{g k^2}
\eea 
We choose the bare quartic tensorial interaction to be local, as in the main text, and of the form: 
\be R_{\alpha \beta,
  \gamma \delta}
=\frac{\mu_0}{2}(\delta_{\alpha\gamma}\delta_{\beta\delta}+\delta_{\alpha\delta}\delta_{\beta\gamma})
+ \frac{\lambda_0}{2} \delta_{\alpha\beta}\delta_{\gamma\delta}.
\label{bareR} \quad , \quad \mu_0=u \quad , \quad \lambda_0=2 v
\ee
where we use the natural notations in the context of membranes. 

\subsection{SCSA equations: dressed propagator and dressed vertex}

The dressed tangent correlator is denoted as
\be
\langle t^i_\alpha(\kb)  t^j_\beta(\kb) \rangle = \delta_{ij} {\cal G}_{\alpha \beta}(\kb) (2 \pi)^d \delta^d(\kb+\kb') \quad , \quad  {\cal G}^{-1}_{\alpha \beta}(\kb) = G^{-1}_{\alpha \beta}(\kb) + \sigma_{\alpha \beta}(\kb) 
\ee 
where $\sigma_{\alpha \beta}(\kb)$ is the self-energy. 
The SCSA gives two self-consistent
Dyson type equations for the self-energy and the dressed vertex, which allow to determine the dressed propagator ${\cal G}_{\alpha \beta}(\kb)$. The infrared behavior of the resulting dressed propagator 
(e.g. the exponent $\eta$, see below) is by construction exact to first order in the
large $d$ expansion, i.e. to order $O(1/d)$ in any $D$ dimension. 
Similarly the dressed vertex is exact as $d \to \infty$. 
The SCSA equations "improve" on the large-$d$ expansion by
making it self-consistent through the use of dressed propagators and vertices.
It amounts to a resumation of a $1/d$ expansion. It thereby provides predictions
for any $D,d$. Since it accounts for the physics of
screening of the interaction by fluctuations, it is known to give accurate results for membranes
in physical dimensions $d=3$ \cite{LRReview}. 

The first SCSA "self-energy" equation takes the form 
\bea
\sigma_{\alpha \gamma}(\kb) = {2 \over d} \int_q 
\tilde{R}_{\alpha \beta, \gamma \delta}(\q) 
{\cal G}_{\beta \delta}(\kb - \q)  \label{firstsc}  
\eea
where everywhere repeated indices are summed over.
Note that in principle it should be symmetrized in $\alpha,\gamma$,
however the only two-index tensors which appear
here are symmetric. 
The complete Dyson equation for the self-energy contains an
additional (UV divergent) $k$-independent "tadpole" diagram contribution.
The integral in \eqref{firstsc} takes also some value at $k=0$. Both contributions have been
subtracted by tuning the bare "mass" term (reduced temperature) $\tau$ to its
critical value (so that $\tau_R=0$). This is the standard
procedure when dealing with a critical theory, where a parameter
(distance to crumpling transition parameters) must be tuned.
Henceforth, the integral in the r.h.s. of \eqref{firstsc}  
is to be understood with its value at $k=0$ substracted. 

In \eqref{firstsc} $\tilde{R}_{\alpha \beta, \gamma \delta}(\q) $ is the (momentum dependent)
dressed quartic vertex. It is a four index tensor which is 
symmetric in $\alpha \leftrightarrow \beta$, in $\gamma \leftrightarrow \delta$ and under the
exchange $(\alpha,\beta) \leftrightarrow (\gamma,\delta)$. It is determined by the 
second SCSA equation, which reads
\bea \label{secondsc}
\tilde{R}_{\alpha \beta, \gamma \delta}(\q) &=& R_{\alpha \beta, \gamma \delta}(\q) 
- R_{\alpha \beta, \alpha' \beta'}(\q) \Pi_{\alpha' \beta', \gamma' \delta'}(\q) \tilde{R}_{\gamma' \delta', \gamma \delta}(\q)
\eea
where $\Pi_{\alpha \beta, \gamma \delta}(\q)$ is the four index vacuum polarization tensor, which is also a symmetric tensor 
\be
\Pi_{\alpha \beta, \gamma \delta}(\q)
=  {\rm sym}_{\alpha \beta} {\rm sym}_{\delta \gamma}  \int_p \Gc_{\alpha \gamma}(\pb) \Gc_{\beta \delta}(\q-\pb)
\ee
where ${\rm sym}_{\alpha \beta} A_{\alpha \beta}= \frac{1}{2} (A_{\alpha \beta} + A_{\beta \alpha})$.

To solve these equations we first parameterize the momentum-dependent tensors $\Pi$ and $\tilde R$ in an appropriate basis of projectors. We now recall how this is done
(see \cite{ALGolubovic} for the original construction of these tensors, and \cite{LRReview} and \cite{OurBuckling} 
for applications within the SCSA).

\subsection{Projectors and tensor multiplication, inversion of the second SCSA equation }

Here we consider four index tensors, such as $R_{\alpha \beta,\gamma \delta}(\qb)$ 
and $\Pi_{\alpha \beta,\gamma \delta}(\qb)$
introduced above, which are 
symmetric in $\alpha \leftrightarrow \beta$, in $\gamma \leftrightarrow \delta$ and in 
$(\alpha,\beta) \leftrightarrow (\gamma,\delta)$. The product of such tensors
is defined as $(T \cdot T')_{\alpha \beta,\gamma \delta}=T_{\alpha \beta,\gamma' \delta'}
T'_{\gamma' \delta',\gamma \delta}$, the identity being $
I_{\alpha \beta,\gamma \delta} = {1\over 2}(\delta_{\alpha\gamma}\delta_{\beta\delta}
+\delta_{\alpha\delta}\delta_{\beta\gamma})$. We recall the definition \cite{LRReview}
of the five "projectors"
$W_i$, $i=1,\ldots,5$, which span the space of such four index tensors
\bea
&& (W_3)_{\alpha\beta,\gamma\delta}(\qb)
={1 \over {D-1}} P^T_{\alpha \beta} P^T_{\gamma \delta}\;,\;\;\;
(W_5)_{\alpha\beta,\gamma\delta}(\q)=P^L_{\alpha \beta}P^L_{\gamma\delta}\;,\\
&& 
(W_4)_{\alpha\beta,\gamma\delta}(\q) = (W_{4a})_{\alpha\beta,\gamma\delta}(\q) 
+ (W_{4b})_{\alpha\beta,\gamma\delta}(\q)\;,\\
&& (W_{4a})_{\alpha\beta,\gamma\delta}(\q) = {1 \over {\sqrt{D-1}}} P^T_{\alpha \beta} P^L_{\gamma \delta}\;,
\quad 
(W_{4b})_{\alpha\beta,\gamma\delta}(\q) ={1 \over {\sqrt{D-1}}} P^L_{\alpha \beta} P^T_{\gamma \delta}\;,\\
&& (W_2)_{\alpha\beta,\gamma\delta}(\q)=
{1 \over 2}(P^T_{\alpha \gamma} P^L_{\beta \delta } 
+ P^T_{\alpha \delta} P^L_{\beta \gamma} + P^L_{\alpha \gamma}
P^T_{\beta \delta } + P^L_{\alpha \delta} P^T_{\beta \gamma})\;, \\
&& 
W_1(\q)={1\over 2}(\delta_{\alpha\gamma}\delta_{\beta\delta}
+\delta_{\alpha\delta}\delta_{\beta\gamma})
- W_3(\q) - W_5(\q) - W_2(\q)\;,
\label{relationW1} \\
&& W_1(\q) + W_3(\q) = {1\over 2}(P^T_{\alpha\gamma}(\q) P^T_{\beta\delta}(\q)
+P^T_{\alpha\delta}(\q) P^T_{\beta\gamma}(\q))
\eea 
where $P^T_{\alpha \beta}=\delta_{\alpha \beta} -
q_{\alpha}q_{\beta}/q^2$ and $P^L_{\alpha
  \beta}=q_{\alpha}q_{\beta}/q^2$ are the standard transverse and
longitudinal projection operators associated to $\q$. The first two
projectors $W_1, W_2$ are mutually orthogonal and orthogonal to the
other three. Note that while $R$, being symmetric, can be expressed in
terms of the symmetric tensors $W_i$, $i=1,..5$, we will need at some
intermediate stages of the calculations some products (such as $\Pi*R$
see below), which are not symmetric. Hence we introduced $W_4^a$ and
$W_4^b$, which together with $W_i$, $i=1,2,3$ and $W_5$ make the
representation complete under tensor multiplication. The rules for the
tensor multiplication $T''=T'*T$ of the tensors $T =
\sum_{i=1}^3 w_i W_i + w_{4a} W_{4a} + w_{4b} W_{4b} + w_{5} W_{5}$
and $T' = \sum_{i=1}^3 w'_i W_i + w'_{4a} W_{4a} + w'_{4b} W_{4b} +
w'_{5} W_{5}$ are 
\bea
w_1''=w'_1 w_1\;,\;\;\; w_2'' = w'_2 w_2\;,\;\;\;
\begin{pmatrix}
w_3''&w_{4a}''\\ w_{4b}''&w_5'' 
\end{pmatrix}
=
\begin{pmatrix}
w'_3&w_{4a}' \\
w_{4b}' &w'_5 
\end{pmatrix}
\begin{pmatrix}
w_3&w_{4a}\\
w_{4b}&w_5
\end{pmatrix}\;,
\label{3.5}
\eea 
with $T'' = \sum_{i=1}^3 w''_i W_i + w''_{4a} W_{4a} + w''_{4b}
W_{4b} + w''_{5} W_{5}$.  \\

Then we define the decomposition of our tensors of interest on this basis (suppressing
the indices here)

\be
\tilde R(\q) = \sum_{i=1}^5 \tilde w_i(\q) W_i(\q)\;,
 \quad , \quad 
\Pi(\q) = \sum_{i=1}^5 \pi_i(\q) W_i(\q)\;,
\label{(3.10)}
\ee

To perform the calculations we need the inversion formula, which allows us to obtain the coefficients
(correcting the obvious misprint in (C19) of \cite{LRReview})
\bea \label{inversion} 
\pi_1(\q) &=& \frac{2}{(D-2) (D+1)} (W_1)_{\alpha \beta,\gamma\delta}  \Pi_{\alpha \beta, \gamma \delta}(\q) \\
\pi_2(\q) &=& \frac{1}{D-1} (W_2)_{\alpha \beta,\gamma\delta}  \Pi_{\alpha \beta, \gamma \delta}(\q) \\
(D-1) \pi_3(\q) &=& P^T_{\alpha \beta}(\q) \Pi_{\alpha \beta, \gamma \delta}(\q) P^T_{\gamma \delta}(\q)\;,\\
\pi_5(\q) &=& P^L_{\alpha \beta}(\q) \Pi_{\alpha \beta, \gamma \delta}(\q) P^L_{\gamma \delta}(\q)\;,\\
\sqrt{D-1}\pi_4(\q)&=& 
P^L_{\alpha \beta}(\q) \Pi_{\alpha \beta, \gamma \delta}(\q) P^T_{\gamma \delta}(\q)
=
P^T_{\alpha \beta}(\q) \Pi_{\alpha \beta, \gamma \delta}(\q) P^L_{\gamma \delta}(\q)\;.
\eea 
Using \eqref{relationW1} the formula for $\pi_1(\q)$ reduces to
\bea \label{inversionpi1} 
\pi_1(\q) &=& \frac{2}{(D-2) (D+1)}  \left[ {1\over 2}(\delta_{\alpha\gamma}\delta_{\beta\delta}
+\delta_{\alpha\delta}\delta_{\beta\gamma})  \Pi_{\alpha \beta, \gamma \delta}(\q) 
- \pi_3(\q) - \pi_5(\q) - (D-1) \pi_2(\q) \right]
\eea

The bare vertex \eqref{bareR}, which is local and $\q$
independent, can also be written in this general basis, as
\bea
&& R = \sum_{i=1}^5 w_iW_i(\q) ,\;\;  w_1= w_2 = \mu_0,\;\;
w_3= \frac{1}{2}(D-1)\lambda_0+ \mu_0,\;\;
w_4=\frac{1}{2} \sqrt{D-1}\lambda_0,\;\;
w_5=\frac{1}{2} \lambda_0+ \mu_0\;.
\label{barew}
\eea
Note that the $\q$ dependence cancels and that the
two eigenvalues of the matrix formed by the $w_i$,
$i=3,4,5$, are then $\mu_0$, and $\mu_0 + \frac{1}{2} D \lambda_0$.
\\

We can now formally solve the second SCSA equation \eqref{secondsc} and find the renormalized couplings $\tilde w_i(\q)$ as functions of the $\pi_i(\q)$ as
\bea \label{tildew}
&& \tilde{w}_1(\q)={ w_1 \over {1 + w_1 \pi_1(\q)}}\;,\;\;\;
\tilde{w}_2(\q)={ w_2 \over {1 + w_2 \pi_2(\q)}}\;, \\
&& 
\begin{pmatrix} \tilde{w}_3(\q) & \tilde{w}_4(\q) 
\\ \tilde{w}_4(\q) & \tilde{w}_5(\q) \end{pmatrix} 
=
\begin{pmatrix}  w_3 & w_4 \\ w_4 & w_5 \end{pmatrix} 
\left( 
\begin{pmatrix}  1 & 0 \\ 0 & 1  \end{pmatrix}   +
\begin{pmatrix} \pi_3(\q) & \pi_4(\q) \\ \pi_4(\q) & \pi_5(\q) \end{pmatrix}
\begin{pmatrix} w_3&w_4 \\ w_4&w_5 \end{pmatrix}\right)^{-1}\;\;.
\eea
These can be substituted into \eqref{sigma} to express the self-energy as
\bea \label{sigma2} 
\sigma_{\alpha \gamma}(\kb) = {2 \over d}  \sum_{i=1}^5 \int_q \tilde w_i(\q)
(W_i)_{\alpha \beta, \gamma \delta}(\q) 
{\cal G}_{\beta \delta}(\kb - \q) 
\eea


\subsection{Ansatz for the propagator at the critical point}

We now look for a solution of the SCSA equations which describes the
critical manifold, i.e., for a propagator which behaves as a power law at small momentum.
A priori one could choose
\be
{\cal G}^{-1}_{\alpha \beta}(\kb) \simeq Z_\kappa k^{2 - \eta} P^L_{\alpha \beta}(\kb) + Z_g k^{2 - \eta_g} P^T_{\alpha \beta}(\kb)   \quad , \quad
{\cal G}_{\alpha \beta}(\kb) \simeq \frac{P_{\alpha \beta}^L(\kb)}{Z_\kappa k^{2-\eta}} 
+ \frac{P_{\alpha \beta}^T(\kb)}{Z_g k^{2-\eta_g}}
\ee 
i.e. two different exponents for the transverse and longitudinal components.
However we will restrict our search to the case $\eta_g=\eta$, but
take into account that there are a priori two 
distinct amplitudes $Z_\kappa$ and $Z_g$ for each component.
While in SCSA each is non-universal, their ratio may be universal
and in this section we will call $s$ their ratio
\be 
s= \frac{Z_\kappa}{Z_g} 
\ee 
since it corresponds to a "dressed" (or renormalized) version of $\kappa/g$. 
Anticipating that $\eta>0$ we see that at small $k$ we can neglect the bare inverse 
propagator compared to the dressed one, hence our ansatz for the
self-energy is given by
\be \label{ansatzsigma} 
\sigma_{\alpha \beta}(\kb) 
\simeq k^{2 - \eta}  \left( Z_\kappa P^L_{\alpha \beta}(\kb) + Z_g P^T_{\alpha \beta}(\kb)  \right) 
\ee

\subsection{Dressed quartic couplings} 

We start with the calculation of the "vacuum polarization bubble" $\pi_(\q)$, presenting here
only the results and leaving the quite technical derivation to 
Section III below. The key result is that at small $q$
$\pi_i(\q)$  diverge as (under the condition that $4-D-2 \eta>0$) 
\bea
\pi_i(\q) \simeq Z_\kappa^{-2} a_i(\eta,D,s) q^{-(4-D-2\eta)}\;. 
\label{piq}
\eea
For the amplitudes $a_i=a_i(\eta,D)$ we find
\bea \label{ais} 
&& a_1= 2 A \left(\frac{2 (\eta -2)^2
   s^2 (D+\eta -1)}{D+\eta
   -2}-\frac{2 (\eta -2) (2 \eta -5)
   s (s-1) (D+\eta -1)}{(D+1)
   (D+\eta
   -2)}+(s-1)^2\right) \\
&& a_2=   \frac{2 A
   (\eta -2) \left(s \left(D^2+D (3
   \eta -7)+2 (\eta
   -2)^2\right)-\left(s^2 (D (D+\eta
   -3)-2 \eta
   +3)\right)-1\right)}{D+\eta -2} \nn \\
&& a_3 =   A
   \left(\frac{4 (\eta -2)^2 s^2
   (D+\eta -1)}{D+\eta -2}+\frac{4
   (\eta -2) s (s-1) (D+\eta
   -1)}{D+\eta -2}+D
   (s-1)^2+(s-1)^2\right) \nn \\
&& a_4 =  A
   \sqrt{D-1} (s-1)^2 (D+2 \eta
   -3) \nn \\
&& a_5 =    
     \frac{A}{D+\eta
   -2} 
\bigg(D^2 (\eta -2)
   (s-5) (s-1)+D^3 (s-1)^2 \nn
   \\
   && -D
   \left(-8 (\eta -4) \eta +s^2+4
   (\eta -4) \eta  s+14
   s-31\right)+(\eta -2) \left(4
   \eta ^2+4 \eta  (s-4)-s
   (s+2)+11\right) \bigg) \nn
   \eea 
   where $A$ is defined by (using the same
notations as in \cite{LRReview}) 
\begin{equation} 
A = \frac{\Pi(\eta,D)}{D^2-1} \quad , \quad 
\Pi(\eta,D) = (D^2-1)
 {\g(2-\eta-\frac{D}{2})\g(\frac{D}{2}+\frac{\eta}{2})^2
  \over4\fpi\g(2- \frac{\eta}{2})^2 \g(D+\eta )} \label{Pieteeta2new} 
\ee 
A useful check of this result is that for $s=0$ we recover the amplitudes $a_i$
obtained for the crumpling transition displayed in Eq. (121) of \cite{LRReview}. 
   
Now let us look for solutions such that the bare couplings $w_1$ and $w_2$ 
are non zero and that the matrix $\begin{pmatrix}  w_3 & w_4 \\ w_4 & w_5 \end{pmatrix}$
is invertible. This is the case for the bare vertex considered in the main text,
\eqref{barew} if we assume $\mu_0>0$ and $2 \mu_0 + D \lambda>0$. 
Note that this excludes some of the (unstable) fixed points discussed
in the RG section (see e.g. \eqref{FP0}, \eqref{FPs}), that
are not of interest to us here.
\\

From \eqref{tildew} in the limit $\q \to 0$ we then find
\bea
&& \tilde w_1(\q) \simeq \frac{1}{\pi_1(\q)}\;,\quad \tilde w_2(\q) \simeq \frac{1}{\pi_2(\q)}
\quad , \quad \begin{pmatrix} \tilde{w}_3(\q) & \tilde{w}_4(\q)\;,\\
\tilde{w}_4(\q) & \tilde{w}_5(\q) \end{pmatrix} 
\simeq \begin{pmatrix} \pi_3(\q) & \pi_4(\q) \\ \pi_4(\q) &
  \pi_5(\q) \end{pmatrix}^{-1}\;.
\label{inverse1} 
\eea 
Hence the dressed couplings are "soft" (screened), i.e. they vanish at small $q$ as
\be 
\tilde w_i(\q) \simeq Z_\kappa^2 \,  c_i(\eta,D,s)  \, q^{\eta_u}  \quad , \quad \eta_u=4-D - 2 \eta 
\ee
where the $c_i= c_i(\eta,D,s)$ are obtained from the $a_i=a_i(\eta,D,s)$ given in \eqref{ais} 
as
\be \label{ac} 
c_1= \frac{1}{a_1} \quad , \quad 
c_2= \frac{1}{a_2}
\quad , \quad 
\begin{pmatrix} c_3 & c_4 \\
c_4 & c_5 \end{pmatrix} 
\simeq \begin{pmatrix} a_3 & a_4 \\ a_4 &
  a_5 \end{pmatrix}^{-1}
\ee 
\\

\subsection{Self-energy}

We can now insert this result for $\tilde w_i(\q)$ into the first SCSA equation for the self-energy \eqref{sigma2}, obtaining
\be \label{sigma3} 
 \sigma_{\alpha \gamma}(\kb) = {2 \over d}  \sum_{i=1}^5 \int_q \tilde w_i(\q)
(W_i)_{\alpha \beta, \gamma \delta}(\q) 
{\cal G}_{\beta \delta}(\kb - \q) \simeq Z_\kappa^2 \sum_{i=1}^5 c_i(\eta,D,s)   
\sigma^i_{\alpha \gamma}(\kb)  
\ee
where the $\sigma^i_{\alpha \gamma}(\kb)$ are the following integrals
\be \label{sigmai0} 
\sigma^i_{\alpha \gamma}(\kb) \simeq
{2 \over d}   \int_q  \, 
(W_i)_{\alpha \beta, \gamma \delta}(\q) 
\frac{|\q|^{4-D - 2 \eta}}{|\kb - \q|^{2-\eta}} \left(  \frac{P_{\beta \delta}^L(\kb-\q)}{Z_\kappa} 
+ \frac{P_{\beta \delta}^T(\kb-\q)}{Z_g}  \right)
\ee
These momentum integrals are computed in Section IV. The result can be written as follows
\bea
\sigma^i_{\alpha \gamma}(\kb)    
= \frac{2}{d} k^{2-\eta} \frac{1}{Z_\kappa} 
( b_i^{(1)}(\eta,D,s) P_{\beta \delta}^T(\kb) + b_i(\eta,D,s) P_{\beta \delta}^L(\kb) )
\eea
where the amplitudes $b^1_i=b_i^{(1)}(\eta,D,s)$ and
$b_i = b_i(\eta,D,s)$ are given by
\bea \label{theb} 
&& 
b_1^1=-\frac{B (D-2) (D+1) \left(D^2 ((3
   \eta -4) s-1)+D^3 s-D (\eta +2
   s-2)-2 (\eta -2) ((2 (\eta -2)
   \eta +3) s-\eta )\right)}{2 (D-1)
   (D+\eta -2)}
   \\
   &&
 b_2^1 =  \frac{B
   \left(D^2 (-\eta 
   (s+2)+s+4)-D^3+D \left(\eta ^2-3
   \eta -2 (\eta -2)^2 s+4\right)+2
   \left((\eta -2)^3+(2 \eta -3)
   (\eta -2) s\right)\right)}{D+\eta
   -2}
 \nn  \\ &&
   b_3^1 =   \frac{B \left(D^2 (5-3 \eta )
   s+D^3 (-s)+D (\eta -1) s+D+(\eta
   -2) (-2 \eta +(4 (\eta -2) \eta
   +5) s+1)\right)}{(D-1) (D+\eta
   -2)}
 \nn   \\  &&
b_4^1 =   \frac{2 B (2
   \eta -3)
   (s-1)}{\sqrt{D-1}}
 \nn   \\ && 
b_5^1 =   B
   (D (s-2)-2 \eta +s+1) \nn
\eea
as well as
\bea 
 && b_1= -\frac{1}{2} B (D-2) (D+1) (D s+2
   \eta  s-2) \\
&& b_2=   
   -\frac{B
   (D-1) \left(D^2-s \left(D ((\eta
   -7) \eta +8)+2 (\eta
   -2)^3\right)+2 \eta
   -4\right)}{D+\eta -2} \nn \\
   && 
   b_3=
   B (-2 s
   (D+\eta )+D+s+1) \nn \\
 &&   b_4=
     -2 B \sqrt{D-1}
   (2 \eta -3) (s-1) \nn \\
  &&  b_5=
      \frac{B
   \left(D^2 (s-2)-D (2 \eta -3)
   (\eta  (s-2)-s+5)+(\eta -2) (2
   \eta  (2 \eta
   +s-8)-s+11)\right)}{D+\eta
   -2} \nn 
\eea 
where we have defined, as in \cite{LRReview} 
\bea \label{Sigma} 
B = \frac{\Sigma(\eta,D)}{D^2-1}  
\quad , \quad \Sigma(\eta,D) = \frac{\left(D^2-1\right) \Gamma
   (2-\eta) \Gamma
   \left(\frac{D}{2}+\frac{\eta}{2}\right
   ) \Gamma
   \left(\eta/2 \right
   )}{4 (4 \pi)^{D/2} \Gamma
   \left(2-\frac{{\eta}}{2}\right
   ) \Gamma
   \left(\frac{D}{2}+\eta\right) \Gamma
   \left(\frac{D}{2}-\frac{{\eta}}{2}+2 \right)} 
\eea 
As a useful check for $s=0$ we recover the amplitudes $b_i$
obtained for the crumpling transition in Eq. (124) of \cite{LRReview}.

\subsection{Self-consistency conditions}

Now we are ready to derive the final self-consistency conditions. 
To this end we equate the self-energy obtained by computing the
integrals in \eqref{sigma3}, \eqref{sigmai0} with our ansatz for
the self-energy \eqref{ansatzsigma},
\be \label{sigmasigma} 
\sigma_{\alpha \beta}(\kb) 
\simeq k^{2 - \eta}  \left( Z_\kappa P^L_{\alpha \beta}(\kb) + Z_g P^T_{\alpha \beta}(\kb)  \right) 
\simeq \frac{2}{d} 
Z_\kappa k^{2-\eta}  \sum_{i=1}^5 c_i(\eta,D,s)   
( b_i^{(1)}(\eta,D,s) P_{\beta \delta}^T(\kb) + b_i(\eta,D,s) P_{\beta \delta}^L(\kb) )
\ee 
We observe that the non-universal amplitude $Z_\kappa$ cancels, giving 
two self-consistency conditions
\bea \label{scF} 
&& \frac{d}{2} = F(\eta, D, s) \quad , \quad F(\eta, D, s)  = \sum_{i=1}^5 c_i(\eta,D,s)   b_i(\eta,D,s) \\
&& \frac{d}{2} = F_1(\eta, D, s) \quad , \quad F_1(\eta, D, s)  = s 
 \sum_{i=1}^5 c_i(\eta,D,s)   b^1_i(\eta,D,s) 
\eea 
where the first equation comes from identifying the longitudinal part
and the second the transverse part of the self-energy in \eqref{sigmasigma}. 
Using the relations \eqref{ac} between the $c_i(\eta,D,s)$
and the $a_i(\eta,D,s)$ one obtains the explicit expressions
for the functions $F_1=F_1(\eta,D,s)$ and $F=F(\eta,D,s)$ 
(with the dependence on $\eta,D,s$ suppressed)
\be  \label{FF} 
F_1 =  {b_1 \over a_1 } + \frac{b_2}{a_2} + { b_3 a_5 -b_4 a_4
+b_5 a_3 \over {a_3 a_5 - a_4^2 }}  \quad , \quad 
F_1 =  s \left( {b^1_1 \over a_1 } + \frac{b^1_2}{a_2} + { b^1_3 a_5 -b^1_4 a_4
+b^1_5 a_3 \over {a_3 a_5 - a_4^2 }} \right) 
\ee 
The expressions of the $a_i=a_i(\eta,D,s)$
and the $b_i=b_i(\eta,D,s)$ are given in \eqref{ais}, \eqref{Pieteeta2new}
and in \eqref{theb}, \eqref{Sigma}.

A priori the two equations \eqref{scF} determine the possible fixed point values 
of the pair $(\eta,s)$ for any given $D<4$.
However, as we will show below, and as is confirmed by a numerical search,
there appears to be only three possible 
solutions corresponding to $s=0,1,+\infty$. Let us first discuss each of these
special cases. Then, in the next subsection, we give an interpretation
of these three solutions as fixed points of an RG flow. 
\\

{\bf Crumpling transition $s=0$}. In the limit $g \to +\infty$, i.e. $s \to 0$ 
the tangent fields are purely longitudinal and one expects to
recover the crumpling transition. Setting $s=0$ in the 
first equation in \eqref{scF} gives
\be \label{finalcrumplingD} 
\frac{d}{2}  =
\frac{D (D+1) (D+\eta -4) (D+2 \eta
   -4) (2 D+2 \eta -3) \Gamma
   (2-\eta ) \Gamma
   \left(2-\frac{\eta }{2}\right)
   \Gamma \left(\frac{\eta
   }{2}\right) \Gamma (D+\eta )}{4
   (\eta -2) (D+\eta -1) (D+2 \eta
   -5) \Gamma
   \left(-\frac{D}{2}-\eta +2\right)
   \Gamma \left(\frac{1}{2} (D-\eta
   +4)\right) \Gamma
   \left(\frac{D}{2}+\eta \right)
   \Gamma \left(\frac{D+\eta
   }{2}\right)}
\ee 
which coincides with the SCSA equation which 
determines the exponent $\eta=\eta_{cr}(D,d)$ for the crumpling transition,
\cite{LRprl}, see in \cite{LRReview}. In $D=2$ it reduces to the root of a cubic equation
(with the smallest positive root as the physical solution)
\be \label{Crumple2D} 
\frac{d}{2}  =
\frac{12 (\eta -1)^2 (2 \eta +1)}{(\eta -4) \eta  (2
   \eta -3)}
\ee
which leads to the prediction $\eta_{cr}(D=3,d=2) =0.5352..$ 
for physical membranes \cite{LRprl}.
Let us recall that at large $d$ it gives
\bea \label{CDCrumpling} 
&& \eta_{cr}(D,d) \simeq \frac{C(D)}{d} + O(1/d^2)
\quad , \quad C(D) = \frac{(D-4)^2 (2 D-3) \Gamma (D+2)}{2 (5-D) (D-1)
   \Gamma \left(2-\frac{D}{2}\right) \Gamma
   \left(\frac{D}{2}+2\right) \Gamma
   \left(\frac{D}{2}\right)^2}\;,\nonumber
\eea
where $C(D)$ is exact by construction and agrees with the straight $1/d$ expansion
\cite{PaczuskiCrumplingLarged,ALGolubovic}. Expanding in $D=4-\epsilon$ one
obtains 
\bea 
\eta_{cr}(D,d) \simeq \frac{25}{3 d} \epsilon^3 + O(\epsilon^4) \quad , \quad 
\eea
This  result, which is $O(\epsilon^3)$, is consistent with the recent three loop RG calculation
in Ref. \cite{Mouhanna}. Since the SCSA builds on a large $d$ limit,
it does not see the disappearance of the critical point at $d=d_c \approx 219$,
found in the RG to $O(\epsilon)$, leading to a runaway flow,
often interpreted as a first order transition. The solution
of the SCSA describes instead a second-order critical crumpling transition
which survives for any $d$.
Interestingly, it was
obtained in Ref. \cite{Mouhanna} that to the next order, $d_c = 218.20 - 448.25 \epsilon$
decreases quite fast as $D$ is lowered, suggesting that the crumpling
critical point may survive to physical dimensions, $D=2$, $d=3$.

It is important to note that strictly for $s=0$ the second
SCSA equation in \eqref{scF} {\it must be dropped, i.e.
does not apply}. This is despite the fact that 
the r.h.s. of \eqref{sigmasigma} contains a correction
to the transverse part of the self-energy, proportional
to $F_1(\eta,D,s)/s=  \sum_{i=1}^5 c_i(\eta,D,s)   b^1_i(\eta,D,s)$,
which, as one can check, remains finite for $s \to 0$. This correction
is immaterial for $s=0$, since in that limit the tangent field is purely longitudinal,
and hence any term proportional to $\vec t_\alpha(\q) P^T_{\alpha \beta}(\q) 
\cdot \vec t_\beta(\q)$ identically vanishes. 
An analogous remark applies to the limit $s \to +\infty$, which is considered below.

\bigskip

{\bf Isotropic critical point, "tattered" $s=1$}.
This solution corresponds to $Z_\kappa=Z_g$, i.e. to an isotropic 
propagator with $\kappa=g$ in a
renormalized sense. It describes a tattered membrane at criticality.
For $s=1$ one finds
\be 
F_1(\eta,D,s=1) = F(\eta,D,s=1) 
\ee 
The two equations \eqref{scF} reduce to a single one,
\be 
\frac{d}{2} = -\frac{(D+1) (D-\eta +2) (D+2 \eta
   -4) (D+2 \eta -2) \Gamma (2-\eta
   ) \Gamma \left(2-\frac{\eta
   }{2}\right) \Gamma
   \left(\frac{\eta }{2}\right)
   \Gamma (D+\eta )}{8 (\eta -2)^2
   (D+\eta -1) \Gamma
   \left(-\frac{D}{2}-\eta +2\right)
   \Gamma \left(\frac{1}{2} (D-\eta
   +4)\right) \Gamma
   \left(\frac{D}{2}+\eta \right)
   \Gamma \left(\frac{D+\eta
   }{2}\right)}
\ee
which can equivalently be written in a slightly shorter form as
\be 
\frac{d}{2} = \frac{(D+1) 2^{D+\eta -4} (D+2 \eta
   -4) \csc \left(\frac{\pi  \eta
   }{2}\right) \Gamma (2-\eta ) \sin
   \left(\frac{1}{2} \pi  (D+2 \eta
   )\right) \Gamma \left(\frac{1}{2}
   (D+\eta -1)\right)}{\sqrt{\pi }
   (2-\eta) \Gamma
   \left(\frac{1}{2} (D-\eta
   +2)\right)}
   \ee
We can now look at the large $d$ expansion of $\eta$, by expanding
the r.h.s. for small $\eta$, obtaining 
\be
 \eta = \frac{2}{d} C(D) + O(\frac{1}{d^2})  \quad , \quad C(D)=
-\frac{(D-4) (D-2) (D+1) (D+2)
   \Gamma (D)}{16 (D-1) \Gamma
   \left(2-\frac{D}{2}\right) \Gamma
   \left(\frac{D}{2}\right)^2 \Gamma
   \left(\frac{D+4}{2}\right)}
   =  \frac{2^{D-4} (D-4) (D+1) \sin \left(\frac{\pi  D}{2}\right) \Gamma
   \left(\frac{D-1}{2}\right)}{\pi ^{3/2} \Gamma \left(\frac{D}{2}+1\right)} 
\ee
We find that $C(D)$ vanishes for $D=2$ and $D=4$, and is positive in-between,
with a maximum a bit below $D=3$, with
\be
 C(3) = \frac{8}{3 \pi^2} \quad , \quad  C(D) 
   = \frac{3 (D-2)}{4}+O\left((D-2)^2\right) \quad , \quad 
   C(D=4-\epsilon) =
    \frac{5}{8} \epsilon^2+O\left(\epsilon^3\right)
\ee
The vanishing of $\eta$ in $D=2$ to this order is consistent
with a lower critical dimension of the tattered membrane 
critical point equal to $2$. However, the stability of the ordered phase
of the tattered membrane remains an open problem
that we leave for future studies.

The above results can be compared with the predictions by Bray
for the $O(d)$ model at first order in $O(1/d)$. We find
\be 
C(D) = \frac{D+1}{2} C_{O(d)}(D) \quad , \quad 
C_{O(d)}(D)= \frac{2 (4-D) \Gamma (D-2)}{D \Gamma
   \left(2-\frac{D}{2}\right) \Gamma
   \left(\frac{D}{2}-1\right)^2
   \Gamma \left(\frac{D}{2}\right)}
\ee

As already discussed in the RG section we expect our isotropic "tattered membrane" fixed point for $s=1$ to be
directly related to the so-called "chiral" critical fixed point of the model $O(n) \times O(m)$ 
studied in e.g. \cite{VicariNM2001}, provided one identifies $n=d$ and $m=D$ the space
dimension. With this correspondence, to leading order in $1/n$ \cite{VicariNM2001},
their result for $\eta$ translates into 
\be 
\eta = \frac{2}{d} \frac{D+1}{2} C_{O(d)}(D) + O(\frac{1}{d^2}) 
\ee 
which agrees with our prediction. They have pushed their analysis to second order in $1/n$.
For general $D$ it leads to a complicated expression, which simplifies for $D=3$.
Setting $m=D=3$ in their result (Eq. (4.43) there) gives
\be 
\eta =\frac{16}{3 \pi^2 d}  - \frac{608}{27 \pi^4 d^2} + O(1/d^3) 
\ee 
This can be directly compared to the SCSA, which gives a result for general $d$.
Expanding this result to the next order in $1/d$ we find for $D=3$
\be 
 \eta = \frac{2}{d} C(3) - \frac{14 C(3)^2}{3 d^2}= 
 \frac{16}{3 \pi^2 d}  - \frac{896}{27 \pi^4 d^2} + O(1/d^3) 
\ee 
While the leading term agrees, the next order correction in the SCSA is an approximation,
but gives a value reasonably close to the exact result at second order in $O(1/d^2)$.

In $D=3$ and finite $d$, the SCSA gives the equation at the isotropic fixed point, together
with their numerical solutions
\bea 
&& \frac{d}{2} = 
\frac{8 (1-2 \eta ) \eta  \cot
   \left(\frac{\pi  \eta }{2}\right)
   \cot (\pi  \eta )}{(3-\eta)
   (2-\eta)} \\
&& 
\eta_{D=3,d=1} \approx 0.250337 \quad , \quad 
\eta_{D=3,d=2} \approx  0.177214 \quad , \quad
\eta_{D=3,d=3} \approx  0.136135 \quad , \quad
\eta_{D=3,d=4} \approx  0.11
\eea
In exactly $D=2$ (or as $D \to 2^+$) the SCSA naively gives 
a finite limit for $\eta$ for $d\leq 3$. Indeed the function $F(\eta,D=2^+,s=1)$ 
is a non-trivial fonction of $\eta$ decreasing from $3/2$ at $\eta=0$
to $0$ at $\eta=1$. However at $D=2$ this branch is not continuously connected
to $\eta=0$ at $d=\infty$, hence it may be a spurious solution.

Finally, around $D=4$ we note that the SCSA gives the same result 
$\eta \simeq \frac{5}{4 d} \epsilon^2$
as the straight $1/d$ expansion, 
consistent with the $O(\epsilon^2)$ calculation in \cite{VicariNM2001},
see Eq. \eqref{etaVicari}. 

\bigskip

{\bf Transverse critical fixed point, $s=+\infty$}.
We also find another fixed point for $s=+\infty$ corresponding
to $\kappa \to +\infty$. In that case the $\vec t_\alpha$ vector field is
purely transverse. In $D=2$ since 
$P^T_{\alpha,\beta}(\q)=\q^T_\alpha \q^T_\beta$,
where $\q^T$ is orthogonal to $\q$, 
there is an exact duality in momentum space, $\q \to \q^T$ which
maps this case onto $\vec t_\alpha$ vector field which 
are purely longitudinal, i.e. onto the membrane and
the crumpling transition. This mapping extends
to the low-temperature phase, and thus predicts that this "transverse" model
exhibits an ordered phase in $D=2$, dual of the flat phase 
of crystalline membranes.

{\bf Duality for $D=2$ and arbitrary $s$}. Before studying $s=+\infty$ let us comment on this duality
within SCSA for general $s$, and that it relates $s \to 1/s$. 
Indeed, one can check on the explicit formula that the
two functions which appear in \eqref{scF} are related according to
\bea \label{duality}
F_1(\eta,D=2,s) = F(\eta,D=2,1/s) 
\eea 
This is a non-trivial identity, which provides a good check of the
calculations. It is a consequence of
the above duality $\q \leftrightarrow \q^T$ in momentum space. 

{\bf Absence of other SCSA solutions} One application of this formula is that in $D=2$ it is now simple to exclude solutions of SCSA for finite $s$ apart from $s=1$. Indeed Eqs. \eqref{scF} and \eqref{duality} imply that 
\be \frac{d}{2} = F_1(\eta,D=2,s) = F_1(\eta,D=2,1/s) 
\ee 
so one can check whether $F_1$ intersects $d/2$ for two different values of $s$.
But a plot of $F_1$ for fixed $\eta \in [0,1]$ shows that it is 
a monotonic function of $s$.
To further exclude solutions at finite $s \neq 1$ in 
$D=3$ one can instead plot both $F_1(\eta,D,s), F(\eta,D,s)$
for fixed $\eta \in [0,1]$
and check that it has a single intersection at $s=1$.

Let us now return to the SCSA fixed point at $s=+\infty$. Again 
one of the two equations in \eqref{scF} drops out,
this time only the second equation remain, so $\eta$ is determined by
\bea 
1 = \frac{2}{d} F_1(\eta,D,+\infty)  
\eea 
The r.h.s. is complicated for general $D$. 
For $D=2$ one recovers exactly the equation \eqref{Crumple2D} 
which determines the crumpling exponent, hence $\eta = \eta_{cr}(D=2,d)$.

For $D=3$ it gives
\be 
\frac{d}{2}  = \frac{8 \eta  (1-2 \eta) \left(24
   \eta ^6-40 \eta ^5-321 \eta
   ^4+410 \eta ^3+1119 \eta ^2-1042
   \eta -990\right) \cot
   \left(\frac{\pi  \eta }{2}\right)
   \cot (\pi  \eta )}{(\eta -5)
   (\eta -3) (\eta +3) (2 \eta +1)
   (3 \eta -5) \left(2 \eta ^3-3
   \eta ^2-6 \eta +14\right)}
\ee 
which can be compared with the equation which determines $\eta$ for
the crumpling transition in $D=3$, from \eqref{finalcrumplingD} 
\be 
\frac{d}{2} = \frac{24 \eta  (1-2 \eta) (2 \eta
   +3) \cot \left(\frac{\pi  \eta
   }{2}\right) \cot (\pi  \eta
   )}{(\eta -5) (\eta -3) (2 \eta
   +1)}
\ee 
and one can see that for general $D>2$ the two fixed points are quite different.
This leads to the values 
\bea
\eta_{cr} = 0.329074 \quad , \quad \eta_{tr} = 0.299348   \quad , \quad D=3, d=1 \\
\eta_{cr} = 0.273121 \quad , \quad \eta_{tr} = 0.233722 \quad , \quad D=3, d=2 \\
\eta_{cr} = 0.23632 \quad , \quad \eta_{tr} = 0.192301 \quad , \quad D=3, d=3 \\
\eta_{cr} = 0.209223 \quad , \quad \eta_{tr} = 0.163235 \quad , \quad D=3, d=4 
\eea
hence the $\eta=\eta_{tr}$ exponent of the transverse FP is smaller
than the one for crumpling critical point.
The large $d$ expansion of the SCSA result gives $\eta = \frac{2}{d} C_{\rm tr}(D)  + O(1/d^2)$ with
\be \label{Ctr} 
C_{\rm tr}(D)= 
\frac{(4-D) (D+1) \left(4 D^8-22
   D^7+43 D^6-62 D^5+228 D^4-623
   D^3+802 D^2-464 D+96\right)
   \Gamma (D)}{4 (D-1)^2 (2 D-1) (3
   D-5) (3 D-2) \left(D^2-3
   D+3\right) \Gamma
   \left(2-\frac{D}{2}\right) \Gamma
   \left(\frac{D}{2}\right)^2 \Gamma
   \left(\frac{D+4}{2}\right)}
\ee 
to be compared with the analogous result for crumpling in \eqref{CDCrumpling}.
One finds $C_{\rm tr}(2)=1$ which is the same as for crumpling. 
For $D=3$ one finds $C_{\rm tr}(3)=176/(35 \pi^2) \approx 0.509501$, significantly smaller than
$C(3)=48/(5 \pi^2) \approx 0.972683$ for the crumpling transition.

Finally around $D=4$ the SCSA gives for the transverse FP
\be 
\eta =  \frac{16 \epsilon^2}{9 d} + O(\epsilon^3) 
\ee 
which coincides with the exact $O(1/d)$ expansion,
and is distinct from the $\eta$ exponents for crumpling and for the isotropic 
critical points.

{\it Lower critical dimension}. Finally, we can obtain the lower critical dimension
of the transverse fixed point. It is defined by the relation
$2 - \eta(D_{lc},d)=D_{lc}$. Inserting the relation $\eta=2-D$ inside
the equation $F_1(\eta,D,+\infty)=d/2$ we obtain that
$D=D_{lc}(d)$ is the root of
\be 
d = \frac{D \left(D^4-D^3-7 D^2+5
   D+8\right)}{2 (D-2) (D-1)
   \left(D^2-D-1\right)}
\ee 
One finds (one must choose the branch
continuously related to $D=D_{lc}(+\infty)=2$)
\bea 
&& D_{lc}(d)=  2 -\frac{2}{d} -\frac{8}{d^2}-\frac{26}{d^3} + O(1/d^4) \\
&& D_{lc}(40) =  1.94462 ~,~ 
D_{lc}(20) =  1.87777 ~,~
D_{lc}(10) = 1.75041 ~,~ \\
&& D_{lc}(4) = 1.68255 ~,~
D_{lc}(3) = 1.67661 ~,~
D_{lc}(2) = 1.67159 ~,~
D_{lc}(1) = 1.66732 ~,~
D_{lc}(0) = 1.66364 
\eea 
The function $D_{lc}(d)$ varies mostly for $d$ of order $10-20$.
Let us recall that for the crumpling transition has a much simpler expression $D_{lc}= 2 d/(1+d)$.

\subsection{RG version of SCSA and flow of $s=\kappa/g$} 

To understand why there are only three critical fixed points discussed above,
we now use SCSA to derive the RG flow of the parameter $s=s(\ell)=\kappa/g$
at scale $k=\Lambda e^{-\ell}$. 

As was explained in details in \cite{OurBuckling} (see section C.3. of the Supp. Mat. there)
one can recast the SCSA equations into an RG flow. It can be done so as
to recover either (i) only the RG functions to first order in $1/d$, or (ii) within the full self-consistent scheme. 
We will only consider here the first approach, which is exact to $O(1/d)$. 

Refering to \cite{OurBuckling} for details, one defines within the SCSA
the natural dimensionless coupling for the RG as
\be \label{hatw} 
\hat w_i := \tilde w_i(q) Z_\kappa^{-2} q^{-\epsilon} 
\ee
In terms of these couplings we obtain the RG equation for $d=+\infty$, exact for any $\epsilon=4-D$,
by taking a derivative $\partial_\ell=- q \partial_q$ on both sides 
of \eqref{tildew} and using \eqref{piq}. This gives
\bea \label{rglarge} 
&& \partial_\ell \hat w_i = \epsilon \hat w_i - \epsilon a_i(0,D,s) \hat w_i^2  \quad , \quad i=1,2 \\
&& \partial_\ell \begin{pmatrix} \hat{w}_3 & \hat{w}_4 \\
\hat{w}_4 & \hat{w}_5\end{pmatrix} = \epsilon  \begin{pmatrix} \hat{w}_3 & \hat{w}_4 \\
\hat{w}_4 & \hat{w}_5\end{pmatrix} -  \epsilon \begin{pmatrix} \hat{w}_3 & \hat{w}_4 \\
\hat{w}_4 & \hat{w}_5\end{pmatrix}
 \begin{pmatrix} a_3(0,D,s) & a_4(0,D,s) \\
a_4(0,D,s) & a_5(0,D,s) \end{pmatrix}
 \begin{pmatrix} \hat{w}_3 & \hat{w}_4 \\
\hat{w}_4 & \hat{w}_5\end{pmatrix}
\eea 
where the functions $a_i(\eta,D,s)$ are those defined in \eqref{piq} and given in \eqref{ais},
\eqref{Pieteeta2new}. Here, since $\eta = O(1/d)$, they have to be evaluated at $\eta=0$,
where they reach finite values. Whenever $s$ can be considered as fixed, the 
fixed point of these RG equations is $\hat w_i=\hat w_i^*$ with
\be \label{fp2} 
\hat w_i^* = \frac{1}{a_i(0,D,s)} \quad , \quad i=1,2 \quad , \quad \begin{pmatrix} \hat{w}^*_3 & \hat{w}^*_4 \\
\hat{w}^*_4 & \hat{w}^*_5\end{pmatrix} =  \begin{pmatrix} a_3(0,D,s) & a_4(0,D,s) \\
a_4(0,D,s) & a_5(0,D,s) \end{pmatrix}^{-1}
\ee

However, $s$, defined as the renormalized value of
$\kappa/g$ has a non-trivial RG flow.
To obtain this flow we
apply $- k \partial_k$ to the equation for the
self-energy as explained in \cite{OurBuckling}.
Here it is convenient to define 
the transverse and longitudinal components
of the self-energy, within the SCSA, and 
write them as
\bea 
&& \sigma_T(k) \simeq Z_g k^{2-\eta} = Z_g(k) k^2  \\
&& \sigma_L(k) \simeq Z_\kappa k^{2-\eta} = Z_\kappa(k) k^2 
\eea 
We then define the two "$\eta$ functions" or anomalous dimensions of
$\kappa$ and $g$ respectively as
\bea 
&& \eta_g = - k \partial_k \log Z_g(k) = - \frac{1}{Z_g(k)} k \partial_k  (\sigma_T(k)/k^2) \\
&& = 
- \frac{1}{Z_g(k)} k \partial_k (  \frac{2}{d} Z_\kappa(k)  \sum_{i=1}^5 \hat w_i   b_i^1(\eta,D,s) ) 
\simeq  \frac{2}{d}  s \sum_{i=1}^5 \hat w_i  
\tilde b_i^1(D,s) \quad , \quad \tilde b_i^1(D,s)  = \lim_{\eta \to 0} 
(\eta  b_i^1(\eta,D,s) ) )  
\eea 
Similarly we define
\bea 
&& \eta_\kappa = - k \partial_k \log Z_\kappa(k) \simeq 
\frac{2}{d}   \sum_{i=1}^5 \hat w_i  
\tilde b_i(D,s) \quad , \quad \tilde b_i(D,s)  = \lim_{\eta \to 0} 
(\eta  b_i^1(\eta,D,s) ) )  
\eea
where we substitute the fixed point values for $\hat w_i$ in \eqref{fp2} 
for a given $s$, and we work to first order in $O(1/d)$. 
Using the definitions in \eqref{scF} it can also be written as
\bea 
&& \eta_g =  \frac{2}{d}  f_1(D,s) \quad , \quad  f_1(D,s)  = \lim_{\eta \to 0}  \eta F_1(\eta,D,s) \\
&& \eta_\kappa =  \frac{2}{d}  f(D,s) \quad , \quad  f(D,s)  = \lim_{\eta \to 0}  \eta F(\eta,D,s) 
\eea 
We can now write the flow of $s=\kappa/g$ as
\be 
\partial_\ell s = (\eta_\kappa - \eta_g ) \, s = \frac{2}{d} \left[ f(D,s) - f_1(D,s) \right] \, s  \label{flows} 
\ee 
Its expression as a function of general $s,D$ is quite cumbersome, so we focus on
the main features. 

First we find that $f(D,s=1)=f_1(D,s=1)$ for any $D$ in \eqref{flows} 
so that for $\kappa=g$, $s$ does not flow, hence it corresponds to a fixed point
which describes the critical point for an isotropic 
tattered membrane. One obtains
the explicit expressions in $D=3$, near $D=2$ and near $D=4$
\bea 
&& \frac{d}{2} (\eta_\kappa-\eta_g) = -\frac{16 (s-1) \left(495 s^5+377
   s^4+238 s^3+246 s^2+43
   s+9\right)}{15 \pi ^2 (s+1)^2 (3
   s+1) (7 s+1) \left(5 s^2+2
   s+1\right)}   \quad , \quad D=3 \\
&& = -\frac{(D-2) (s-1) \left(3 s^2-2
   s+3\right)}{8 s (s+1)} + O((D-2)^2) \\
   && =
-\frac{8 (s-1) s \left(49 s^4+2
   s^3+21 s^2+8 s+4\right) \epsilon^2}{3 \left(3 s^2+1\right)
   \left(7 s^2+4 s+1\right)^2} + O(\epsilon^3) \quad , \quad D=4 -\epsilon 
\eea  
One sees that the flow of $s$ vanishes in $D=2$ for any fixed $s>0$
and finite, as in fact both exponents $\eta_\kappa$ and $\eta_g$ vanish
in that limit. However the limits $s \to 0$ and $D \to 2$,
as well as $s \to +\infty$ and $D \to 2$ do not commute,
as the points $s=0,+\infty$ are special (and dual to each others
in $D=2$, as discussed above). 
\\

We now extract the behavior of the flow of $s$ around its three fixed points
$s=0,1,+\infty$. 

(i) Linearizing \eqref{flows} around $s=1$, one finds, for general $D$, up to $O(1/d^2)$ terms
\bea 
&& \eta_\kappa-\eta_g =  \omega_1(D,d) (s-1) +O\left((s-1)^2\right) \\
&& \omega_1(D,d)= - \frac{2}{d} 
\frac{(4-D) (D-2) \left(D^3-D^2+4
   D-8\right)  \Gamma (D)}{16
   (D-1)^2 \Gamma
   \left(2-\frac{D}{2}\right) \Gamma
   \left(\frac{D}{2}+2\right) \Gamma
   \left(\frac{D}{2}\right)^2}  
\eea 
This defines an eigenvalue exponent $\omega_1=\omega_1(D,d)$, such that $s-1$ flows as $k^{-\omega_1}$
for $k \to 0$.
We find that $\omega_1(D,d)$ vanishes for $D=2$ as 
$\omega_1(D,d) \simeq - (D-2)/(2 d)$, and for $D=4$ 
as $\omega_1(D=4-\epsilon,d) \simeq - 7 \epsilon^2/(9 d)$,
and is negative in between
with $\omega_1(D=3,d) = - 44/(15 d \pi^2)$. 
Hence $s$ flows to $1$ at large scale, and the isotropic tattered membrane 
fixed point $s=1$ is stable.

(ii) Linearizing \eqref{flows} around $s=0$ one finds, up to $O(1/d^2)$ 
\bea
&& \eta_\kappa-\eta_g =  \omega_0(D,d) + O(s) \quad , \quad 
\omega_0(D,d) = \frac{2}{d} 
\frac{(D-4)^2 D (D+1) (2 D-3)
   \Gamma (D)}{4 (5-D) (D-1) \Gamma
   \left(2-\frac{D}{2}\right) \Gamma
   \left(\frac{D}{2}+2\right) \Gamma
   \left(\frac{D}{2}\right)^2} \\
&&    \omega_0(D=3,d)=\frac{96}{5 d \pi ^2}  \quad , \quad \omega_0(D=2,d)=\frac{2}{d}
   \quad , \quad \omega_0(D=4-\epsilon,d) \simeq
  \frac{25 \epsilon^3}{3 d}
\eea
It turns out that to this order in $1/d$ this exponent is nothing but the crumpling exponent
\be 
\omega_0(D,d) =  \eta_{cr}(D,d) 
\ee 
This is because $\eta_g$ vanishes at $s=0$. Since
$\omega_0(D,d)>0$, it shows that any initial
small transverse perturbation of the tangent fields
with $s=\kappa/g>0$ grows as $k^{- \omega_0(D,d)}$
at large scale. This gives information on
how these "tattering" or defects destroy the
purely longitudinal constraint of an elastic membrane at large scale.

(iii) Finally, linearizing \eqref{flows} around $s=+\infty$ one finds, up to $O(1/d^2)$ 

\bea
&& \partial_\ell s =  \omega_{\infty}(D,d)  s  \quad , \quad \partial_\ell \left(\frac{1}{s} \right) = - \omega_{\infty}(D,d) \frac{1}{s} \\
&& \omega_{\infty}(D,d) = - \frac{2}{d} C_{\rm tr}(D) 
\eea 
where $C_{\rm tr}(D)$ was given in \eqref{Ctr}, with its value in $D=3$ and
$D=2$ were given, as well at its expansion for $D=4-\epsilon$.
The exponent $\omega_{\infty}(D,d)$
is negative, which means that any small positive
value of $1/s=g/\kappa>0$ grows as $k^\omega_{\infty}(D,d)$.
In $D=2$ the behavior around $s=0$ and $s=+\infty$ are
dual of each others (by the transformation $s \to 1/s$).
In general dimensions $D<4$ both of these fixed points
are unstable towards the isotropic one, $s=1$.

\subsection{Limit from the SCSA to the RG around $D=4$} 

In this section we show how the SCSA recovers the
results from the RG to $O(\epsilon)$ performed
in Section I near $D=4-\epsilon$.

%

We will first identify the fixed points. 
Let us consider the amplitudes $a_i(\eta,D,s)$ given in \eqref{ais}.
Since the SCSA gives that $\eta$ is
at most of order $\eta = O(\epsilon^2)$
at any of the fixed points, we can 
set $\eta$ to zero in the formula \eqref{ais},
and expand these amplitudes
to leading order at small $\epsilon$.
Similarly, we know that to one-loop order
$s$ does not flow so we can consider
any fixed value of $s$.
We obtain
\be
a_1=a_2=2 A (s (7 s+4)+1) \;,\;\;a_3=A (s (17 s+2)+5) \;,\;\;a_4=\sqrt{3}
   A (s-1)^2 \;,\;\;a_5= 3 A (s (5 s+2)+1)
\label{epsAn} 
\ee
where from \eqref{Pieteeta2new} one has
\begin{equation}
A = \frac{1}{192 \pi ^2 \epsilon}\;.
\end{equation}
We now use Eqs. \eqref{piq} for $\pi_i(\q)$ and 
\eqref{inverse1} for the dressed couplings $\tilde w_i(\q)$.
In terms of the dimensionless couplings $\hat w_i=\tilde w_i(\q) Z_\kappa^{-2} q^{-\epsilon}$ 
defined in \eqref{hatw} we obtain to this order in $\epsilon$, near $D=4$,
%

\bea
&& \hat w_1 =  \hat w_2 
\simeq  \frac{1}{2 A (s (7 s+4)+1)}\\
&& \begin{pmatrix} \hat{w}_3 & \hat{w}_4 \\ \hat{w}_4 & \hat{w}_5 \end{pmatrix} 
= \frac{1}{A}
   \frac{1}{
   \left(3 s^2+1\right) (s (7
   s+4)+1)}
   \left(
\begin{array}{cc}
 \frac{1}{4} (s (5 s+2)+1) &
   -\frac{(s-1)^2}{4 \sqrt{3}} \\
 -\frac{(s-1)^2}{4 \sqrt{3}} &
   \frac{1}{12} (s (17 s+2)+5) \\
\end{array}
\right)
   \eea

%
Remarkably, comparing with \eqref{barew} and 
\eqref{bareR}, we see that the dressed
vertex can again be parameterized by only two renormalized Lam\'e moduli
(as in the bare theory with local elasticity). In terms of the
dimensionless couplings it reads
\begin{equation}
\hat w_1=\hat w_2
= 8 \pi^2 \hat u \quad \hat w_3=
 8 \pi^2 (3 \hat v + \hat u)
\quad \hat w_4 = 
  8 \pi^2 \sqrt{3} \hat  v 
\quad \hat w_5 = 
 8 \pi^2 ( \hat  v+ \hat v) u
\end{equation} 
where the factor $8 \pi^2$ comes from the definition of 
the dimensionless RG couplings \eqref{defuhat}. Using simple
algebra we exactly recover for $\hat u$ and $\hat v$ 
the values at the one-loop stable fixed point 
(the third fixed point) in \eqref{FPs}, It is important to
note that since $\eta$ vanishes quite fast (at least as $\epsilon^2$)
at the fixed points of interest, the SCSA to $O(\epsilon)$
is indistinguishable from the straight $O(1/d)$ expansion. 
This is why we only recover the large $d$ limit of
the RG equations of Section I.

%

By similar manipulations, it is easy to show that the RG version of the 
SCSA, i.e. the Eqs. \eqref{rglarge}, exactly recover
 the large $d$ limit of the one-loop RG equations 
for the coupling constants $\hat u$ and $\hat v$
obtained in Section I, namely the Eqs. \eqref{RGlargeds}.

\section{III. Computation of the vacuum polarization $\pi_i(\qb)$}

In this section we present all the details of the computation of the the vacuum polarization integrals.
Here and in the following section it is more convenient 
to parameterize the dressed propagator as
\bea \label{formG} 
{\cal G}_{\alpha \beta}(\kb)  \simeq ( \frac{1}{Z_1} \delta_{\alpha \beta} + \frac{1}{Z_2} P_{\alpha \beta}^L(\kb)
) k^{-2 + \eta}  \quad , \quad \frac{1}{Z_1}=\frac{1}{Z_g} \quad , \quad 
\frac{1}{Z_2}=\frac{1}{Z_\kappa}- \frac{1}{Z_g}
\eea 
We will express all the $\pi_i(\q)$ using the amplitudes $Z_1$ and $Z_2$
and their ratio, 
and at the end transform back in terms of $Z_\kappa$, $Z_g$ and $s=Z_\kappa/Z_g$. 

We now insert the form \eqref{formG} for the propagator inside the vacuum polarization
integral. 
It then splits into four terms
\be
\Pi_{\alpha \beta, \gamma \delta}(\p)
=  {\rm sym}_{\alpha \beta} {\rm sym}_{\delta \gamma}  \int_q \Gc_{\alpha \gamma}(\q) \Gc_{\beta \delta}(\p-\q) = \Pi^{11}_{\alpha \beta, \gamma \delta}(\p)
+ \Pi^{12}_{\alpha \beta, \gamma \delta}(\p) 
+ \Pi^{21}_{\alpha \beta, \gamma \delta}(\p) + \Pi^{22}_{\alpha \beta, \gamma \delta}(\p)
\ee
with, denoting $2 a=2 b = 2 -\eta$,
\bea
&& \Pi^{11}_{\alpha \beta, \gamma \delta}(\p)
= {1 \over Z_1^2} {\rm sym}_{\alpha \beta} {\rm sym}_{\delta \gamma}  \delta_{\alpha \gamma} 
\delta_{\beta \delta} 
\int_q \frac{1}{(\pb + \q)^{2a} q^{2 b}}  \\
&& \Pi^{21}_{\alpha \beta, \gamma \delta}(\p)
= {1 \over Z_1 Z_2} {\rm sym}_{\alpha \beta} {\rm sym}_{\delta \gamma}  \delta_{\beta \delta} 
\int_q q_\alpha q_\gamma \frac{1}{(\pb + \q)^{2a} q^{2 b+2}}  \\
&& \Pi^{12}_{\alpha \beta, \gamma \delta}(\p)
= {1 \over Z_1 Z_2} {\rm sym}_{\alpha \beta} {\rm sym}_{\delta \gamma}  \delta_{\alpha \gamma} 
\int_q (p_\beta+q_\beta) 
(p_\delta+q_\delta)  \frac{1}{(\pb + \q)^{2a+2} q^{2 b}}  \\
&& 
= {1 \over Z_1 Z_2} {\rm sym}_{\alpha \beta} {\rm sym}_{\delta \gamma}  \delta_{\alpha \gamma} 
\int_q q_\beta q_\delta  \frac{1}{(\pb + \q)^{2b} q^{2 a+2}} \\
&&  \Pi^{22}_{\alpha \beta, \gamma \delta}(\p)
= {1 \over Z_2^2} {\rm sym}_{\alpha \beta} {\rm sym}_{\delta \gamma}  \delta_{\beta \delta} 
\int_q q_\alpha q_\beta q_\gamma q_\delta \frac{1}{(\pb + \q)^{2a+2} q^{2 b+2}} 
\eea
We have kept $a,b$ arbitrary but here since $\eta_g=\eta$ 
all integrals have $a=b = (2 -\eta)/2$ and one finds that $\Pi^{21}_{\alpha \beta, \gamma \delta}(\p)=\Pi^{12}_{\alpha \beta, \gamma \delta}(\p)$. Hence we have
\be
\Pi_{\alpha \beta, \gamma \delta}(\p)= \Pi^{11}_{\alpha \beta, \gamma \delta}(\p)
+ 2 \Pi^{12}_{\alpha \beta, \gamma \delta}(\p) + \Pi^{22}_{\alpha \beta, \gamma \delta}(\p)
\ee

Let us define the function
\bea \label{defH}
H_{n,m}(a,b) =  {\g(a+b-D/2-n)\g(D/2-a+n)\g(D/2-b+m-n)
  \over2^n \fpi\g(a)\g(b)\g(D-a-b+m)}
\eea 
Then (see e.g. the Appendices of \cite{LRReview}) the following integrals can be expressed as
\bea \label{3integrals} 
&& I(a,b) =\int_q{1\over(\p+\q)^{2a} q^{2b}} = p^{D-2 a - 2 b} H_{0,0}(a,b) \\
&& I_{\alpha_1}(a,b) =\int_q{q_{\alpha_1}\over(\p+\q)^{2a} q^{2b}}= -  p^{D-2a-2b+1 } \hat p_{\alpha_1} H_{0,1}(a,b) \\
&& I_{\alpha_1 \alpha_2}(a,b) = \int_q{q_{\alpha_1}q_{\alpha_2}\over(\p+\q)^{2a} q^{2b}} =
p^{D-2a-2b+2}\left[\delta_{\alpha_1 \alpha_2} H_{1,2}(a,b) +
 \hat p_{\alpha_1} \hat p_{ \alpha_2} H_{0,2}(a,b) \right] 
\eea
\\

Let us start with $\Pi^{11}$ and determine its components $\pi_i^{11}$.
We obtain, with $2 a=2 b = 2 -\eta$
\bea
\Pi^{11}_{\alpha \beta, \gamma \delta}(\p)
= {1 \over 2 Z_1^2} (\delta_{\alpha \gamma} \delta_{\beta \delta} + \delta_{\alpha \delta} \delta_{\beta \gamma} )
p^{D-2 a - 2 b} H_{0,0}(a,b)
\eea 
Using the inversion formula \eqref{inversion} we find
\bea
&& (D-1) \pi^{11}_3(\p) = {1 \over 2 Z_1^2} p^{D-2 a - 2 b} H_{0,0}(a,b) 
P^T_{\alpha \beta}(\p)  (\delta_{\alpha \gamma} \delta_{\beta \delta} + \delta_{\alpha \delta} \delta_{\beta \gamma} ) P^T_{\gamma \delta}(\p) 
= {1 \over Z_1^2} (D-1) p^{D-2 a - 2 b} H_{0,0}(a,b) 
\eea 
\bea
&&  \pi^{11}_5(\p) = {1 \over 2 Z_1^2} p^{D-2 a - 2 b} H_{0,0}(a,b) 
P^L_{\alpha \beta}(\p)  (\delta_{\alpha \gamma} \delta_{\beta \delta} + \delta_{\alpha \delta} \delta_{\beta \gamma} ) P^L_{\gamma \delta}(\p) 
= {1 \over Z_1^2} p^{D-2 a - 2 b} H_{0,0}(a,b) 
\eea 
\bea
&&  \sqrt{D-1}\pi^{11}_4(\p) = {1 \over 2 Z_1^2} p^{D-2 a - 2 b} H_{0,0}(a,b) 
P^L_{\alpha \beta}(\p)  (\delta_{\alpha \gamma} \delta_{\beta \delta} + \delta_{\alpha \delta} \delta_{\beta \gamma} ) P^T_{\gamma \delta}(\p) = 0
\eea 
Next
\bea
\pi_2^{11}(\p) &=& \frac{1}{D-1} (W_2)_{\alpha \beta,\gamma\delta}  \Pi^{11}_{\alpha \beta, \gamma \delta}(\p) \\
&& = \frac{1}{2(D-1)} {1 \over 2 Z_1^2} p^{D-2 a - 2 b} H_{0,0}(a,b)  ( 
P^T_{\alpha \gamma} P^L_{\beta \delta } 
+ P^T_{\alpha \delta} P^L_{\beta \gamma} + P^L_{\alpha \gamma}
P^T_{\beta \delta } + P^L_{\alpha \delta} P^T_{\beta \gamma})
 (\delta_{\alpha \gamma} \delta_{\beta \delta} + \delta_{\alpha \delta} \delta_{\beta \gamma} ) \\
 && =  {1 \over Z_1^2} p^{D-2 a - 2 b} H_{0,0}(a,b) 
\eea

Finally, using \eqref{inversionpi1},
one finds
\bea
\pi_1^{11}(\p) &=& \frac{2}{(D-2) (D+1)}  
{1 \over Z_1^2} p^{D-2 a - 2 b} H_{0,0}(a,b) 
[ {1\over 2}(\delta_{\alpha\gamma}\delta_{\beta\delta}
+\delta_{\alpha\delta}\delta_{\beta\gamma})   
{1 \over 2} (\delta_{\alpha \gamma} \delta_{\beta \delta} + \delta_{\alpha \delta} \delta_{\beta \gamma} )
- 2 
- (D-1)  ] \\
&& = \frac{2}{(D-2) (D+1)}  
{1 \over Z_1^2} p^{D-2 a - 2 b} H_{0,0}(a,b) 
[ \frac{D (D+1)}{2} 
- 2 
- (D-1)  ] = {1 \over Z_1^2} p^{D-2 a - 2 b} H_{0,0}(a,b) 
\eea 

Hence to summarize
\bea
(\pi_1^{11}(\p),\pi_2^{11}(\p),\pi_3^{11}(\p),\pi_4^{11}(\p),\pi_5^{11}(\p)) = 
 {1 \over Z_1^2} p^{D-2 a - 2 b} H_{0,0}(a,b)  (1,1,1,0,1) 
\eea 
\\

Next we calculate $\Pi^{21}$ and determine its components $\pi_i^{21}$.
One has 
\bea
&& \Pi^{21}_{\alpha \beta, \gamma \delta}(\p)
= {1 \over Z_1 Z_2} {\rm sym}_{\alpha \beta} {\rm sym}_{\delta \gamma}  \delta_{\beta \delta} 
I_{\alpha \gamma}(a,b+1)  \\
&& =
{1 \over Z_1 Z_2} p^{D-2 a - 2 b} 
{\rm sym}_{\alpha \beta} {\rm sym}_{\delta \gamma}  \delta_{\beta \delta} 
\left[\delta_{\alpha \gamma} H_{1,2}(a,b+1) +
 \hat p_{\alpha} \hat p_{\gamma} H_{0,2}(a,b+1) \right] \\
 && = {1 \over Z_1 Z_2} p^{D-2 a - 2 b}  [ \frac{1}{2} 
 (\delta_{\alpha \gamma} \delta_{\beta \delta} + \delta_{\alpha \delta} \delta_{\beta \gamma} )
 H_{1,2}(a,b+1)  + \frac{1}{4} (  \hat p_{\alpha} \hat p_{\gamma}  \delta_{\beta \delta} 
 + \hat p_{\beta} \hat p_{\gamma}  \delta_{\alpha \delta}
 + \hat p_{\alpha} \hat p_{\delta}  \delta_{\beta \gamma} 
 + \hat p_{\beta} \hat p_{\delta}  \delta_{\alpha \gamma} ) H_{0,2}(a,b+1)] \nonumber 
\eea
Using \eqref{inversion} we obtain
\bea
&& (D-1) \pi^{21}_3(\p) = P^T_{\alpha \beta}(\p) P^T_{\gamma \delta}(\p)
{1 \over Z_1 Z_2} p^{D-2 a - 2 b}  [ \frac{1}{2} 
 (\delta_{\alpha \gamma} \delta_{\beta \delta} + \delta_{\alpha \delta} \delta_{\beta \gamma} )
 H_{1,2}(a,b+1)  \\
 && + \frac{1}{4} (  \hat p_{\alpha} \hat p_{\gamma}  \delta_{\beta \delta} 
 + \hat p_{\beta} \hat p_{\gamma}  \delta_{\alpha \delta}
 + \hat p_{\alpha} \hat p_{\delta}  \delta_{\beta \gamma} 
 + \hat p_{\beta} \hat p_{\delta}  \delta_{\alpha \gamma} ) H_{0,2}(a,b+1)] \nn \\
&& = (D-1) {1 \over Z_1 Z_2} p^{D-2 a - 2 b}   H_{1,2}(a,b+1) 
\eea 
Next
\bea
\pi_5^{21}(\p) &=& P^L_{\alpha \beta}(\p) P^L_{\gamma \delta}(\p) {1 \over Z_1 Z_2} p^{D-2 a - 2 b}  [ \frac{1}{2} 
 (\delta_{\alpha \gamma} \delta_{\beta \delta} + \delta_{\alpha \delta} \delta_{\beta \gamma} )
 H_{1,2}(a,b+1)  \\
 && + \frac{1}{4} (  \hat p_{\alpha} \hat p_{\gamma}  \delta_{\beta \delta} 
 + \hat p_{\beta} \hat p_{\gamma}  \delta_{\alpha \delta}
 + \hat p_{\alpha} \hat p_{\delta}  \delta_{\beta \gamma} 
 + \hat p_{\beta} \hat p_{\delta}  \delta_{\alpha \gamma} ) H_{0,2}(a,b+1)] \nn \\
 && = {1 \over Z_1 Z_2} p^{D-2 a - 2 b} [ H_{1,2}(a,b+1) + H_{0,2}(a,b+1)] 
\eea
Next
\bea
\sqrt{D-1}\pi^{21}_4(\p)&=& 
P^L_{\alpha \beta}(\p) P^T_{\gamma \delta}(\p) {1 \over Z_1 Z_2} p^{D-2 a - 2 b}  [ \frac{1}{2} 
 (\delta_{\alpha \gamma} \delta_{\beta \delta} + \delta_{\alpha \delta} \delta_{\beta \gamma} )
 H_{1,2}(a,b+1)  \\
 && + \frac{1}{4} (  \hat p_{\alpha} \hat p_{\gamma}  \delta_{\beta \delta} 
 + \hat p_{\beta} \hat p_{\gamma}  \delta_{\alpha \delta}
 + \hat p_{\alpha} \hat p_{\delta}  \delta_{\beta \gamma} 
 + \hat p_{\beta} \hat p_{\delta}  \delta_{\alpha \gamma} ) H_{0,2}(a,b+1)] = 0
\eea
Next
\bea
&& \pi_2^{21}(\p) = \frac{1}{D-1} 
(W_2)_{\alpha \beta,\gamma\delta} 
\Pi^{21}_{\alpha \beta, \gamma \delta}(\p) \\
&& = \frac{1}{2(D-1)} ( 
P^T_{\alpha \gamma} P^L_{\beta \delta } 
+ P^T_{\alpha \delta} P^L_{\beta \gamma} + P^L_{\alpha \gamma}
P^T_{\beta \delta } + P^L_{\alpha \delta} P^T_{\beta \gamma})
{1 \over Z_1 Z_2} p^{D-2 a - 2 b}  [ \frac{1}{2} 
 (\delta_{\alpha \gamma} \delta_{\beta \delta} + \delta_{\alpha \delta} \delta_{\beta \gamma} )
 H_{1,2}(a,b+1)  \\
 && + \frac{1}{4} (  \hat p_{\alpha} \hat p_{\gamma}  \delta_{\beta \delta} 
 + \hat p_{\beta} \hat p_{\gamma}  \delta_{\alpha \delta}
 + \hat p_{\alpha} \hat p_{\delta}  \delta_{\beta \gamma} 
 + \hat p_{\beta} \hat p_{\delta}  \delta_{\alpha \gamma} ) H_{0,2}(a,b+1)] \nn \\
 && = {1 \over Z_1 Z_2} p^{D-2 a - 2 b}  [ H_{1,2}(a,b+1)  + \frac{1}{2} H_{0,2}(a,b+1)]
\eea
and finally we find,
\bea
&& \pi_1^{21}(\p) = \frac{2}{(D-2) (D+1)}  [ 
- \pi_3(\p) - \pi_5(\p) - (D-1) \pi_2(\p)  \\
&& 
+ {1\over 2}(\delta_{\alpha\gamma}\delta_{\beta\delta}
+\delta_{\alpha\delta}\delta_{\beta\gamma}) {1 \over Z_1 Z_2} p^{D-2 a - 2 b}  [ \frac{1}{2} 
 (\delta_{\alpha \gamma} \delta_{\beta \delta} + \delta_{\alpha \delta} \delta_{\beta \gamma} )
 H_{1,2}(a,b+1) \nn  \\
 && + \frac{1}{4} (  \hat p_{\alpha} \hat p_{\gamma}  \delta_{\beta \delta} 
 + \hat p_{\beta} \hat p_{\gamma}  \delta_{\alpha \delta}
 + \hat p_{\alpha} \hat p_{\delta}  \delta_{\beta \gamma} 
 + \hat p_{\beta} \hat p_{\delta}  \delta_{\alpha \gamma} ) H_{0,2}(a,b+1)] \nn \\
 && = \frac{2}{(D-2) (D+1)} {1 \over Z_1 Z_2} p^{D-2 a - 2 b}  [-    \bigg( H_{1,2}(a,b+1)  
 + H_{1,2}(a,b+1) + H_{0,2}(a,b+1) \\
 && + (D-1)  [ H_{1,2}(a,b+1)  + \frac{1}{2} H_{0,2}(a,b+1) \bigg)
 + \frac{D(D+1)}{2}  H_{1,2}(a,b+1)  + \frac{D+1}{2} H_{0,2}(a,b+1) ] \nn \\
&& = {1 \over Z_1 Z_2} p^{D-2 a - 2 b} [ H_{1,2}(a,b+1) + \frac{1}{D+1} H_{0,2}(a,b+1)]
\eea 

Hence to summarize
\bea
&& (\pi_1^{21}(\p),\pi_2^{21}(\p),\pi_3^{21}(\p),\pi_4^{21}(\p),\pi_5^{21}(\p)) \\
&& =
 {1 \over Z_1 Z_2} p^{D-2 a - 2 b}
 [ H_{1,2}(a,b+1)  (1,1,1,0,1) + 
 H_{0,2}(a,b+1) (\frac{1}{D+1},\frac{1}{2},0,0,1) ] \nn
\eea 
\\

Next we calculate $\Pi^{22}$ and determine its components $\pi_i^{22}$. We obtain
\bea
&& \pi^{22}_3(\p) = Z_2^{-2} (D+1) H_{2,4}(a+1,b+1) p^{D-2 a- 2b} \\
&& \sqrt{D-1} \pi^{22}_4(\p) = Z_2^{-2}  (D-1) (1 - 2 a - 2 b + D) H_{2,4}(a+1,b+1)  p^{D-2a-2b} \\
&& \pi_5^{22}(\p) = Z_2^{-2}  p^{D-2a-2b} \bigg[H_{1,2}(a+1,b+1) + H_{0,2}(a+1,b+1) - 6 H_{1,3}(a+1,b+1) \\
&& 
- 2 H_{0,3}(a+1,b+1) 
+3  H_{2,4}(a+1,b+1)
 + 6  H_{1,4}(a+1,b+1)
 + H_{0,4}(a+1,b+1) \bigg]  \nn \\
 && \pi_2^{22}(\p) = Z_2^{-2} 
 \left[2 H_{2,4}(a+1,b+1) + 2 H_{1,4}(a+1,b+1) + \frac{1}{2} H_{1,2}(a+1,b+1)
- 2 H_{1,3}(a+1,b+1)\right] p^{D-2a-2b} \nn \\
&& \pi_1^{22}(\p) = 2 Z_2^{-2}  H_{2,4}(a+1,b+1) p^{D-2a-2b}
\eea

Let us now put together the pieces. Defining $A=H_{2,4}(a+1,b+1)$, we obtain
\bea
&& \pi_1^{11}(\p) + 2 \pi_1^{21}(\p) + \pi_1^{22}(\p) \\
&& = p^{D-2a-2b} \big[
2 Z_2^{-2}  H_{2,4}(a+1,b+1) +
 {2 \over Z_1 Z_2} [ H_{1,2}(a,b+1) + \frac{1}{D+1} H_{0,2}(a,b+1)] +
 {1 \over Z_1^2} H_{0,0}(a,b) \big] \nn
\eea 

\bea
&& \pi_2^{11}(\p) + 2 \pi_2^{21}(\p) + \pi_2^{22}(\p) \\
&& = p^{D-2a-2b}  \bigg[ Z_2^{-2} 
 [2 H_{2,4}(a+1,b+1) + 2 H_{1,4}(a+1,b+1) + \frac{1}{2} H_{1,2}(a+1,b+1)
- 2 H_{1,3}(a+1,b+1)] \nn 
\\
&& + {2 \over Z_1 Z_2}  [ H_{1,2}(a,b+1)  + \frac{1}{2} H_{0,2}(a,b+1)]
+ {1 \over Z_1^2} H_{0,0}(a,b) \bigg] \nn
\eea 

\bea
&& \pi_3^{11}(\p) + 2 \pi_3^{21}(\p) + \pi_3^{22}(\p) \\
&& = p^{D-2a-2b}  \bigg[ 
Z_2^{-2} (D+1) H_{2,4}(a+1,b+1) +
{2 \over Z_1 Z_2}   H_{1,2}(a,b+1) +
 {1 \over Z_1^2} H_{0,0}(a,b) \bigg] \nn
\eea

\be
 \pi_4^{11}(\p) + 2 \pi_4^{21}(\p) + \pi_4^{22}(\p)  =   Z_2^{-2}  \sqrt{D-1}  (1 - 2 a - 2 b + D) H_{2,4}(a+1,b+1)  p^{D-2a-2b}
\ee

\bea
&& \pi_5^{11}(\p) + 2 \pi_5^{21}(\p) + \pi_5^{22}(\p) \\
&& = p^{D-2a-2b}  \bigg[ Z_2^{-2}   \big[H_{1,2}(a+1,b+1) + H_{0,2}(a+1,b+1) - 6 H_{1,3}(a+1,b+1) \nn \\
&& - 2 H_{0,3}(a+1,b+1) 
+3  H_{2,4}(a+1,b+1)
 + 6  H_{1,4}(a+1,b+1)
 + H_{0,4}(a+1,b+1) \big]  \nn \\
&& + {2 \over Z_1 Z_2} [ H_{1,2}(a,b+1) + H_{0,2}(a,b+1)] 
+ {1 \over Z_1^2} H_{0,0}(a,b) \bigg] \nn 
\eea
\\
%

Now, from the explicit formula for the function $H_{n,m}$ in \eqref{defH}, setting
$a=b=(2-\eta)/2$, and reexpressing $Z_1,Z_2$ as functions of $Z_\kappa,Z_g$
using \eqref{formG}, as well as
and $Z_2/Z_1=s/(1-s)$, where $s=Z_\kappa/Z_g$, we obtain 
the formula \eqref{piq} for $\pi_i(\q)$ together with the formula \eqref{Pieteeta2new} 
and \eqref{ais} for the amplitudes $a_i(\eta,D,s)$. 

%

\section{IV. Calculation of the self-energy integrals}

In this section we compute the integrals \eqref{sigmai0} arising in the
evaluation of the self-energy, but using a
different parametrization of the propagator as in \eqref{formG}, i.e. we compute
(using the invariance $\q \to - \q$)
\be \label{sigmai0new} 
\sigma^i_{\alpha \gamma}(\kb) \simeq
{2 \over d}   \int_q  \, 
(W_i)_{\alpha \beta, \gamma \delta}(\q) 
\frac{|\q|^{4-D - 2 \eta}}{|\kb - \q|^{2-\eta}} 
\left( \frac{1}{Z_1} \delta_{\beta \delta} + \frac{1}{Z_2} P_{\beta \delta}^L(\kb+ \q)
\right) \quad , \quad \frac{1}{Z_1}=\frac{1}{Z_g} \quad , \quad 
\frac{1}{Z_2}=\frac{1}{Z_\kappa}- \frac{1}{Z_g}
\ee
For convenience the result will be first parameterized as follows
\be \label{sigmaii}
\sigma^i_{\alpha \gamma}(\kb)    =
\frac{2}{d} k^{2-\eta} \frac{1}{Z_2} ( b_i^{(1)} \delta_{\alpha \gamma} 
+ 
b_i^{(2)}  \hat k_\alpha \hat k_\gamma )
\ee
and then at the very end will be translated in the form
\bea \label{sigmajj}
\sigma^i_{\alpha \gamma}(\kb)    
= \frac{2}{d} k^{2-\eta} \frac{1}{Z_\kappa} 
( b_i^{(1)}(\eta,D,s) P_{\beta \delta}^T(\kb) + b_i(\eta,D,s) P_{\beta \delta}^L(\kb) )
\eea
which will give the amplitudes $b_i^{(1)}(\eta,D,s)$ and 
$b_i(\eta,D,s)$ displayed in \eqref{theb}. Note that they
differ from the $b_i^{(1)}$ and $b_i^{(2)}$ above by
some simple factors, and some linear combinations.
We denote everywhere in this calculation
\bea
2a = 2-\eta \quad , \quad 2b=D+2 \eta-4
\eea 
We start with
\bea
&& \sigma^3_{\alpha \gamma}(\kb) = {2 \over d} \frac{1}{D-1}  \int_q 
P^T_{\alpha \beta}(\q) P^T_{\gamma \delta}(\q) 
( \frac{1}{Z_1} \delta_{\beta \delta} + \frac{1}{Z_2} 
P_{\beta \delta}^L(\kb + \q) ) \frac{q^{4-D-2 \eta} }{(\kb + \q)^{2-\eta}} 
= \sigma^{31}_{\alpha \gamma}(\kb) + \sigma^{32}_{\alpha \gamma}(\kb) \\
&& \sigma^{31}_{\alpha \gamma}(\kb) = {2 \over d} \frac{1}{D-1} \frac{1}{Z_1}  \int_q 
P^T_{\alpha \gamma}(\q)  \frac{1}{(\kb + \q)^{2a} q^{2b}} \\
&& \sigma^{32}_{\alpha \gamma}(\kb) = {2 \over d} \frac{1}{D-1}  \frac{1}{Z_2}  k_\beta k_\delta    \int_q 
P^T_{\alpha \beta}(\q) P^T_{\gamma \delta}(\q) 
 \frac{1}{(\kb + \q)^{2a+2} q^{2b}}
\eea
We use, in addition to \eqref{3integrals}
\bea
&& I_{\alpha_1 \alpha_2 \alpha_3}(a,b) =
-k^{D-2a-2b+3}\left[
(\hat{k}_{\alpha_1} \delta_{\alpha_2\alpha_3}+
  \hat{k}_{\alpha_2} \delta_{\alpha_3\alpha_1}+
  \hat{k}_{\alpha_3} \delta_{\alpha_1\alpha_2}) H_{1,3}(a,b)
+ \hat{k}_{\alpha_1}\hat{k}_{\alpha_2}\hat{k}_{\alpha_3} H_{0,3}(a,b) \right] \nonumber  
\eea
and 
\bea
&& I_{\alpha_1\alpha_2\alpha_3\alpha_4}(a,b)
= k^{D-2a-2b+4}\bigg[
(\delta_{\alpha_1 \alpha_2} \delta_{\alpha_3
    \alpha_4} + \delta_{\alpha_1 \alpha_3} \delta_{\alpha_2 \alpha_4}
  + \delta_{\alpha_1 \alpha_4} \delta_{\alpha_2 \alpha_3} ) H_{2,4}(a,b) \\
&& 
 + ( \hat{k}_{\alpha_1}\hat{k}_{\alpha_2}\delta_{\alpha_3\alpha_4}+
  \hat{k}_{\alpha_1}\hat{k}_{\alpha_3}\delta_{\alpha_2\alpha_4}+
  \hat{k}_{\alpha_1}\hat{k}_{\alpha_4}\delta_{\alpha_2\alpha_3}+\hat{k}_{\alpha_2}\hat{k}_{\alpha_3} \delta_{\alpha_1\alpha_4}+
  \hat{k}_{\alpha_2}\hat{k}_{\alpha_4} \delta_{\alpha_1\alpha_3}+
  \hat{k}_{\alpha_3}\hat{k}_{\alpha_4} \delta_{\alpha_1\alpha_2} )
 H_{1,4}(a,b) \nn
\\
&&  +\hat{k}_{\alpha_1}\hat{k}_{\alpha_2}\hat{k}_{\alpha_3}\hat{k}_{\alpha_4} H_{0,4}(a,b) \bigg] \nonumber  
\eea

We have
\bea
\sigma^{31}_{\alpha \gamma}(\kb) = {2 \over d} \frac{1}{D-1} \frac{1}{Z_1} k^{D-2 a - 2 b}
\bigg( (H_{0,0}(a,b) -  H_{1,2}(a,b+1) ) \delta_{\alpha \gamma} - 
 \hat k_{\alpha} \hat k_{\gamma} H_{0,2}(a,b+1) \bigg)
\eea
and
\bea
&&  \sigma^{32}_{\alpha \gamma}(\kb) = {2 \over d} \frac{1}{D-1}  \frac{1}{Z_2}  
k_\beta k_\delta   
 \bigg(I_{\alpha \beta \gamma \delta}(a+1,b+2) - I_{\alpha \beta}(a+1,b+1) \delta_{\gamma \delta} 
 - \delta_{\alpha \beta}   I_{\gamma \delta}(a+1,b+1) + \delta_{\alpha \beta} 
 \delta_{\gamma \delta}  I(a+1,b) \bigg) \nonumber \\
&& = {2 \over d} \frac{1}{D-1}  \frac{1}{Z_2}  k^{D-2 a - 2 b} \bigg(
\hat k_\alpha \hat k_\gamma
\big( H_{0,0}(a+1,b) - 2 H_{1,2}(a+1,b+1) - 2 H_{0,2}(a+1,b+1) 
\\
&& + H_{0,4}(a+1,b+2) + 2 H_{2,4}(a+1,b+2) + 5 H_{1,4}(a+1,b+2)  \big) \nn \\
&& + \delta_{\alpha \gamma} ( H_{2,4}(a+1,b+2) + H_{1,4}(a+1,b+2) )
\bigg) \nn
 \eea
We find that
\bea
\hat k_\alpha \hat k_\gamma \sigma^{32}_{\alpha \gamma}(\kb) =
{2 \over d} \frac{1}{D-1}  \frac{1}{Z_2}  k^{D-2 a - 2 b}  (D^2-1) 
H_{2,4}(a+1,b+2) = {2 \over d}  \frac{1}{Z_2} b_3^{\rm crumpling} k^{2-\eta}
\eea 
with $b_3^{\rm crumpling}=(D+1) \frac{\Sigma(\eta,D)}{D^2-1}$,
where we recall the definition \cite{LRReview} 
\bea
&&\Sigma(\eta,D) = \frac{\left(D^2-1\right) \Gamma
   (2-\eta) \Gamma
   \left(\frac{D}{2}+\frac{\eta}{2}\right
   ) \Gamma
   \left(\eta/2 \right
   )}{4 (4 \pi)^{D/2} \Gamma
   \left(2-\frac{{\eta}}{2}\right
   ) \Gamma
   \left(\frac{D}{2}+\eta\right) \Gamma
   \left(\frac{D}{2}-\frac{{\eta}}{2}+2 \right)} = (D^2-1) H_{2,4}(a+1,b+2) 
\eea 
With this we find
\bea
&& b_3^{(2)} = \frac{1}{D-1} \left( \frac{(D-2) D (D+\eta )}{D+\eta -2}+2 (\eta -1) 
- \frac{Z_2}{Z_1}  \frac{2 (\eta -2)^2 (D+2 \eta -2)}{D+\eta -2}
\right) \frac{\Sigma(\eta,D)}{D^2-1}  \\
&& b_3^{(1)} = \frac{1}{D-1}  \left( \frac{D+(5-2 \eta ) \eta -2}{D+\eta -2}
- \frac{Z_2}{Z_1} 
\frac{(D+2 \eta -2) \left(D^2+D (\eta -3)+(7-2 \eta )
   \eta -6\right)}{D+\eta -2} \right) \frac{\Sigma(\eta,D)}{D^2-1}
\eea 
Note that the sum simplifies into
\bea
b_3^{(1)}  + b_3^{(2)}  = \frac{1}{D-1} \left(D^2-1 - \frac{Z_2}{Z_1}  (D-1) (-2 + D + 2 \eta) \right)
\frac{\Sigma(\eta,D)}{D^2-1}
\eea 

Next, we have

\bea
&& \sigma^4_{\alpha \gamma}(\kb) = {2 \over d} \frac{1}{\sqrt{D-1}}  \int_q 
(P^T_{\alpha \beta}(\q) P^L_{\gamma \delta}(\q) + P^L_{\alpha \beta}(\q) P^T_{\gamma \delta}(\q))
( \frac{1}{Z_1} \delta_{\beta \delta} + \frac{1}{Z_2} 
P_{\beta \delta}^L(\kb + \q) ) \frac{q^{4-D-2 \eta} }{(\kb + \q)^{2-\eta}} 
= \sigma^{41}_{\alpha \gamma}(\kb) + \sigma^{42}_{\alpha \gamma}(\kb) \nn \\
&& \sigma^{41}_{\alpha \gamma}(\kb)  = 
{2 \over d} \frac{1}{\sqrt{D-1}}  \frac{1}{Z_2}   \int_q 
P^T_{\alpha \beta}(\q) P^L_{\gamma \delta}(\q) 
k_{\beta}  (k_\delta + q_\delta) \frac{1}{(\kb + \q)^{2a+2} q^{2b}}  \\
&& \sigma^{42}_{\alpha \gamma}(\kb)  = 
{2 \over d} \frac{1}{\sqrt{D-1}}  \frac{1}{Z_2}   \int_q 
 P^L_{\alpha \beta}(\q) P^T_{\gamma \delta}(\q) (k_{\beta} + q_\beta) k_\delta 
\frac{1}{(\kb + \q)^{2a+2} q^{2b}} 
\eea
We have
\be
 \sigma^{42}_{\alpha \gamma}(\kb)  = 
{2 \over d} \frac{1}{\sqrt{D-1}}  \frac{1}{Z_2} [  k_{\beta}  k_\delta \int_q 
P^T_{\alpha \beta}(\q) P^L_{\gamma \delta}(\q)  \frac{1}{(\kb + \q)^{2a+2} q^{2b}} 
+ k_{\delta} \int_q 
P^T_{\gamma \delta}(\q) q_\alpha \frac{1}{(\kb + \q)^{2a+2} q^{2b}} ] 
\ee
as well as
\bea
&& \sigma^{41}_{\alpha \gamma}(\kb)  = 
{2 \over d} \frac{1}{\sqrt{D-1}}  \frac{1}{Z_2} [  k_{\beta}  k_\delta \int_q 
P^T_{\alpha \beta}(\q) P^L_{\gamma \delta}(\q)  \frac{1}{(\kb + \q)^{2a+2} q^{2b}} 
+ k_{\beta} \int_q 
P^T_{\alpha \beta}(\q) q_\gamma \frac{1}{(\kb + \q)^{2a+2} q^{2b}} ] \\
&& = {2 \over d} \frac{1}{\sqrt{D-1}}  \frac{1}{Z_2} [  k_{\beta}  k_\delta
(\delta_{\alpha \beta} I_{\gamma \delta}(a+1,b+1) - I_{\alpha \beta \gamma \delta}(a+1,b+2) )
+ k_\beta \delta_{\alpha \beta} I_{\gamma}(a+1,b) - 
k_\beta I_{\alpha \beta \gamma}(a+1,b+1) ] \nonumber \\
&& = 
{2 \over d} \frac{1}{\sqrt{D-1}}  \frac{1}{Z_2} k^{D-2a-2b} [  \hat k_{\beta}  \hat k_\delta
\delta_{\alpha \beta}   \left[\delta_{\gamma \delta} H_{1,2}(a+1,b+1) +
 \hat k_{\gamma} \hat k_{ \delta} H_{0,2}(a+1,b+1) \right] \\
&& - \hat k_\beta \delta_{\alpha \beta} 
  \hat k_{\gamma} H_{0,1}(a+1,b) \nn \\
&& +
\hat k_\beta \left[
(\hat{k}_{\alpha} \delta_{\beta \gamma}+
  \hat{k}_{\beta} \delta_{\alpha \gamma}+
  \hat{k}_{\gamma} \delta_{\alpha \beta}) H_{1,3}(a+1,b+1)
+ \hat{k}_{\alpha}\hat{k}_{\beta}\hat{k}_{\gamma} H_{0,3}(a+1,b+1) \right]  \nn \\
&& -  \hat k_{\beta}  \hat k_\delta \bigg[
(\delta_{\alpha \beta} \delta_{\gamma \delta} + \delta_{\alpha \gamma} \delta_{\beta \delta}
  + \delta_{\alpha \delta} \delta_{\beta \gamma} ) H_{2,4}(a+1,b+2) \nn \\
&& 
 + ( \hat{k}_{\alpha}\hat{k}_{\beta}\delta_{\gamma \delta}+
  \hat{k}_{\alpha}\hat{k}_{\gamma}\delta_{\beta \delta}+
  \hat{k}_{\alpha}\hat{k}_{\delta}\delta_{\beta \gamma}+\hat{k}_{\beta}\hat{k}_{\gamma} \delta_{\alpha \delta}+
  \hat{k}_{\beta}\hat{k}_{\delta} \delta_{\alpha \gamma}+
  \hat{k}_{\gamma}\hat{k}_{\delta} \delta_{\alpha \beta} )
 H_{1,4}(a+1,b+2) \nn 
\\
&&  +\hat{k}_{\alpha}\hat{k}_{\beta}\hat{k}_{\gamma}\hat{k}_{\delta} H_{0,4}(a+1,b+2) \bigg] \nonumber  
\eea
This leads to 
\bea 
&& \sigma^{41}_{\alpha \gamma}(\kb)   =  {2 \over d} \frac{1}{\sqrt{D-1}}  \frac{1}{Z_2} k^{D-2a-2b} \bigg[  \hat k_\alpha \hat k_{\gamma} [ H_{1,2}(a+1,b+1) + H_{0,2}(a+1,b+1) - H_{0,1}(a+1,b) ] \\
&& +
\delta_{\alpha \gamma}  H_{1,3}(a+1,b+1) +  \hat k_\alpha \hat k_{\gamma} ( 2 H_{1,3}(a+1,b+1)
+ H_{0,3}(a+1,b+1) ) \\
&& 
- \delta_{\alpha \gamma} ( H_{2,4}(a+1,b+2)  + H_{1,4}(a+1,b+2) )
-  \hat k_\alpha \hat k_{\gamma} ( 2 H_{2,4}(a+1,b+2) + 5 H_{1,4}(a+1,b+2) + H_{0,4}(a+1,b+2) ) \bigg] \nonumber  
\eea
One observes that $\sigma^{42}_{\alpha \gamma}(\kb)  = \sigma^{41}_{\alpha \gamma}(\kb)$,
and
hence finds
\bea
&& b_4^1 = - \frac{2}{\sqrt{D-1}}  (2 \eta-3) \frac{\Sigma(\eta,D)}{D^2-1} \\
&& b_4^2 = \frac{2}{\sqrt{D-1}} D (2 \eta -3) \frac{\Sigma(\eta,D)}{D^2-1}
\eea 
with which we recover $b_4^1+b_4^2=2 \sqrt{D-1}  (2 \eta-3) \frac{\Sigma(\eta,D)}{D^2-1} = 
b_4^{\rm crumpling}$. 
Next one has
\bea
&& \sigma^5_{\alpha \gamma}(\kb) = {2 \over d}  \int_q 
P^L_{\alpha \beta}(\q) P^L_{\gamma \delta}(\q) 
( \frac{1}{Z_1} \delta_{\beta \delta} + \frac{1}{Z_2} 
P_{\beta \delta}^L(\kb + \q) ) \frac{q^{4-D-2 \eta} }{(\kb + \q)^{2-\eta}} 
= \sigma^{51}_{\alpha \gamma}(\kb) + \sigma^{52}_{\alpha \gamma}(\kb) \\
&& \sigma^{51}_{\alpha \gamma}(\kb) = {2 \over d}  \frac{1}{Z_1}  \int_q 
P^L_{\alpha \gamma}(\q)  \frac{1}{(\kb + \q)^{2a} q^{2b}} = {2 \over d}  \frac{1}{Z_1}   
I_{\alpha \gamma}(a,b+1) \nn  \\
&& = 
{2 \over d}  \frac{1}{Z_1}  k^{D-2a-2b}\left[\delta_{\alpha \gamma} H_{1,2}(a,b+1) +
 \hat k_{\alpha} \hat k_{ \gamma} H_{0,2}(a,b+1) \right] 
\eea
as well as
\bea
&& \sigma^{52}_{\alpha \gamma}(\kb) = {2 \over d}  \frac{1}{Z_2}      \int_q 
P^L_{\alpha \beta}(\q) P^L_{\gamma \delta}(\q) (k_\beta + q_\beta)(k_\delta + q_\delta)
 \frac{1}{(\kb + \q)^{2a+2} q^{2b}} \\
 && =  {2 \over d}  \frac{1}{Z_2}   \bigg[ k_\beta k_\delta  \int_q 
P^L_{\alpha \beta}(\q) P^L_{\gamma \delta}(\q) 
 \frac{1}{(\kb + \q)^{2a+2} q^{2b}} + 
 k_\delta  \int_q P^L_{\gamma \delta}(\q) q_\alpha
 \frac{1}{(\kb + \q)^{2a+2} q^{2b}} + 
 k_\beta  \int_q P^L_{\alpha \beta}(\q) q_\gamma
 \frac{1}{(\kb + \q)^{2a+2} q^{2b}} \nonumber \\
&& + \int_q 
q_{\alpha} q_{\gamma} \frac{1}{(\kb + \q)^{2a+2} q^{2b}} \bigg] \\
&& =  {2 \over d}  \frac{1}{Z_2}   \bigg[ k_\beta k_\delta I_{\alpha \beta \gamma \delta}(a+1,b+2)
+  k_\delta  I_{\alpha \gamma \delta}(a+1,b+1)
+  k_\beta  I_{\alpha \beta \gamma}(a+1,b+1) + I_{\alpha \gamma}(a+1,b) \bigg] \\
&& =  {2 \over d}  \frac{1}{Z_2}  k^{D-2a-2b} \bigg[ 
 \delta_{\alpha \gamma} ( H_{2,4}(a+1,b+2)  + H_{1,4}(a+1,b+2) )
\\
&& + \hat k_\alpha \hat k_{\gamma} ( 2 H_{2,4}(a+1,b+2) + 5 H_{1,4}(a+1,b+2) + H_{0,4}(a+1,b+2) ) \nn
\\
&& - 
\delta_{\alpha \gamma}  2 H_{1,3}(a+1,b+1) -  \hat k_\alpha \hat k_{\gamma} ( 4 H_{1,3}(a+1,b+1)
+ 2 H_{0,3}(a+1,b+1) ) \nn \\
&& + \delta_{\alpha \gamma} H_{1,2}(a+1,b) +
 \hat k_{\alpha} \hat k_{ \gamma} H_{0,2}(a+1,b) \bigg] \nn 
\eea
One obtains
\bea
&& b_5^1 =  \left(1 -2 D-2 \eta  + \frac{Z_2}{Z_1} (2 - D - 2 \eta) \right) \frac{\Sigma(\eta,D)}{D^2-1} \\
&& b_5^2 =  \left( 2 \left(\eta  \left(\frac{D}{D+\eta -2}+2 \eta
   -7\right)+5\right) + \frac{Z_2}{Z_1} \frac{2 (\eta -2)^2 (D+2 \eta -2)}{D+\eta -2} \right) \frac{\Sigma(\eta,D)}{D^2-1} 
\eea 
One checks that
\bea
b_5^1 + b_5^2|_{Z_1=+\infty} = 
\frac{-2 D^2+4 D \eta ^2-16 D \eta +15 D+4 \eta ^3-24
   \eta ^2+43 \eta -22}{D+\eta -2} \times \frac{\Sigma(\eta,D)}{D^2-1}  = b_5^{\rm crumpling}
\eea

Now we calculate

\bea
&& \sigma^2_{\alpha \gamma}(\kb) = {2 \over d} \frac{1}{2} \int_q 
(P^T_{\alpha \gamma}(\q) P^L_{\beta \delta }(\q) 
+ P^T_{\alpha \delta}(\q) P^L_{\beta \gamma}(\q) + P^L_{\alpha \gamma}(\q)
P^T_{\beta \delta }(\q) + P^L_{\alpha \delta}(\q) P^T_{\beta \gamma})(\q)
\\
&& \times ( \frac{1}{Z_1} \delta_{\beta \delta} + \frac{1}{Z_2} 
P_{\beta \delta}^L(\kb + \q) ) \frac{1}{(\kb + \q)^{2a} q^{2b}}
= \sigma^{21}_{\alpha \gamma}(\kb) + \sigma^{22}_{\alpha \gamma}(\kb) \nn
\eea
First 
\bea
&& \sigma^{21}_{\alpha \gamma}(\kb)  = {2 \over d} \frac{1}{2 Z_1} \int_q 
(P^T_{\alpha \gamma}(\q) 
+ (D-1) P^L_{\alpha \gamma}(\q)  )  \frac{1}{(\kb + \q)^{2a} q^{2b}}
= {2 \over d} \frac{1}{2 Z_1} (\delta_{\alpha \gamma} I(a,b) 
+ (D-2) I_{\alpha \gamma}(a,b+1) ) \\
&& = 
 {2 \over d} \frac{1}{2 Z_1} k^{D-2 a - 2 b} (\delta_{\alpha \gamma} 
  H_{0,0}(a,b)
 + (D-2) (\delta_{\alpha \gamma} H_{1,2}(a,b+1) +
 \hat k_{\alpha} \hat k_{ \gamma} H_{0,2}(a,b+1) )) \nn
\eea
and
\bea
&& \sigma^{22}_{\alpha \gamma}(\kb)  = {2 \over d} \frac{1}{2 Z_2} \int_q 
(P^T_{\alpha \gamma}(\q) P^L_{\beta \delta }(\q) 
+ P^T_{\alpha \delta}(\q) P^L_{\beta \gamma}(\q) \\
&& + P^L_{\alpha \gamma}(\q)
P^T_{\beta \delta }(\q) + P^L_{\alpha \delta}(\q) P^T_{\beta \gamma}(\q) )
(k_\beta + q_\beta)(k_\delta + q_\delta) 
  \frac{1}{(\kb + \q)^{2a+2} q^{2b}} \nn \\
  && = {2 \over d} \frac{1}{2 Z_2}  \bigg[ k_\beta k_\delta [A_{\alpha \beta \gamma \delta} + 3 perm ]
  + \int_q 
P^T_{\alpha \gamma}(\q)  \frac{1}{(\kb + \q)^{2a+2} q^{2b-2}} 
\\
&& + 2 k_\beta 
\int_q P^T_{\alpha \gamma}(\q) q_\beta  \frac{1}{(\kb + \q)^{2a+2} q^{2b}} 
 + 2 {\rm sym}_{\alpha \gamma} k_\delta 
\int_q P^T_{\alpha \delta}(\q) q_\gamma
 \frac{1}{(\kb + \q)^{2a+2} q^{2b}} \nn
\eea
where we have defined
\bea
A_{\alpha \beta \gamma \delta}  = \int_q 
P^T_{\alpha \gamma}(\q) P^L_{\beta \delta }(\q) 
  \frac{1}{(\kb + \q)^{2a+2} q^{2b}} = \delta_{\alpha \gamma} I_{\beta \delta}(a+1,b+1) 
  - I_{\alpha \beta \gamma \delta}(a+1,b+2) 
\eea 
One has
\bea
&& k_\beta k_\delta [A_{\alpha \beta \gamma \delta} + 3 perm ] =
\delta_{\alpha \gamma} k_\beta k_\delta I_{\beta \delta}(a+1,b+1)
\\
&& + k^2 I_{\alpha \gamma}(a+1,b+1)
+ 2 {\rm sym}_{\alpha \gamma} k_\alpha k_\beta I_{\beta \gamma}(a+1,b+1)
- 4 k_\beta k_\delta I_{\alpha \beta \gamma \delta}(a+1,b+2) \nn
\eea 
Hence we find
\bea
&& \sigma^{22}_{\alpha \gamma}(\kb)  = {2 \over d} \frac{1}{2 Z_2} \big[
\delta_{\alpha \gamma} k_\beta k_\delta I_{\beta \delta}(a+1,b+1)
\\
&& + k^2 I_{\alpha \gamma}(a+1,b+1)
+ 2 {\rm sym}_{\alpha \gamma} k_\alpha k_\beta I_{\beta \gamma}(a+1,b+1)
- 4 k_\beta k_\delta I_{\alpha \beta \gamma \delta}(a+1,b+2) \nn 
\\
&& + \delta_{\alpha \gamma} I(a+1,b-1) 
- I_{\alpha \gamma}(a+1,b) \nn \\
&& + 2 k_\beta  \delta_{\alpha \gamma} I_\beta(a+1,b) 
+ 2 {\rm sym}_{\alpha \gamma} k_\alpha I_\gamma(a+1,b)
- 4 k_\beta  I_{\alpha \gamma \beta}(a+1,b+1) \nn
\eea 
which leads to 
\bea
&& \sigma^{22}_{\alpha \gamma}(\kb)  = {2 \over d} \frac{1}{2 Z_2} k^{D-2a-2b} \big[
\delta_{\alpha \gamma} (H_{1,2}(a+1,b+1) + H_{0,2}(a+1,b+1) )
\\
&& + \delta_{\alpha \gamma} (H_{1,2}(a+1,b+1) - H_{1,2}(a+1,b)
+ H_{0,0}(a+1,b-1) - 2   H_{0,1}(a+1,b) ) \nn \\
&& +
 \hat k_{\alpha} \hat k_{ \gamma} \big( H_{0,2}(a+1,b+1) - H_{0,2}(a+1,b)  
 + 2 H_{1,2}(a+1,b+1) + 2 H_{0,2}(a+1,b+1) - 2 H_{0,1}(a+1,b) \big) \nn \\
&& +
\delta_{\alpha \gamma}  4 H_{1,3}(a+1,b+1) +  \hat k_\alpha \hat k_{\gamma} ( 8 H_{1,3}(a+1,b+1)
+ 4 H_{0,3}(a+1,b+1) )  \nn \\
&& 
- 4 \delta_{\alpha \gamma} ( H_{2,4}(a+1,b+2)  + H_{1,4}(a+1,b+2) )
-  4 \hat k_\alpha \hat k_{\gamma} ( 2 H_{2,4}(a+1,b+2) + 5 H_{1,4}(a+1,b+2) + H_{0,4}(a+1,b+2) ) 
\bigg] \nonumber
\eea 
We find
\bea
&& b_2^1 = \left( -D^2-\frac{(D-2) D}{D+\eta -2}-(D+8) \eta +3 D+2 \eta^2+8 
+ \frac{Z_2}{Z_1} \frac{(D+2 \eta -2) \left(-D^2-(D+3) \eta +3 D+\eta
   ^2+2\right)}{D+\eta -2} \right) \times \frac{\Sigma(\eta,D)}{D^2-1} \nonumber 
\\
&& b_2^2 = \left( \frac{D^2 (2 \eta -3)-D (\eta -1) \eta -2 (\eta -3)
   (\eta -2) (\eta -1)}{D+\eta -2}
   + \frac{Z_2}{Z_1} \frac{(D-2) (\eta -2)^2 (D+2 \eta -2)}{D+\eta -2} \right) \times \frac{\Sigma(\eta,D)}{D^2-1} 
\eea 
and we can check that
\bea
b_2^1 + b_2^2|_{Z_1=+\infty} = -\frac{(D-1) \left(D^2+2 \eta -4\right)}{D+\eta -2} \times \frac{\Sigma(\eta,D)}{D^2-1} 
= b_2^{\rm crumpling}
\eea 

Finally, let us denote
\bea \label{sum} 
\tilde b_1 = b_1 + b_3 + b_5 + b_2 
\eea 
Using the inversion formula \eqref{inversion},
it can be extracted from 
\bea
&& \sigma^1_{\alpha \gamma}(\kb) = {2 \over d} \frac{1}{2} \int_q 
(\delta_{\alpha\gamma}\delta_{\beta\delta}
+\delta_{\alpha\delta}\delta_{\beta\gamma}) ( \frac{1}{Z_1} \delta_{\beta \delta} + \frac{1}{Z_2} 
P_{\beta \delta}^L(\kb + \q) ) \frac{1}{(\kb + \q)^{2a} q^{2b}}
= \sigma^{11}_{\alpha \gamma}(\kb) + \sigma^{12}_{\alpha \gamma}(\kb)
\eea
with
\bea
&& \sigma^{11}_{\alpha \gamma}(\kb) = {2 \over d} \frac{1}{2 Z_1}   
\delta_{\alpha \gamma} (D+1) k^{D-2 a - 2 b} H_{0,0}(a,b)
\eea
and
\bea
&& \sigma^{12}_{\alpha \gamma}(\kb) = {2 \over d} \frac{1}{2 Z_2} 
k^{D-2 a - 2 b} (\delta_{\alpha\gamma} H_{0,0}(a,b)
+  \int_q (k_\alpha + q_\alpha)  (k_\gamma + q_\gamma) 
\frac{1}{(\kb + \q)^{2a+2} q^{2b}} )
\\
&& = {2 \over d} \frac{1}{2 Z_2} 
k^{D-2 a - 2 b} (\delta_{\alpha\gamma} H_{0,0}(a,b)
+ \hat k_\alpha \hat k_\gamma H_{0,0}(a+1,b)
- 2 \hat k_\alpha \hat k_\gamma H_{0,1}(a+1,b) 
+ \delta_{\alpha \gamma} H_{1,2}(a+1,b) +
 \hat k_{\alpha} \hat k_{ \gamma} H_{0,2}(a+1,b) ) \nonumber \\
&& = {2 \over d} \frac{1}{2 Z_2} 
k^{D-2 a - 2 b} (\delta_{\alpha\gamma} H_{0,0}(a,b)
+  I_{\alpha \gamma}(b,a+1) ) \\
&& = {2 \over d} \frac{1}{2 Z_2} 
k^{D-2 a - 2 b} (\delta_{\alpha\gamma} H_{0,0}(a,b)
+ \delta_{\alpha \gamma} H_{1,2}(b,a+1) +
 \hat k_{\alpha} \hat k_{ \gamma} H_{0,2}(b,a+1) )
\eea
We checked that these are two equivalent expressions. 

Hence
\bea
&& \tilde b^1_1 = \left( -\frac{(D-\eta +4) (D+2 \eta -4) (D+2 \eta -2)}{2
   (D+\eta -2)} -  \frac{Z_2}{Z_1} \frac{(D+1) (D-\eta +2) (D+2 \eta -4) (D+2 \eta -2)}{2
   (D+\eta -2)}
   \right) \times \frac{\Sigma(\eta,D)}{D^2-1} \nonumber \\
&& \tilde b^2_1 =  \frac{1}{2} (D+2 \eta -4) (D+2 \eta -2)   \times \frac{\Sigma(\eta,D)}{D^2-1} 
\eea
One finds
\bea
\tilde b^1_1 + \tilde b^2_1|_{Z_1=+\infty} = \frac{(\eta -3) (D+2 \eta -4) (D+2 \eta -2)}{D+\eta -2} \times \frac{\Sigma(\eta,D)}{D^2-1} 
\eea 
which recovers exactly the $\tilde b_1^{\rm crumpling}$. 

Reexpressing all the results using $Z_\kappa$, $Z_g$ and $s=Z_\kappa/Z_g$,
using the relation \eqref{sum} between $\tilde b_1^j$ and $b_1^j$ and
expressing the $b_i^{(1,2)}$ in terms of $b^1(\eta,D,s)$ and $b(\eta,D,s)$
one finally obtains the amplitudes \eqref{theb}.

\section{Remark}

Let us recall, as mentioned in Section II, that  in \eqref{sigmaii}, \eqref{sigmajj} 
an isotropic, $\delta_{\alpha \gamma}$ component is always generated 
for the inverse propagator of the vector field $\vec t_\alpha$,
even in the $s \to 0$ limit when the propagator is purely longitudinal. 
Here we note that this is innocuous at the $s=0$ fixed point, up to boundary terms. 
Namely, for $\vec t_\alpha = \partial_\alpha \vec r$, by integration
by part we have 
\bea
&& \int d^D x (\partial_\alpha \vec t_\beta)^2  = 
\int d^D x (\partial_\alpha \partial_\beta \vec r)^2 = 
\int d^D x (\partial^2_\alpha \vec r)^2 + \int d^D x ~ [ \partial_\alpha (\partial_\beta \vec r \partial_{\alpha} \partial_{ \beta} \vec r) -
\partial_\beta (\partial_\beta \vec r \partial_\alpha^2 \vec r) ] \\
&& = \int d^D x (\partial_\alpha \vec t_\alpha)^2 
+ \int d^D x ~ [ \partial_\alpha (\vec t_\beta \cdot \partial_{\alpha} \vec t_{ \beta} ) -
\partial_\beta (\vec t_\beta  \cdot \partial_\alpha \vec t_\alpha) ]
\eea

%
%
\end{widetext}


\begin{thebibliography}{}

\bibitem{NP} {\it Fluctuations in membranes with crystalline and
    hexatic order}, D.~R.~Nelson and L.~Peliti, {\it J.~Phys.} (Paris)
  {\bf 48}, 1085 (1987).

\bibitem{AL} {\it Fluctuations of Solid Membranes}, J.~A.~Aronovitz
  and T.~C.~Lubensky, {\it Phys.~Rev.~Lett.} {\bf 60}, 2634 (1988);
    
\bibitem{CrumplingBucklingGuitter} {\it Crumpling and Buckling
    Transitions in Polymerized Membranes}, E.~Guitter, F.~David,
  S.~Leibler, and L.~Peliti, {\em Phys. Rev. Lett.} {\bf 61} 2949
  (1988).

\bibitem{GDLP} {\it Crumpling transition in elastic membranes:
    renormalization group treatment}, F.~David and E.~Guitter, {\it
    Europhys.~Lett.} {\bf 5}, 709 (1988); {\it Thermodynamical
    behavior of polymerized membranes}, E.~Guitter, F.~David,
  S.~Leibler, and L.~Peliti, {\it J.~Phys.} (Paris) {\bf 50}, 1789
  (1989).

\bibitem{LRprl} {\it Self-consistent theory of polymerized membranes},
  P.~Le Doussal and L.~Radzihovsky, {\it Phys.~Rev.~Lett.}  {\bf 69},
  1209 (1992).

\bibitem{LRrapid} {\it Flat glassy phases and wrinkling of polymerized
    membranes with long-range disorder}, P.~Le Doussal and
  L.~Radzihovsky, {\it Phys.~Rev. B} {\bf 48} {\it Rapid Comm.} 3548
  (1993).

\bibitem{GuitterMC} {\it Stretching and buckling of polymerized
    membranes: a Monte Carlo study}, E. Guitter, S. Leibler,
  A. C. Maggs, and F. David, {\em J Phys} {\bf 51}, 1055-1060 (1990).

\bibitem{Jerusalem} For a review, and extensive references, see the
  articles in {\it Statistical Mechanics of Membranes and Interfaces},
  2nd edition, edited by D.~R.~Nelson, T.~Piran, and S.~Weinberg
  (World Scientific, Singapore, 1989).
  
\bibitem{Bensimon} {\it Wrinkling transition in partially polymerized
    vesicles}, M. Mutz, D. Bensimon, and M. J. Brienne, {\em
    Phys. Rev. Lett.} {\bf 67} 923 (1991).

\bibitem{NRlett} {\it Polymerized Membranes with Quenched Random
    Internal Disorder}, D.~R.~Nelson and L.~Radzihovsky, {\it
    Europhys.~Lett.}  {\bf 16}, 79 (1991).

\bibitem{RNpra} {\it Statistical mechanics of randomly polymerized
    membranes}, L.~Radzihovsky and D.~R.~Nelson, {\it Phys.~Rev.~A}
  {\bf 44}, 3525 (1991).


\bibitem{ML} {\it Curvature disorder in tethered membranes: A new flat
    phase at $T=0$}, D.~C.~Morse and T.~C.~Lubensky, {\it
    Phys.~Rev.~A} {\bf 46}, 1751 (1992).

\bibitem{RTtubule} A generalization to anisotropic in-plane elasticity
  was considered and extensively explored in {\it A New Phase of
    Tethered Membranes: Tubules}, Leo Radzihovsky and John Toner, {\em
    Phys. Rev. Lett.} {\bf 75}, 4752 (1995); {\it Elasticity, Shape
    Fluctuations and Phase Transitions in the New Tubule Phase of
    Anisotropic Tethered Membranes}, {\em Phys. Rev. E} {\bf 57},
  1832-1863 (1998).
     
\bibitem{LRReview} {\it Anomalous elasticity, fluctuations and
    disorder in elastic membranes}, P. Le Doussal and L. Radzihovsky,
  arXiv:1708.05723, {\em Annals of Physics} {\bf 392}, 340-410 (2018).


\bibitem{Geim2004} {\it Electric Field Effect in Atomically Thin
    Carbon Films}, K. S. Novoselov, A. K. Geim, S. V. Morozov,
  D. Jiang, Y. Zhang, S. V. Dubonos, I. V. Grigorieva, and
  A. A. Firsov, {\em Science} {\bf 306}, 666 (2004).


\bibitem{GeimMacDonald} {\it Graphene: Exploring carbon flatland},
  Andrey K. Geim and Allan H. MacDonald, {\em Physics Today}, August
  (2007).

\bibitem{reviewRMPGraphene} {\it The electronic properties of
    graphene}, A. H. Castro Neto, F. Guinea, N. M. R. Peres,
  K. S. Novoselov, and A. K. Geim {\em Rev. Mod. Phys.} {\bf 81}, 109
  (2009).

\bibitem{suspendGrapheneNature2007} {\it The structure of suspended
    graphene sheets}.  J. C. Meyer, A. K. Geim, M. I. Katsnelson,
  K. S. Novoselov, T. J. Booth, and S. Roth, {\em Nature} {\bf 446},
  60 (2007).

\bibitem{Hohenberg} {\em Existence of Long-Range Order in One and Two
    Dimensions}, P. Hohenberg, {\em Phys. Rev.} {\bf 158}, 383 (1967).

\bibitem{MerminWagner} {\it Absence of Ferromagnetism or
    Antiferromagnetism in One- or Two-Dimensional Isotropic Heisenberg
    Models}, N. D. Mermin and H. Wagner, {\em Phys. Rev. Lett.} {\bf
    17}, 1133 (1966).

\bibitem{Coleman} {\it There are no Goldstone bosons in two
    dimensions}, S. Coleman, {\em Commun. Math. Phys.} {\bf 31}, 259
  (1973).


\bibitem{criticalMatterLR} {\it Critical Matter}, Leo Radzihovsky
  arXiv:2306.03142, chapter to be published in World Scientific, as
  "50 years of the renormalization group", dedicated to the memory of
  Michael E. Fisher, edited by Amnon Aharony, Ora Entin-Wohlman, David
  Huse, and Leo Radzihovsky.
    
    
\bibitem{Gazit} {\it Structure of physical crystalline membranes
    within the self-consistent screening approximation}, D. Gazit,
  {\em Phys Rev E} {\bf 80} (4), 041117 (2009).

    
\bibitem{MouhannaTwoLoopFlat} {\em The flat phase of polymerized
    membranes at two-loop order}, O. Coquand, D. Mouhanna, S. Teber,
  arXiv:2003.13973, {\em Phys. Rev. E} {\bf 101}, 062104 (2020).


\bibitem{Burmistrovlarged} {\it Absolute Poisson's ratio and the
    bending rigidity exponent of a crystalline two-dimensional
    membrane}, Saykin, D. R., Gornyi, I. V., Kachorovskii, V. Y.,
  Burmistrov, I. S.  Annals of Physics, 414, 168108 (2020).

\bibitem{MouhannaThreeLoopFlat} {\it Three-loop order approach to flat
    polymerized membranes}, Metayer, S., Mouhanna, D., Teber,
  S. Physical Review E, 105(1), L012603 (2022).

\bibitem{simulationsGraphene} {\it Scaling behavior and strain dependence
  of in-plane elastic properties of graphene}, J. H. Los, A. Fasolino,
  M. I. Katsnelson, {\em Phys. Rev. Lett.} {\bf 116}, 015901 (2016).

\bibitem{experimentElasticModuli} {\it Increasing the elastic modulus
    of graphene by controlled defect creation}, G. L\'opez-Pol\'in,
  C. G\'omez-Navarro, V. Parente, F. Guinea, M. I. Katsnelson,
  F. P\'erez-Murano, J. G\'omez-Herrero, {\em Nature Physics} {\bf
    11}, 26-31 (2015).

\bibitem{NelsonKantorCrumplingTransition} {\it Crumpling transition in
    polymerized membranes}, Y. Kantor and D. R.  Nelson, {\it
    Phys. Rev. Lett.} {\bf 58}, 2774 (1987)

\bibitem{KantorNelson} {\it Phase transitions in flexible polymeric
    surfaces}, Y.~Kantor and D.~R.~Nelson, {\it Phys. Rev.~A} {\bf
    38}, 4020 (1987).
 

  
\bibitem{PKN} {\it Landau theory of the crumpling transition},
  M. Paczuski, M. Kardar and D.R. Nelson, {\it Phys. Rev. Lett.}  {\bf
    60}, 2638 (1988).


\bibitem{ALGolubovic} {\it Fluctuations and lower critical dimensions
    of crystalline membranes}, J.~A.~Aronovitz, L.~Golubovi\'c, and
  T.~C.~Lubensky, {\it J.~Phys.} (Paris) {\bf 50}, 609 (1989).
  
\bibitem{PaczuskiCrumplingLarged} {\it Renormalization-group analysis
    of the crumpling transition in large $d$}, M.  Paczuski and
  M. Kardar, {\it Phys.  Rev.  A}, {\bf 39}, 6086 (1989).

\bibitem{Mouhanna1} {\it Crumpling transition and flat phase of
    polymerized phantom membranes} J.-P. Kownacki, D. Mouhanna, {\em
    Phys. Rev. E} {\bf 79}, 040101 (2009).
    
\bibitem{MouhannaCrumpling}
{\it First order phase transitions in polymerized phantom membranes},
K. Essafi, J.-P. Kownacki, D. Mouhanna, arXiv:1402.0426
{\em Phys. Rev. E} {\bf 89}, 042101 (2014).

\bibitem{OurBuckling} {\it Thermal buckling transition of crystalline
    membranes in a field}, P. Le Doussal and L. Radzihovsky, {\it
    Phys.  Rev. Lett.}, {\bf 127}, 015702 (2021). See Supp. Mat. in
  ArXiv:2102.08970.

\bibitem{perforatedCrumple} {\it Thermal crumpling of perforated
    two-dimensional sheets}, D. Yllanes, S. S.  Bhabesh, D. R.
  Nelson, M. J.  Bowick, {\it Nature Comm.}, {\bf 8}, 1381 (2017).

\bibitem{VicariNM2001} See, {\it Large-n critical behavior of
    $O (n) \times O (m)$ spin models}, A. Pelissetto, P. Rossi,
  E. Vicari, {\it Nuclear Physics B}, {\bf 607}(3), 605-634 (2001),
  and references therein.
 
\bibitem{spinorBEC} {\it Degenerate quantum gases with
  spin-orbit coupling: a review}, H. Zhai,  {\it Reports on Progress in Physics},
  {\bf 78}(2), 026001 (2015).

\bibitem{MolecularBECChoi}. {\it Finite-momentum superfluidity and
    phase transitions in a p-wave resonant Bose gas}, S. Choi and
  L. Radzihovsky, {\it Phys. Rev. A} {\bf 84}, 043612 (2011).

\bibitem{He3book} {\it The superfluid phases of helium 3},
  D. Vollhardt, P. Wolfle, (2013).. Courier Corporation.

\bibitem{FFLOlr} {\it Fluctuations and phase transitions in
    Larkin-Ovchinnikov liquid-crystal states of a
    population-imbalanced resonant Fermi gas}.  L. Radzihovsky, {\it
    Phys.  Rev. A} {\bf 84}, 023611 (2011).

\bibitem{magnets} {\it A renormalization-group study of helimagnets in
    $D= 2+ \epsilon$ dimensions}, P. Azaria, B. Delamotte, F. Delduc,
  T.  Jolicoeur, {\it Nuclear Physics B} {\bf408}(3), 485-511 (1993).

\bibitem{SpinSmectic} {\it $O(N)$ smectic $\sigma$-model}, Tzu-Chi
  Hsieh, Leo Radzihovsky, arXiv:2310.13046.

\bibitem{AharonyFisher} {\it Critical Behavior of Magnets with Dipolar
    Interactions. I. Renormalization Group near Four Dimensions},
  A. Aharony and M. E. Fisher, {\it Phys. Rev. B} {\bf 8}, 3323 (1973);
  A. Aharony, ibid. {\bf 8}, 3342 (1978).
       
\bibitem{FreyIsotropicDipolar} {\it Renormalized field theory for the
    static crossover in isotropic dipolar ferromagnets}, E. Frey and
  F. Schwabl, {\it Phys.  Rev. B} {\bf 43}(1), 833 (1991).
  
\bibitem{NovelClarkFerroNematic} {\it First-principles experimental
    demonstration of ferroelectricity in a thermotropic nematic liquid
    crystal: Polar domains and striking electro-optics}, X. Chen,
  E. Korblova, D.  Dong, X.  Wei, R. Shao, L. Radzihovsky, M. Glaser,
  J. Maclennan, D. Bedrov, D. Walba, N. A. Clark, N. A. (2020), {\it
    Proceedings of the National Academy of Sciences} {\bf 117}(25),
  14021-14031 (2020).
 
\bibitem{Bray} {\it Self-Consistent Screening Calculation of the
    Critical Exponent $\eta$}, A.~J.~Bray, {\it Phys.~Rev.~Lett.} {\bf
    32}, 1413 (1974).
  
\bibitem{LRslits} P. Le Doussal and L. Radzihovsky, in preparation.
  
\bibitem{KT} {\it Ordering, metastability and phase transitions in
    two-dimensional systems}, J.M. Kosterlitz and D.J. Thouless, {\it
    J. Phys. C: Solid State Phys.} {\bf 6} 1181 (1973);

\bibitem{HNmelting} {\it Theory of two-dimensional melting},
  B.I. Halperin and D.R. Nelson, {\it Phys. Rev. Lett.} {\bf 41}, 121
  (1978).

\bibitem{Youngmelting} {\it Melting and the vector Coulomb gas in two
    dimensions}, A. P. Young, {\it Phys. Rev. B} {\bf 19}, 1855
  (1979).

\bibitem{SM} See supplementary material.

\bibitem{LRelastomer} {\it Nonlinear Elasticity, Fluctuations and
    Heterogeneity of Nematic Elastomers}, Xiangjun Xing and Leo
  Radzihovsky, {\em Annals of Physics} {\bf 323}, 105-203 (2008); {\it
    Phases and Transitions in Phantom Nematic Elastomer Membranes},
  {\em Phys. Rev. E} {\bf 71}, 011802 (2005); {\it Thermal
    fluctuations and anomalous elasticity of homogeneous nematic
    elastomers}, {\em Europhysics Letters} {\bf 61}, 769 (2003); {\it
    Universal Elasticity and Fluctuations of Nematic Gels} {\em
    Phys. Rev. Lett.} {\bf 90}, 168301 (2003).

\bibitem{LubenskyElastomer} {\it Anomalous elasticity of nematic
    elastomers}, O. Stenull and T. C. Lubensky, {\it Europhys. Lett.}
  {\bf 61}, 776 (2003); {\it Phys. Rev. E} {\bf 69}, 021807 (2004).

\bibitem{Mouhanna} {\it Auxiliary fields approach to shift-symmetric
    theories: the $\phi^4$ derivative theory and the crumpled-to-flat
    transition of membranes at two-loop order}, L. Delzescaux,
  C. Duclut, D. Mouhanna, M. Tissier, arXiv:2307.00600, {\it
    Phys. Rev. D} {\bf 108}, L081702 (2023).




%


%
%
%
%
%
%
%
%
%

%
%

%
%
%
%
%
%
%
%
%
%
%
%
%
%
%
%
%

%
%
%
%
%
%
%
%
%
%
%
%
%
%
%
%
%

%
%
%
%
%
%
%
%
%
%
%
%
%
%
%
%
%
%
%
%
%
%
%
%
%
%
%
%
%
%
%
%
%
%
%
%
%
%
%
%
%
%
%
%
%
%
%
%
%
%
%
%
%
%
%
%
%
%
%
%
%
%
%
%
%
%
%

\end{thebibliography}
\end{document}